\documentclass{aa}  
\usepackage{natbib}
\usepackage{graphicx}
\usepackage{footmisc}
\usepackage{float}
\usepackage{caption}
\usepackage[caption=false]{subfig}
\usepackage{amsmath}
\usepackage{nccmath}
\usepackage{color}
\usepackage{array} 
\usepackage{ulem}
\usepackage[flushleft]{threeparttable}
\DeclareTextFontCommand{\textroman}{\fontlibertine}
\usepackage{txfonts}

\hypersetup{draft}
\begin{document}

   \title{Atmospheric characterization of the ultra-hot Jupiter MASCARA-2b/KELT-20b}

   \subtitle{Detection of Ca$\mathrm{II}$, Fe$\mathrm{II}$, Na$\mathrm{I}$, and the Balmer series of $\mathrm{H}$ ($\mathrm{H}\alpha$, $\mathrm{H}\beta$, and $\mathrm{H}\gamma$) with high-dispersion transit spectroscopy}

   \author{N. Casasayas-Barris\inst{1,2} \and E. Pallé\inst{1,2} \and F. Yan\inst{3}  \and G. Chen\inst{1,2,4} \and S. Kohl\inst{5} \and M. Stangret\inst{1,2}   \and  H. Parviainen\inst{1,2} \and Ch. Helling\inst{6,7,8} \and N. Watanabe\inst{9,10}  \and S. Czesla\inst{5} \and A. Fukui\inst{14,1} \and P. Montañés-Rodríguez\inst{1,2} \and E. Nagel\inst{5} \and  N. Narita\inst{9,11,12,1} \and L. Nortmann\inst{1,2} \and G. Nowak\inst{1,2} \and J.H.M.M. Schmitt\inst{5} \and M.R. Zapatero Osorio\inst{13} 
}


   \institute{Instituto de Astrofísica de Canarias, Vía Láctea s/n, E-38205 La Laguna, Tenerife, Spain
              \\
              \email{nuriacb@iac.es}
         \and
             Departamento de Astrofísica, Universidad de La Laguna, Spain
        \and
            Institut für Astrophysik, Georg-August-Universität, Friedrich- Hund-Platz 1, 37077 Göttingen, Germany
         \and
             Key Laboratory of Planetary Sciences, Purple Mountain Observatory, Chinese Academy of Sciences, Nanjing 210008, China
         \and
             Hamburger Sternwarte, Universität Hamburg, Gojenbergsweg 112, 21029 Hamburg, Germany
        \and 
            Centre for Exoplanet Science, University of St Andrews, St Andrews, UK
        \and
             SUPA, School of Physics \& Astronomy, University of St Andrews, St Andrews, KY16 9SS, UK
        \and
             SRON Netherlands Institute for Space Research, Sorbonnelaan 2, 3584 CA Utrecht, NL
        \and 
            National Astronomical Observatory of Japan, 2-21-1 Osawa, Mitaka, Tokyo 181-8588, Japan
        \and 
            SOKENDAI (The Graduate University of Advanced Studies), 2-21-1 Osawa, Mitaka, Tokyo 181-8588, Japan
        \and
             Astrobiology Center, 2-21-1 Osawa, Mitaka, Tokyo 181-8588, Japan
        \and
             JST, PRESTO, 2-21-1 Osawa, Mitaka, Tokyo 181-8588, Japan
        \and
             Centro de Astrobiología (CSIC-INTA), Carretera de Ajalvir km 4, 28850 Torrejón de Ardoz, Madrid, Spain.
        \and 
            Department of Earth and Planetary Science, Graduate School of Science, The University of Tokyo, 7-3-1 Hongo, Bunkyo-ku, Tokyo 113-0033, Japan
        }


   \date{Received Month 00, 2017; accepted Month 00, 2017}

 
  \abstract{
 Ultra hot Jupiters orbit very close to their host star and consequently receive strong irradiation that makes their atmospheric chemistry different from the common gas giants. Here, we study the atmosphere of one of these particular hot planets, MASCARA-2b / KELT-20b, using four transit observations with high resolution spectroscopy facilities. Three of these observations were performed with HARPS-N and one with CARMENES. Additionally, we simultaneously observed one of the transits with MuSCAT2 to monitor possible spots in the stellar surface. At high resolution, the transmission residuals show the effects of Rossiter-McLaughlin and center-to-limb variations from the stellar lines profiles, which we correct to finally extract the transmission spectra of the planet. We clearly observe the absorption features of CaII, FeII, NaI, H$\alpha$ and H$\beta$ in the atmosphere of MASCARA-2b, and indications of H$\gamma$ and MgI at low signal to noise. In the case of NaI, the true absorption is difficult to disentangle from the strong telluric and interstellar contamination. The results obtained with CARMENES and HARPS-N are consistent, measuring an H$\alpha$ absorption depth of $0.68\pm0.05~\%$ and $0.59\pm0.07~\%$, and NaI absorption of $0.11\pm0.04~\%$ and $0.09\pm0.05~\%$ for a $0.75~\mathrm{\AA}$ passband, in both instruments respectively. The H$\alpha$ absorption corresponds to $\sim1.2~R_p$, which implies an expanded atmosphere, as a result of the gas heating caused by the irradiation received from the host star. For H$\beta$ and H$\gamma$ only HARPS-N covers this wavelength range, measuring an absorption depth of $0.28\pm0.06~\%$ and $0.21\pm0.07~\%$, respectively. For CaII, only CARMENES covers this wavelength range measuring an absorption depth of $0.28\pm0.05~\%$, $0.41\pm0.05~\%$ and $0.27\pm0.06~\%$ for CaII $\lambda8498\mathrm{\AA}$, $\lambda8542\mathrm{\AA}$ and $\lambda8662\mathrm{\AA}$ lines, respectively.
 Three additional absorption lines of FeII are observed in the transmission spectrum by HARPS-N (partially covered by CARMENES), measuring an average absorption depth of $0.08\pm0.04~\%$ ($0.75~\mathrm{\AA}$ passband). The results presented here are consistent with theoretical models of ultra hot Jupiters atmospheres, suggesting the emergence of an ionised gas on the day-side of such planets. Calcium and iron, together with other elements, are expected to be singly ionised at these temperatures and be more numerous than its neutral state. The Calcium triplet lines are detected here for the first time in transmission in an exoplanet atmosphere.
 }


   \maketitle
%
\section{Introduction}
\label{intro}

The study of Hot Jupiters atmospheres has revealed a broad composition and structural diversity in gas-rich planets (\citealt{Jensen2012}, \citealt{2012Sing}, \citealt{Crossfield2015}, \citealt{DemingSeager2017}, \citealt{Madhu2016SSRv..205..285M}). When these planets are under extreme conditions, in particular, located very close to their host stars and consequently being strongly irradiated, their temperatures increase and lead to different atmospheric chemistry between the day and night-sides of atmosphere (\citealt{Arcangeli2018}, \citealt{Bell2018}, \citealt{2019arXiv190304565H}) in comparison with cooler planets. This creates a new type of exoplanets called ultra hot Jupiters (hereafter UHJ).

We define UHJs as planets with day-side temperatures higher than $2200~\mathrm{K}$ \citep{Parmentier2018}. One of the most significant differences between hot and ultra hot Jupiters is perhaps the presence of water vapour in their atmospheres, which in hot Jupiters has become a common species but is missing in the day-side of ultra hot Jupiters, even though the ingredients to form water vapor are present.
The high irradiation received by the day-side of such UHJs causes the atmospheric gas temperature to increase to further more than $3000$K, as for for example demonstrated for  WASP-18b \citep{Helling2019}. The dominating gas species in the upper, day-side atmosphere is therefore atomic hydrogen (HI) instead of molecular hydrogen (H$_2$). Water (H$_2$O) is by $\approx$ 5 orders of magnitude less abundant than on the night-side. Such high day-side gas temperatures further result in an increased number of H$^-$ compared to cooler giant gas planets like HD\,189733b. In the case of WASP-18b, in the upper low-pressure atmosphere, the day-side has more than $10$ orders of magnitude more H$^-$ and H$^+$ than the night-side. Therefore, H$^-$ opacity at the day-side plays an important role \citep{Lothringer2018}. 

Several ultra hot Jupiters have been studied: WASP-33b \citep{Haynes2015}, WASP-121b \citep{Evans2017Natur.548...58E}, WASP-103b \citep{Keidberg2018}, WASP-18b (\citealt{Sheppard2017WASP18}, \citealt{Arcangeli2018}, \citealt{Helling2019}) , HAT-P-7b \citep{Armstrong2016}, most of them presenting thermal inversions. Focusing on transmission spectroscopy at high resolution, a study of the atmosphere of KELT-9b, the hottest planet known to date ($T_{eq}=4050~\mathrm{K}$ and day-side temperature of $4600~\mathrm{K}$), by \citet{YanKELT9} shows a detection of an extended atmosphere of H$\alpha$ produced by irradiation. On the other hand, \citet{Hoeijmakers2018} studied the atmosphere of this same planet, detecting Fe, Fe$^+$ and Ti$^+$ and concluding that at this high temperatures metals are preponderantly in their ionic forms. \citet{Cauley2019} further reported the detection of H$\beta$ and the optical MgI triplet, and found that planetary rotational broadening is needed to reproduce the Balmer line transmission profile shapes.

\begin{figure}[h]
\centering
\includegraphics[width=0.5\textwidth]{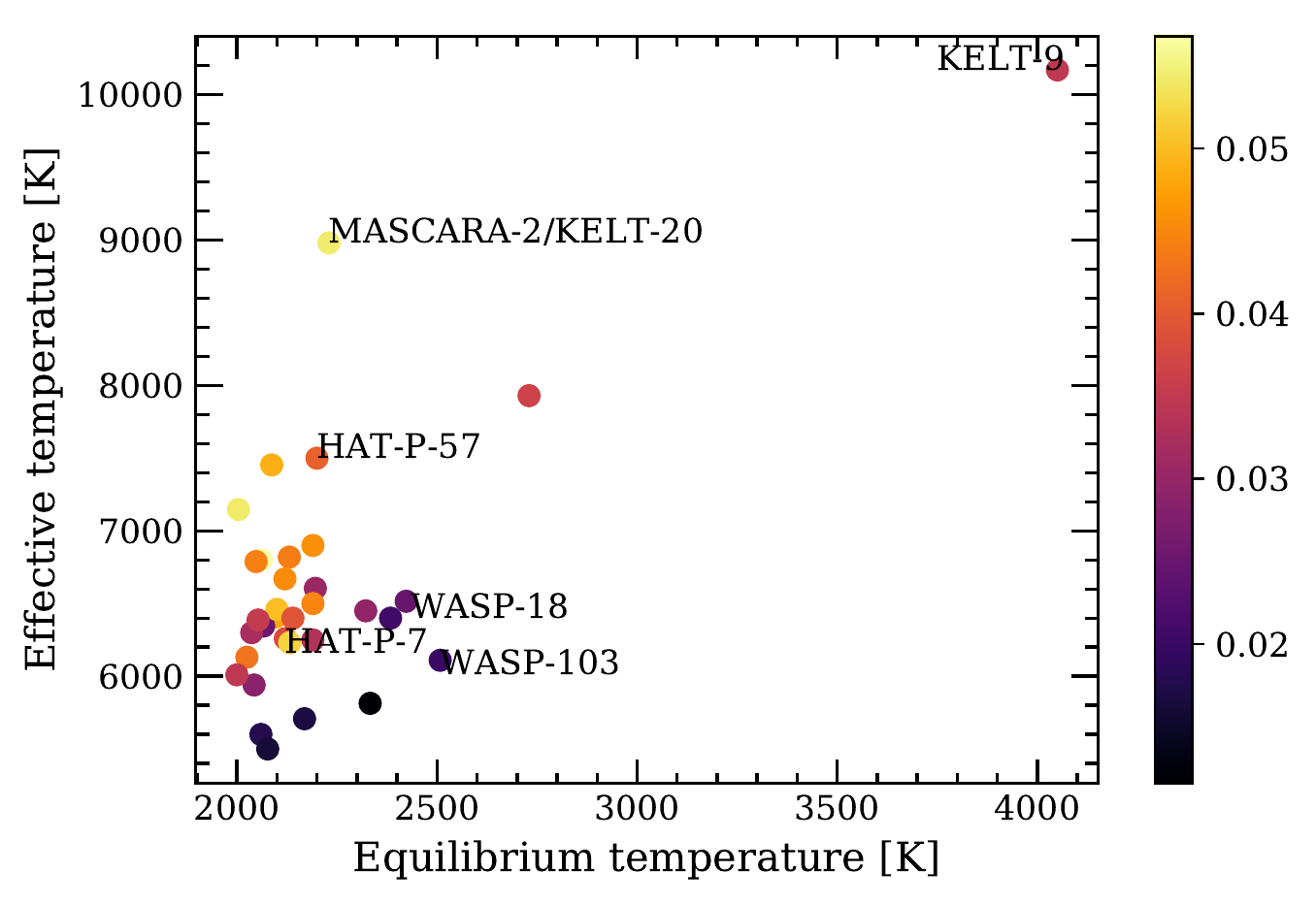}
\caption{Ultra hot Jupiters sample from {\tt exoplanets.eu}. This sample corresponds to all confirmed planets with equilibrium temperatures larger than $2000~\mathrm{K}$, which have been detected via radial-velocity or transits methods. In the horizontal axis we show the equilibrium temperature of the planet, and in the vertical axis the stellar effective temperature. The semi-major axis of each system are color-coded in the side bar (in AU). Some names here mentioned are shown as reference.}
\label{fig:context}
\end{figure}

Here, we present the study of the atmosphere of MASCARA-2b \citep{MASCARA22017Talens}, also known as KELT-20b \citep{2017Lund}, using four transit observations, three of them observed with the HARPS-North spectrograph and one observed with the CARMENES spectrograph. MASCARA-2b is an ultra hot Jupiter with an equilibrium temperature of $2260~\mathrm{K}$, located in a $3.5~\mathrm{day}$ orbit around an A-type star ($m_V=7.6$) with an effective temperature ($T_{\mathrm{eff}}$) of $8980~\mathrm{K}$ and a rotation ($v\sin i$) of $116~\mathrm{km/s}$. MASCARA-2b orbits very close to its host star ($a =0.057~\mathrm{AU}$) and is strongly irradiated. Its mass, because radial-velocity measurements of rapid-rotators are very challenging, has not been measured, and only an upper limit is available: $M_p < 3.510~M_J$. The details about the system are summarized in Table~\ref{tab:Param}. In Figure~\ref{fig:context} our target can be contextualized among the other known ultra hot Jupiters. MASCARA-2b orbits around the second hottest host star of the UHJ population, although its equilibrium temperature is typical for UHJs. Given that the effective temperature of MASCARA-2 is larger than WASP-18, the effect of thermal ionisation of the day-side can be expected to be stronger for MASCARA-2b/KELT-20b due to a stronger irradiation by the host star.

\renewcommand{\thefootnote}{\fnsymbol{footnote}}
\begin{table}[h]
\small
\centering
\caption{Physical and orbital parameters of MASCARA-2b.}
\resizebox{\columnwidth}{!}{
\begin{tabular}{lll}
\\ \hline \hline
\\[-1em]
 Description  & Symbol & Value \\ \hline
 Stellar parameters & & \\
 \\[-1em]
\quad  Identifiers & -&KELT-20, HD~185603 \\
\quad  V-band magnitude &$m_{\mathrm{V}}$& $7.6$\\
\quad  Effective temperature & $T_{\mathrm{eff}}$& $8980^{+90}_{-130}~\mathrm{K}$\\
  \\[-1em]
\quad  Projected rotation speed & $v\sin i_{\star}$ & $114\pm3~\mathrm{km~s^{-1}}$\\
 \quad Surface gravity & $\log g$ & $4.31\pm0.02~\mathrm{cgs}$\\
\quad  Metallicity & [Fe/H] & $-0.02\pm0.07$\\
 \quad Stellar mass & $M_{\star}$ & $1.89^{+0.06}_{-0.05}~\mathrm{M_{\odot}}$\\
 \\[-1em]
\quad  Stellar radius &$R_{\star}$& $1.60\pm0.06~\mathrm{R_{\odot}}$\\ 
 \hline
 \\[-1em]
 Planet parameters & & \\
 \\[-1em]
 \quad Planet mass$^($\footnotemark[1]$^)$ & $M_p$ &  $<3.510~\mathrm{M_{Jup}}$\\
 \quad Planet radius & $R_p$ & $1.83\pm0.07~\mathrm{R_{Jup}}$ \\
 \quad Equilibrium temperature &$T_{eq}$& $2260\pm50~\mathrm{K}$\\ 
 \quad Surface gravity$^($\footnotemark[1]$^)$ &$\log g_P$& $<3.460~\mathrm{cgs}$\\ 
 \\[-1em]
 \hline
 \\[-1em]
 System parameters & & \\
 \\[-1em]
 \quad Right ascension& -&  19$^h$38$^m$38.73$^s$\\
 \quad Declination & -&  +31$^\mathrm{o}$13'09.2"\\
 \quad Epoch& $T_{c}$ & $2457909.5906^{+0.0003}_{-0.0002}~\mathrm{BJD}$\\
  \\[-1em]
 \quad Period$^($\footnotemark[1]$^)$& $P$ & $3.4741070\pm0.0000019~\mathrm{days}$\\
  \\[-1em]
 \quad Transit duration$^($\footnotemark[1]$^)$& $T_{14}$ & $0.14882^{+0.00092}_{-0.00090}~\mathrm{days}$\\
  \\[-1em]
 \quad Ingress/Egress duration$^($\footnotemark[1]$^)$& $\tau$ & $0.01985^{+0.00082}_{-0.00079}~\mathrm{days}$\\
 \\[-1em]
 \quad Semimajor axis$^($\footnotemark[1]$^)$ &$a$& $0.0542^{+0.0014}_{-0.0021}~\mathrm{AU}$\\
 \\[-1em]
 \quad Inclination$^($\footnotemark[1]$^)$ & $i$ & $86.15^{+0.28}_{-0.27}~\mathrm{deg}$\\
  \\[-1em]
 \quad Eccentricity& e & $0$ (fixed)\\
 \quad Systemic velocity &$\gamma$& $-21.07\pm0.03~\mathrm{km~s^{-1}}$\\
 \quad Projected obliquity &$\lambda$& $0.6\pm4~\mathrm{deg}$\\\lasthline
\end{tabular}
}
\\
\begin{tablenotes}
\item Notes. $^($\footnotemark[1]$^)$ From \citet{2017Lund}. All the remaining parameters are taken from \citet{MASCARA22017Talens}. 
\end{tablenotes}
\label{tab:Param}
\end{table}
\renewcommand{\thefootnote}{\arabic{footnote}}

Due to its brightness and the fast rotation of the host star, MASCARA-2b is an ideal target for transmission spectroscopy. Its atmosphere was previously studied in \citet{Casasayas2018}, where only the first HARPS-North transit on 16 August 2017 was analyzed, finding evidence of H$\alpha$ and NaI absorption features and noisy features at H$\beta$ position. Here, combining four transits, we achieve higher signal-to-noise ratio (SNR) being able to clearly distinguish the center-to-limb variation (CLV) and the Rossiter-McLaughlin effect (RME), whose treatment will become very important for atmospheric studies of this kind of planets when observing with ESPRESSO-like facilities.

This paper is organized in seven sections. In Section~\ref{intro} we present a general introduction of transmission spectroscopy and ultra hot Jupiters. In Section~\ref{obs} we detail the observations. Methods to extract the color-photometry light curves and high resolution transmission spectra are explained in Section~\ref{method}. In Section~\ref{sec:modeling} we explain how the CLV and RME are modeled. In Section~\ref{sec:transmission} we present the results obtained in the analysis of high resolution transmission spectra. The analysis of the systematic effects are presented in Section~\ref{sec:tests} and the conclusions in Section~\ref{sec:disc}.

\section{Observations}
\label{obs}

We observed a total of four transits of MASCARA-2b using the HARPS-N and CARMENES high resolution spectrographs. One of these transits was simultaneously observed with MuSCAT2 (Multicolor Simultaneous Camera for studying Atmospheres of Transiting exoplanets 2), a 4-color simultaneous imager, in order to monitor possible stellar activity. One additional epoch was also observed with MuSCAT2 to reproduce the results of the first observation. Details about the observations can be found in Table~\ref{Tab:Obs} and Figure~\ref{fig:observed}.

\subsection{HARPS-N observations}
We observed three transits of MASCARA-2b using the HARPS-North spectrograph mounted on the $3.58$-m TNG telescope, located at the Observatorio del Roque de los Muchachos (ORM, La Palma). The observations were on 16 August 2017, 12 and 19 July 2018. For the 2 last observations we used the GIARPS mode, which permits simultaneous high resolution spectroscopy observations in the near-infrared and in the optical wavelength regions, using GIANO-B and HARPS-North, respectively.  

We exposed continuously before, during and after the transit, using the fiber A on the target and fiber B on the sky in order to monitor possible sky emissions. For the first and second night we used an exposure time of $200$ s while in the third night we increased it to $300$ s, to obtain a higher signal to noise ratio (SNR).

\subsection{CARMENES observations}

One more transit was observed with CARMENES spectrograph (Calar Alto high-Resolution search for M dwarfs with Exoearths with Near-infrared and optical Echelle Spectrographs) located at the Calar Alto Observatory, on 23 August 2017. The strategy of these observations was the same as with HARPS-North. For these observations an exposure time of $192$ s was used, obtaining a SNR of around $100$ near the H$\alpha$ line.

\renewcommand{\thefootnote}{\fnsymbol{footnote}}
\begin{table*}[t]
\centering
\caption{Observing log of the MASCARA-2b transit observations.}
\begin{tabular}{cccccccccccc}
\hline\hline
Night & Tel. & Instrument & Date of     & Start & End  & Filter & $T_\mathrm{exp}$ & $N_\mathrm{obs}$ & Airmass & Aperture \\
      &      & & observation & UT     & UT   &        & [s]              &  & & [px] \\ \hline
\\[-1em]
1 & TNG & HARPS-N & 2017-08-16 & 21:21 & 03:56 & $-$ & 200 & 90 & 1.089$\rightarrow$1.001$\rightarrow$2.089 & Spec.\\ \hline
\\[-1em]
2 & TNG & GIARPS & 2018-07-12 & 21:27 & 05:15 & $-$ & 200 & 116 & 1.604$\rightarrow$1.001$\rightarrow$1.527 & Spec.\\ \hline
\\[-1em]
3 & TNG & GIARPS & 2018-07-19 & 21:25 & 04:23 & $-$ & 300 & 78 & 1.442$\rightarrow$1.000$\rightarrow$1.389 & Spec.\\ \hline
\\[-1em]
4 & CA 3.5~m & CARMENES & 2017-08-23 & 21:09 & 02:36 & $-$ & 192 & 74 & 1.012$\rightarrow$1.006$\rightarrow$1.903 & Spec.\\ \hline
\\[-1em]
2 & TCS & MuSCAT2 & 2018-07-12 & 23:49 & 05:41 & $g$ & 0.7 & 3365 & 1.066$\rightarrow$1.001$\rightarrow$1.767 & 14,23,29\\
2 & TCS & MuSCAT2 & 2018-07-12 & 23:49 & 05:41 & $r$ & 0.9 & 3262 & 1.066$\rightarrow$1.001$\rightarrow$1.767 & 14,14,23\\
2 & TCS & MuSCAT2 & 2018-07-12 & 23:49 & 05:41 & $i$ & 0.9 & 3205 & 1.066$\rightarrow$1.001$\rightarrow$1.767 & 15,23,29\\
2 & TCS & MuSCAT2 & 2018-07-12 & 23:49 & 05:41 & $z_s$ & 0.9 & 3117 & 1.066$\rightarrow$1.001$\rightarrow$1.767 & 14,14,23\\
\hline \\[-1em]
5 & TCS & MuSCAT2 & 2018-09-30 & 20:16 & 00:30 & $g$ & 2 & 2321 & 1.002$\rightarrow$1.001$\rightarrow$1.801 & 29,32,37\\
5 & TCS & MuSCAT2 & 2018-09-30 & 20:16 & 00:30 & $r$ & 2 & 2322 & 1.002$\rightarrow$1.001$\rightarrow$1.801 & 27,32,37\\
5 & TCS & MuSCAT2 & 2018-09-30 & 20:16 & 00:30 & $i$ & 4 & 2322 & 1.002$\rightarrow$1.001$\rightarrow$1.801 & 28,30,35\\
5 & TCS & MuSCAT2 & 2018-09-30 & 20:16 & 00:30 & $z_s$ & 10 & 1320 & 1.002$\rightarrow$1.001$\rightarrow$1.801 & 29,36,40\\
\\[-1em]
\hline\hline
\end{tabular}\\
\label{Tab:Obs}
\end{table*}
\renewcommand{\thefootnote}{\arabic{footnote}}

\subsection{Photometric observations with MuSCAT2}

The transit of MASCARA-2b on 12 July 2018 was observed simultaneously using MuSCAT2 \citep{MUSCAT2} located at the $1.52~\rm{m}$ Carlos Sánchez Telescope (TCS) at the Teide Observatory. Another transit on 30 September 2018 was observed using MuSCAT2 only. Both observations were heavily defocussed to avoid saturation. While the first observation covered the whole transit event, the second observation had to stop at $\sim$43 minutes after the mid-transit point due to low elevation of the target. Both nights were mostly clear. Although defocussed, the profile of point spread function (PSF) was not stable for both observations. The observing log is presented in Table~\ref{Tab:Obs}. 

\section{Methods}
\label{method}

In this section we present the data processing applied to extract the four color light curves observed with MuSCAT2 (photometry) and the transmission spectra using the high resolution observations (spectroscopy).

\subsection{Photometry}
\label{ssec:photom}

Two independent approaches were used for the MuSCAT2 photometry, light curve detrending, and transit modelling
to search for possible spot-crossing events.

The first approach uses the dedicated MuSCAT2 photometry and transit modelling pipeline:
the photometry follows the standard aperture photometry practices, the transit modelling is carried out with
\texttt{PyTransit} \citep{Parviainen2015Pytransit} and the priors for the quadratic limb darkening coefficients are
calculated using \texttt{LDTk} \citep{Parviainen2015LDTK}. The final transit light curve modelling was carried
out jointly for the two nights and all passbands, using a Gaussian process (GP) based systematics model
calculated with \texttt{george} \citep{Ambikasaran2014} with the sky level, airmass, seeing, and CCD centroid shifts as covariates.

The second approach followed closely the approach described in \citet{2014A&A...563A..40C}. The photometry
was carried out with the DAOPHOT package in the NASA IDL Astronomy User's Library, the transit modelling
was done with \texttt{batman} \citep{2015PASP..127.1161K}, \texttt{Celerite} \citep{2017AJ....154..220F} was used to implement a Gaussian process-based systematics model, and the method by \citet{2015MNRAS.450.1879E} was used to obtain priors on the quadratic limb darkening coefficients.

\begin{figure}[h]
\centering
\includegraphics[width=0.5\textwidth]{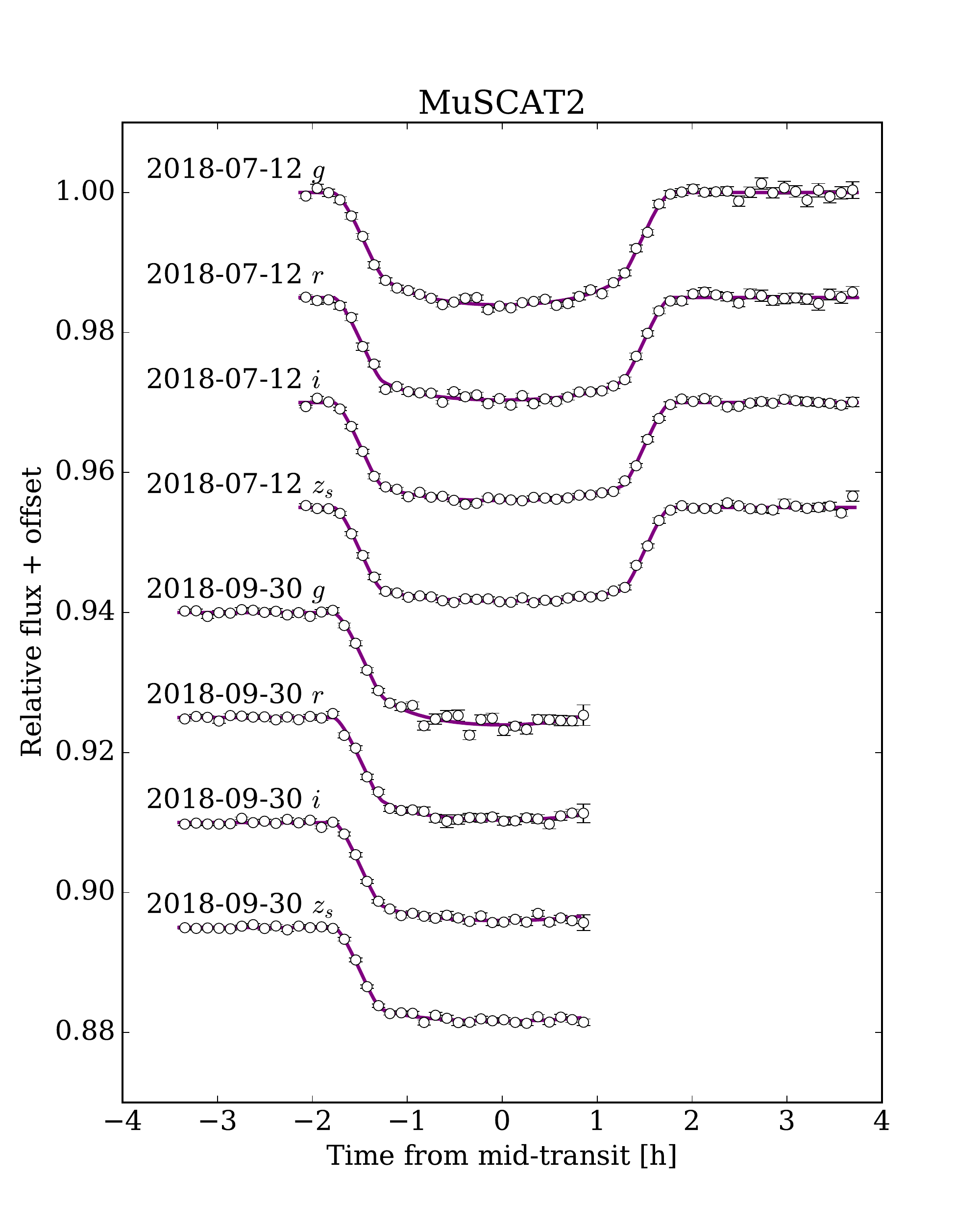}
\caption{The MASCARA-2b transit light curves obtained with MuSCAT2. The data have been binned to 7.2 min intervals for display clarity.}
\label{fig:mct2_lc}
\end{figure}

In both cases, \texttt{emcee} \citep{2013PASP..125..306F} was used to estimate the model parameter posteriors.
The parameter posterior estimates are presented in Table~\ref{Tab:musc2}, and the MuSCAT2 light curves are shown in Figure~\ref{fig:mct2_lc}. The transit light curves do not feature evident spot-crossing events, nor do they
show significant colour-dependent depth variations.

\renewcommand{\thefootnote}{\fnsymbol{footnote}}
\begin{table}[t]
\small
\centering
\caption{Best fit parameters from MuSCAT2 transit light curve.}
\begin{tabular}{lccr}
\hline\hline
Parameter & Symbol & Units & Value \\ \hline
\\[-1em]
Transit epoch & $T_{0}$ & BJD$^($\footnotemark[4]$^)$ & 2457503.120120 $\pm$ 0.00018\\ 
\\[-1em]
Period  & $P$ & days & 3.47410196 $\pm$ 0.00000106\\
\\[-1em]
Inclination & $i$ & deg & 85.61 $\pm$ 0.19 \\ 
\\[-1em]
Radius ratio & $R_p/R_{\star}$ & $-$ & 0.1176 $\pm$ 0.0014 \\ 
\\[-1em]
 & $(R_p/R_{\star})^2$ & $-$ & 0.01382 $\pm$ 0.00033 \\ 
\\[-1em]
Semimajor axis& $a/R_{\star}$ & $-$  & 7.254 $\pm$ 0.096 \\ 
\\[-1em]
\hline
\\[-1em]
Mid-transit 1$^($\footnotemark[1]$^)$ & $T_{mid}^1$ & BJD & 2458312.58566 $\pm$ 0.00022 \\
\\[-1em]
Mid-transit 2$^($\footnotemark[3]$^)$ & $T_{mid}^2$ & BJD & 2458392.48932 $\pm$ 0.00042 \\ 
\\[-1em]
\hline\hline
\end{tabular}\\
\begin{tablenotes}
\item Notes. $^($\footnotemark[4]$^)$ All times are given in BJD TDB. $^($\footnotemark[1]$^)$ Mid-transit time of 12 July 2018. $^($\footnotemark[3]$^)$ Mid-transit time of 30 September 2018.
\end{tablenotes}
\label{Tab:musc2}
\end{table}
\renewcommand{\thefootnote}{\arabic{footnote}}

\subsection{Spectroscopy}
\label{ssec:spectro}

The observations were reduced with the HARPS-North Data reduction Software (DRS), version $3.7$. The DRS extracts the spectra order-by-order, which are then flat-fielded using the daily calibration set. A blaze correction and the wavelength calibration are applied to each spectral order and, finally, all the spectral orders from each two-dimensional echelle spectrum are combined and resampled into a one-dimensional spectrum. The resulting one-dimensional spectra cover the optical range between $3800~\mathrm{\AA}$ and $6900~\mathrm{\AA}$ and has a wavelength step of $0.01~\mathrm{\AA}$. The spectra are referred to the Baricentric rest frame and the wavelengths are given in the air.

All nights present stable seeing at ${\sim 1~\mathrm{arcsec}}$ with the Moon at more than $90~\mathrm{deg}$ from our target. HARPS-N Night 1 and CARMENES observations are observed to be very stable in terms of SNR during the night. For Night 2 and 3, however, we noticed variable SNR and variations of the continuum similar to a bad blaze function correction. This was caused by a problem with the Atmospheric Dispersion Corrector (ADC) of the telescope, that causes PSF distortions and variations of the continuum. For these both nights, re-reducing the data with the DRS did not produced any improvement. The way of correcting this variation is discussed in detail in Section~\ref{subsec:Ha}. On the other hand, during the Night 2 we observe a drop in the SNR of some spectra before the transit, possibly caused by the presence of a cloud. We discard the spectra from 22:58 UT until 23:24 UT ($8$ spectra), where the SNR is lower than $35$ (around $6600~\mathrm{\AA}$). Additionally, for consistency with the SNR established for the cloud region, for this night we also removed the spectra taken at 01:28UT which also present SNR smaller than $35$ (i.e. a total of $9$ spectra are discarded).

\begin{figure*}[h]
\centering
\includegraphics[width=1.0\textwidth]{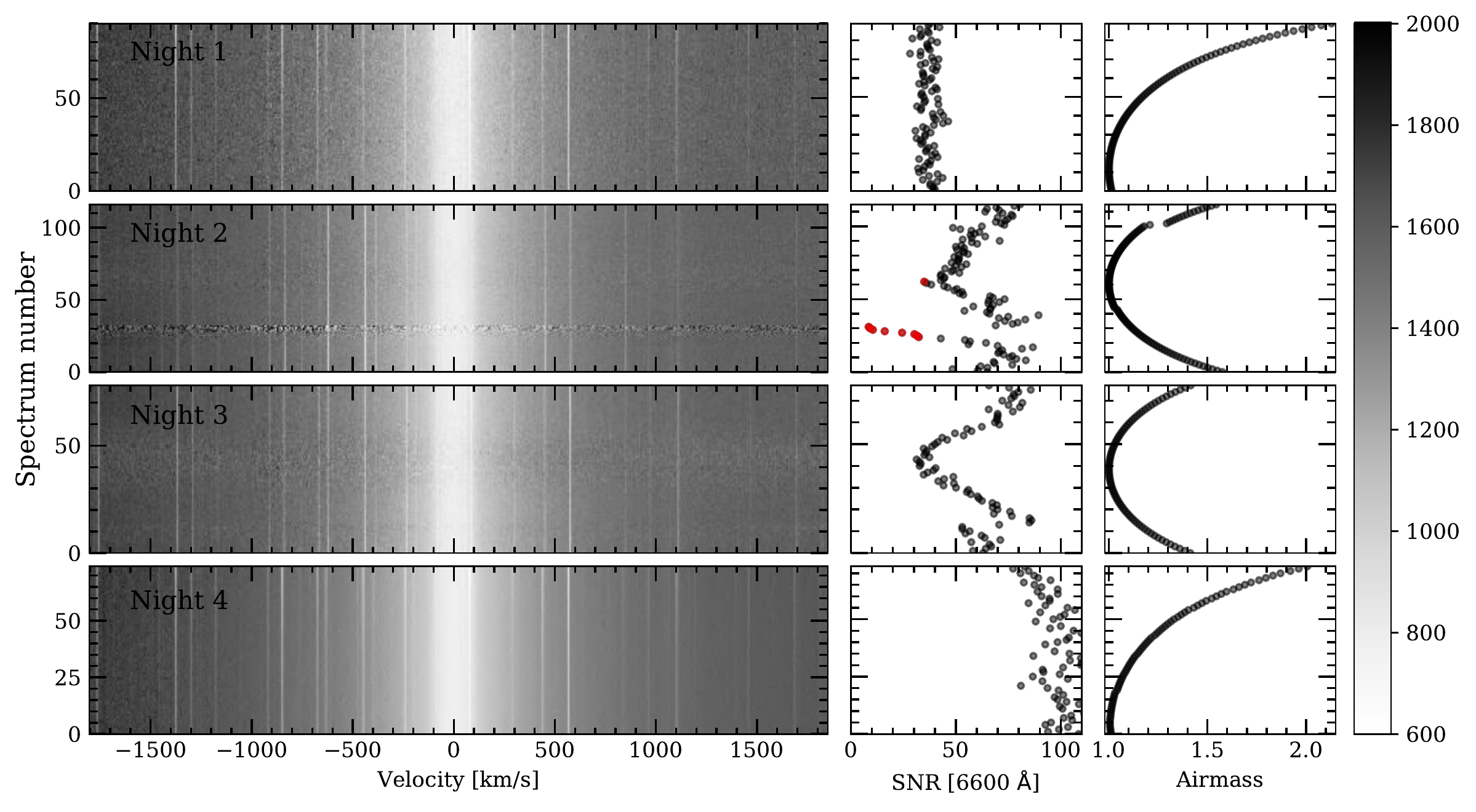}
\caption{Observed spectra around the H$\alpha$ line (left column), for Night 1 (first row), Night 2 (second row), Night 3 (third row) and Night 4 (fourth row). The three first nights were observed with HARPS-N and the last night, with CARMENES. The spectra are normalized to the same continuum level (see color bar) and moved to the stellar rest frame. In the y-axis we show the spectrum number starting from zero and in the x-axis the velocity in $\mathrm{km/s}$. The narrow lines correspond to telluric contamination and the deepest central line (at $0~\mathrm{km/s}$) corresponds to the center of the H$\alpha$ line core. In the second and third columns we show the SNR and the airmass variation during the night, respectively. The SNR per pixel is extracted in the continuum near the H$\alpha$ line, specifically around $6600~\mathrm{\AA}$ as the H$\alpha$ line for this target is very broad due to its rapid rotation. The red SNR values correspond to the spectra which are not considered in our the analysis. In particular, we can observe an stable SNR during Night 1. Night 2 show a drop in the SNR for the presence of a cloud around spectrum number $25$ and two stops of the instrument around spectrum number $50$ and $100$ because GIANO-B suddenly aborted itself and it was necessary to restart the exposure with both HARPS-N and GIANO-B instruments.}
\label{fig:observed}
\end{figure*}

CARMENES observations are processed with the CARMENES pipeline CARACAL (CARMENES Reduction And Calibration; \citet{CARACAL}) which considers bias, flat-relative optimal extraction \citep{flatrelative}, cosmic ray correction, and a wavelength calibration described in \citet{carmcalibration}. The reduced spectra are referred to the terrestrial rest frame and the wavelengths are given in the vacuum. At the beginning of the night some technical problems lead to some very short time exposures with very low SNR which are directly discarded. The spectra, SNR and airmass evolution of each night can be observed in Figure~\ref{fig:observed}.

\subsubsection{Telluric correction}
\label{ssubsec:Telluric}

Taking advantage that the observations were carried out using fiber B monitoring the sky, we checked for possible atmospheric emissions features during the night. We did not observe any telluric emission feature that needs to be corrected, and since subtracting the sky to the target spectra introduces additional noise, we did not apply this correction. 

Following \citet{Allart2017}, we correct the telluric absorption contamination using version 1.2.0 of {\tt Molecfit} (\citealt{Molecfit1} and \citealt{Molecfit2}). {\tt Molecfit} is an ESO tool built to correct Earth atmospheric features in ground-based spectra, which uses a line-by-line radiative transfer model (LBLRTM). HARPS-North spectra are given in the Solar System baricentric rest frame, while Molecfit models are computed in the terrestrial rest frame. For this reason, before fitting the telluric contamination we shift the spectra to the terrestrial rest frame using the Baricentric Earth Radial Velocity (BERV) information. We shift the spectra using the IRAF \citep{IRAF} routine {\tt dopcor}. The parameters used to compute the high-resolution telluric spectrum are detailed in \citet{Allart2017}, and the wavelength regions with strong lines used here are very similar to those used in this same paper, but considering only those regions out of stellar features. 

\begin{figure}[h]
\centering
\includegraphics[width=0.50\textwidth]{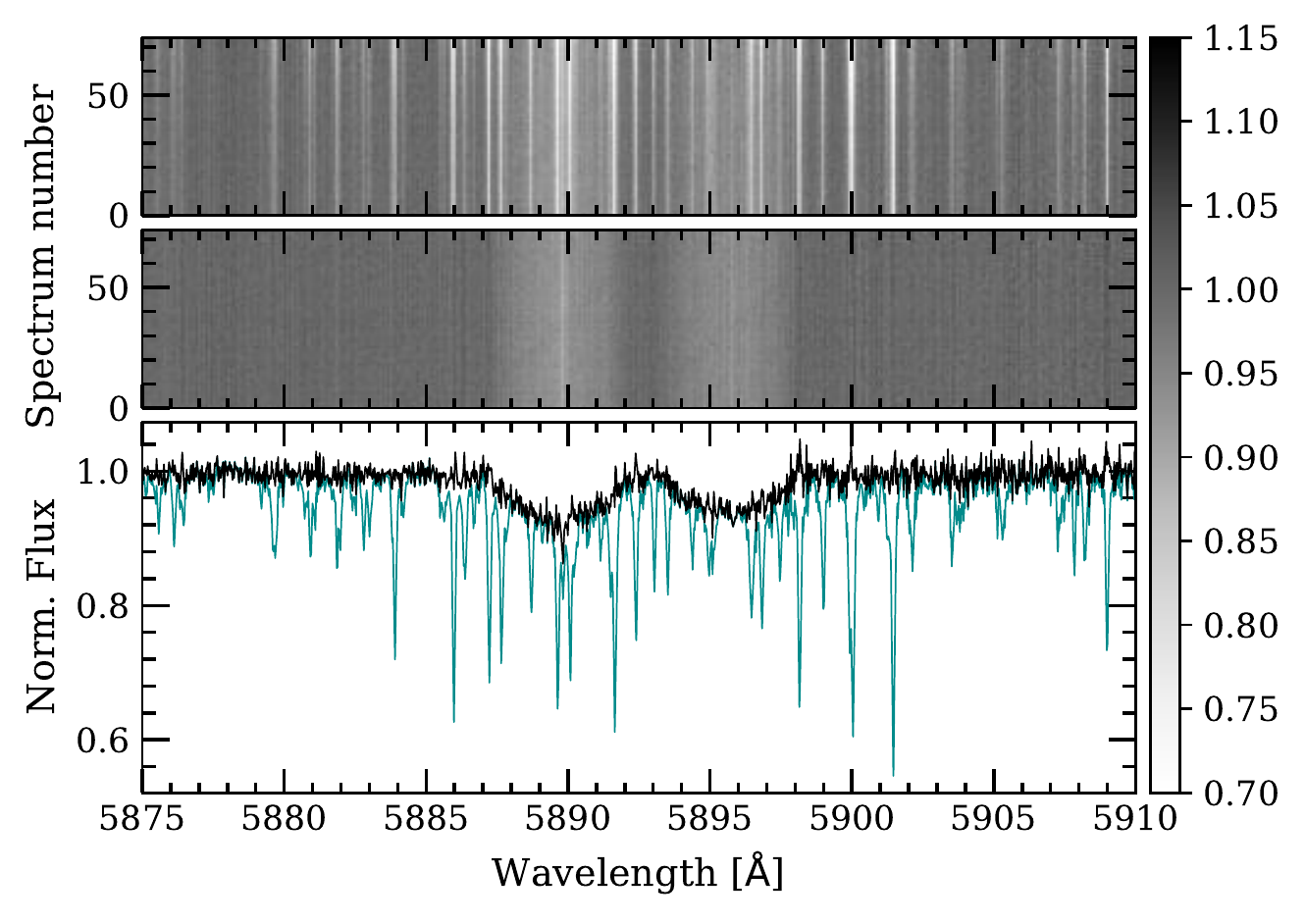}
\caption{Telluric correction of CARMENES observations (Night 4) of MASCARA-2b in the NaI doublet region. In the top panel we present the observed spectra before the telluric correction and in the second panel we show the spectra after correcting the telluric contamination with {\tt Molecfit}. For these two panels the y-axis corresponds to the spectrum number, the x-axis to wavelength in $\mathrm{\AA}$, and the normalized flux is shown in the color bar. In the third panel we present one single spectrum before (blue) and after (black) the telluric correction. The narrow lines observed in the first and third panels are telluric lines, while the broad lines centered at $5890~\mathrm{\AA}$ and $5896~\mathrm{\AA}$ are the stellar sodium lines. In the second panel, where the telluric lines have been subtracted, we can observe the interstellar sodium in the center of the stellar lines and, additionally, the planetary shadow (Rossiter-McLaughlin effect) as a trail moving into the lines. The spectra are presented in the terrestrial rest frame.}
\label{fig:tell}
\end{figure}

In the case of CARMENES data, the spectra after CARACAL reduction are already given in the terrestrial rest frame. The {\tt Molecfit} correction applied is slightly different, and follows the method discussed in \citet{Nortmann2018Science} and \citet{Salz2018He}. The telluric uncorrected and corrected spectra in the atomic sodium (NaI) region for Night 4 obtained with CARMENES are shown in Figure~\ref{fig:tell}.

\subsubsection{General transmission spectrum extraction}
\label{ssubsec:TSextract}

Once the telluric correction has been applied, we cut the spectra in the wavelength region we want to study and normalize them by the continuum. This is done recursively over the whole spectral range covered by both spectrographs to screen for all possible planetary signals. Then, as the spectra after the {\tt Molecfit} correction is referred to the terrestrial rest frame, we shift them to the stellar rest frame considering the system velocity ($\gamma$) and the BERV information. The stellar reflex motion is not considered because the mass of the planet ($M_p$) has not yet been determined, and the radial velocity measurements in rapid-rotating stars are difficult to determine. If we assume $K_s = 322.51~\mathrm{m/s}$ (i.e. the mass upper limit $M_p=3.510~M_J$; \citealt{2017Lund}), the maximum stellar radial-velocity change is $\sim\pm 0.3~\mathrm{km/s}$. Since one pixel in HARPS is around $0.8~\mathrm{km/s}$ and the stellar lines present a large broadening caused by the fast stellar rotation, considering or not considering this reflex motion contribution to the radial velocity correction does not have any significant impact to the results, as presented in \citet{Casasayas2018}. In addition, the spectra could be also affected by stellar activity and instrumental radial-velocity drifts. However, an A-type star like MASCARA-2 is not expected to be active, and the instrumental effects would similarly affect to all spectra, resulting in global wavelength offsets from the rest frame position (which after the daily calibration are expected to be of the order of $10~\mathrm{m/s}$ for the instruments used here). For CARMENES observations, when shifting the spectra to the stellar rest frame, we rebin the data to the same wavelength grid using the IDL routine {\tt rebinw} from the {\tt PINTofALE} package, which ensures flux conservation \citep{rebinwIDL}. 

For the next steps, we follow \citet{2015A&A...577A..62W}: we compute the out-of-transit master spectrum (combination of all the stellar spectra taken when the planet is not transiting) and we then divide each spectrum by this master spectrum to remove the stellar lines. The resulting residuals contain the variations of the stellar lines profile produced by the presence of the planet and additional systematic effects. These variations include the absorption produced by the exoplanet atmosphere, the center-to-limb variation (CLV) and the Rossiter-McLaughlin efect (RME). At this step we find the best fit model of the absorption, CLV and RME, and the resulting best fit parameters are used to compute the best fit CLV and RME model, which is then removed from the residuals (see Section~\ref{sec:modeling} for a detailed explanation of the models). Once these effects have been corrected, we shift the remaining residuals to the planet rest frame and we then combine the in-transit residuals to find the transmission spectrum for each night. 

HARPS-N Night 2 and Night 3 are affected by an ADC problem, presenting weird continuum variations similar to a bad blaze function correction. In order to solve this problem, we tried to re-reduce the raw data with the DRS pipeline, without any improvement. This effect introduces differences in the line wings changing in time, and which are more significant in very broad spectral lines, specially in the H$\alpha$ line. This differences become very important after dividing the spectra by the out-of-transit master spectrum, as the resulting spectra present broad variations that partially hide the narrow residuals we are looking for behind this broad effect. For this reason, after computing the residuals spectra we applied a very broad median filter using {\tt medfilt} from {\tt scipy} library with a kernel size of $1505$ to protect the planetary signals from being diluted and remove only the low frequency variations produced by the variation of the continuum, and the final transmission spectrum was normalised with {\tt UnivariateSplines} from {\tt scipy} in order to remove remaining undulations. This additional correction is only applied to HARPS-N Night 2 and Night 3.

Finally, we note that the parameters used during the whole process are those presented in Table~\ref{tab:Param} except for the epoch ($T_0$), orbital period ($P$), and inclination ($i$), which are from the recalculated parameters of Table~\ref{Tab:musc2}.

\section{Modeling and fitting the CLV and RME}
\label{sec:modeling}

The Rossiter-McLaughlin effect (RME) strongly affects the stellar lines profile of a fast-rotating star such as MASCARA-2b, which has a $v\sin{i} = 116~\rm{km/s}$. 
As the transit progresses, the light from different regions of the stellar disk is blocked by the planet, producing a lack of flux in different parts of the stellar lines which depends on the geometry of the system. On the other hand, the center-to-limb variation (CLV) also affects the stellar lines profile \citep{Szesla2015CLV}, which become broader and/or deeper than the integrated flux spectrum depending on the stellar properties of the region obscured by the planet. 

The stellar spectrum is modeled using {\tt VALD3} \citep{VALD3} line lists and Kurucz {\tt ATLAS9} models, computed with {\tt Spectroscopy Made Easy tool} (SME; \citealt{SME}). As presented in \citet{2017A&A...603A..73Y}, using {\tt SME} we compute the stellar spectrum at $21$ different limb-darkening angles ($\mu$) without considering the rotation broadening. For the stellar model computation we use local thermodynamic equilibrium (LTE) and solar abundance for the hydrogen lines. For the NaI, the solar abundance model strongly differs from the data. For this reason, we fit the NaI abundance using the {\tt SME} tool. In this process, the spectra used to fit the abundance is the combination of all the out-of-transit data of Night 4 with CARMENES, which have higher SNR. After checking the fitted [Na/H] results for a range of [Fe/H] values from literature ([Fe/H]$=-0.02\pm0.07$ in \citealt{MASCARA22017Talens}) we noticed that the NaI abundance change significantly when varying the [Fe/H] value (see below). The most reasonable NaI abundance is obtained in the $1-\sigma$ upper limit value i.e. assuming [Fe/H]$=+0.05$ finding [Na/H]$=0.98$. As we do not consider non-LTE effects that could be important in this stellar type \citep{2004AtypemodelsNLTE}. These [Fe/H], [Na/H] values together with the parameters given in Table~\ref{tab:Param} are those used to compute the stellar models. We note here that for our computation of these models, we need a good description of the line profile and the actual abundance value is not the focus of our analysis. The maximum relative difference obtained between models when assuming [Fe/H]$=-0.29$ \citep{2017Lund} (for which we obtain the [Na/H]$=1.39$) or [Fe/H]$=+0.05$ ([Na/H]$=0.98$) is ${\sim 0.08\%}$.

We divide, then, the stellar disk in cells of size $0.01~R_{\star}~$x$~0.01~R_{\star}$. Each of these cells contain its own properties, such as the rotation velocity due to $v\sin i$, which depends on its distance to the stellar rotation axis (i.e. the spin-orbit angle, $\lambda$), its $\mu$ angle value, and could be obscured by the planet at a given time. At the time of each exposure we can compute the position of the planet in the stellar disk using {\tt EXOFAST} \citep{exofast} and compute the integrated stellar disk flux considering the obscured cells, their proper radial velocity shift caused by the rotation and the correspondent $\mu$ spectrum (which are linearly interpolated using the $21$ $\mu$ angle reference spectra). The only difference between the HARPS-N and CARMENES modeled spectra is the spectral resolution, $\mathcal{R} = 115~000$ and $\mathcal{R} = 94~600$, respectively.

The modeled stellar spectra at different orbital phases depend on the geometry of the system but also the planetary radius, $R_p$. For a particular wavelength, and depending on the amount of material, the atmosphere becomes opaque at a given pressure level or radius of the planet, $R_{\lambda}$. In order to consider this radius change when fitting the data, the modeled stellar spectra is computed for a grid of $R_{\lambda}$, increasing from $0.7~R_p$ ($1~R_p = 1.83~R_J$; \citet{MASCARA22017Talens}) to $2.5~R_p$ in steps of $0.1~R_p$. For a given $R_{\lambda}$, we then calculate the stellar spectrum by linearly interpolating these grid of models. Finally, the modeled residuals are computed dividing all the modeled stellar spectra by the out-of-transit model. 

Finally, and similarly to \citet{YanKELT9}, in the modeled in-transit residuals we add a Gaussian profile to describe the absorption of the planet. This profile, depends on the center of the Gaussian ($v_0$), the full width at half maximum ($FWHM$) and constrast ($h$). The center position is defined as $v_0 = v_p + v_{wind}$ where $v_p = K_p \sin(2\pi\varphi)$ is the planet radial velocity in each in-transit exposure time ($\varphi$ is the orbital phase) and $v_{wind}$ is the radial velocity of the atmospheric wind. 

With this, we perform a Markov chain Monte Carlo (MCMC) analysis to fit the observed residuals using {\tt emcee} \citep{emcee} code. The free parameters are $R_{\lambda}$, $h$, $v_{wind}$, $K_p$ and $FWHM$. We use $14$ walkers and $1$x$10^5$ steps. The best fit $R_{\lambda}$ value is used to compute the CLV and RME model, which is then subtracted to the data, remaining only the true planetary absorption. Before combining the data residuals, this absorption is shifted to the planet rest frame using the best fit $K_p$ value. The MCMC analysis is based on binned spectra ($0.05~\mathrm{\AA}$ and $0.002$ orbital phase bins) and only applied to the fully in-transit data (i.e. excluding the ingress and egress). The planetary absorption spectrum in the ingress and in the egress could present different geometries \citep{YanKELT9}, as it was recently shown in \citet{Salz2018He}.

During the modelling, we observe that in some cases the best-fit value of $R_{\lambda}$ is smaller than $1~R_p$. This is possibly due to the intrinsic error of the stellar models, which are computed considering LTE and solar abundance in case of the hydrogen lines (the NaI abundance is fitted). The non-LTE effects in these stellar-types can become strong \citep{2004AtypemodelsNLTE} and, as calculated previously, the abundance of the lines also influences the final model. For two different values of [Fe/H] inside the error bars presented in \citet{2017Lund} and \citet{MASCARA22017Talens}, we can obtain the same intensity of the RME for two different $R_{\lambda}$. Additionally, if we compare two models computed with the same abundance, one calculated for $0.85~R_p$ and one for $1.0~R_p$, in the mid-transit time their relative difference in the RME centre is ${\sim 0.2\%}$. Thus, there are different assumptions that can produce variations of the final $R_p$ values. However, our main goal here with the model fitting is to remove the strong RME and CLV effects, which are easily observed in the residuals. For this reason, considering the $R_p$ change permits us to find the best model that reproduce the effects, even if the best-fit $R_{\lambda}$ value is not exactly the expected. This is one of the reasons we are not linking the increase of radius ($R_{\lambda}$) with the absorption that we measure ($h$) in the models, and we are leaving both free, independently. We do not expect this $R_{\lambda}$ values to have any effect to our results and we use the models as a tool to discard the RME and CLV effects from the absorption signatures. When comparing models, we observe that in the mid-transit time the RME is more constraining for the $R_{\lambda}$ parameter. In the ingress and egress regions, where the CLV become important, both CLV and RME are constraining $R_{\lambda}$. However, we stress that only the the fully in-transit data is used in the fitting procedure i.e. where the RME is the more constraining for the $R_{\lambda}$ parameter.


\section{Results}
\label{sec:transmission}

We present here the results obtained in the different wavelength regions. In all cases, we have analysed each individual night separately and, additionally, the three HARPS-N data together, sorting the spectra in time from the mid-transit center. We note that in this latter case all the spectra were telluric corrected first. 

Here, we show the 2D maps of the final residuals obtained after applying the differential spectroscopy, the final transmission spectrum and the transmission light curves around several spectral lines detectable in MASCARA-2b's transmission spectrum. The 2D maps of individual nights are presented in the Appendix~\ref{ap:individualTS}, their best-fit values in Table~\ref{Tab:res} and MCMC correlation diagrams in Appendix~\ref{ap:cornHa}. All light curves are also presented in Appendix~\ref{ap:individualTLC}. The absorption depth values measured in the transmission spectra and light curves can be found in Table~\ref{tab:AD_vals}. 

The transmission light curves are obtained calculating the equivalent width of the residuals after dividing each spectrum by the master out-of-transit spectrum, and then moved to the planet rest frame. The integration to obtain the equivalent width is computed using the {\tt scipy} library \citep{scipy} from {\tt Python2.7}. 

Hereafter, in Tables and Figures, we refer to the different nights as HARPS-N 1 (Night 1), HARPS-N 2 (Night 2), HARPS-N 3 (Night 3) and CARMENES 1 (Night 4) in order to not create confusion between instruments. The combined results of all three HARPS-N nights are presented as HARPS-N$x3$.

\subsection{H$\alpha$}
\label{subsec:Ha}

\begin{figure*}[h]
\centering
\includegraphics[width=1\textwidth]{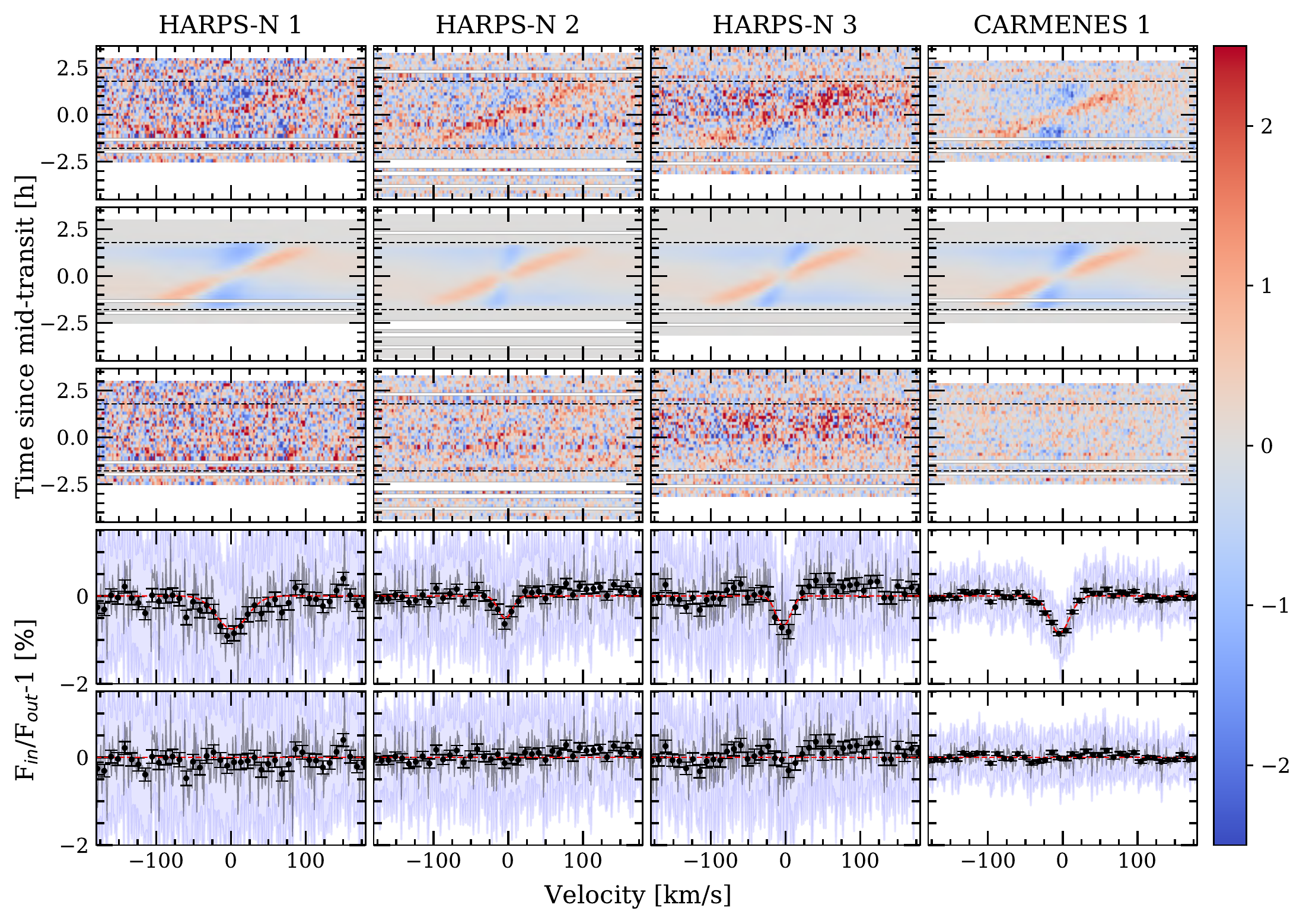}
\caption{Results of the H$\alpha$ line analysis of MASCARA-2b. Each column corresponds to one night. First row (top): residuals after dividing each spectrum by the master-out spectrum. Second row: best fit model describing the first panel residuals, including absorption, CLV and RME. Third row: residuals after subtracting the best fit model (second row) to the data (first row). Fourth row: The transmission spectrum obtained when combining the in-transit residuals from fourth row in the planetary rest frame. The black dots show the transmission spectrum binned in $0.2~\mathrm{\AA}$ intervals and the propagated errors from the photon noise. In light blue we show the standard deviation of the residuals from fourth row. The red lines show the Gaussian profile derived from the best fit parameters of second row. Fifth row: residuals after subtracting the Gaussian profile to the transmission spectrum. In all cases the residuals (rows from 1st to 3rd) are binned in intervals of $0.002$ in orbital phase and $0.05~\mathrm{\AA}$ in wavelength. The relative flux (F$_{in}$/F$_{out}$ -1) in $\%$ is shown in the colour bar. In white we show the time regions for which we do not have information. The horizontal lines show the first and last contact times of the transit.}
\label{fig:Ha_res}
\end{figure*}

 The residuals after computing the ratio between each spectra and the master out-of-transit spectrum, the best-fit models and the transmission spectra around H$\alpha$ ($6562.81~\mathrm{\AA}$) of each individual night can be observed in Figure~\ref{fig:Ha_res}. The best fit values and the 1-$\sigma$ error bars obtained with the MCMC analysis are presented in Table~\ref{Tab:resHa}. 

\renewcommand{\thefootnote}{\fnsymbol{footnote}}
\begin{table*}[h]
\centering
\small
\caption{Best fit parameters and 1-$\sigma$ error bars from the MCMC analysis of the H$\alpha$ line for each individual night and combined HARPS-N observations.}
\begin{tabular}{lcccccc}
\hline\hline
\\[-1em]
 & $h$ & FWHM& K$_p$& v$_{wind}$ &  R$_{\lambda}$ $^($\footnotemark[1]$^)$ & R$_{\lambda}~(h)$ $^($\footnotemark[4]$^)$\\ 
 & [$\%$]& [km/s] &[km/s] & [km/s] &[R$_p$] & [R$_p$]\\
\\[-1em]
\hline

HARPS-N 1 & $-0.75^{+0.10}_{-0.11}$ & $34.9^{+7.7}_{-6.4}$ & $140.3^{+39.0}_{-42.1}$& $1.3^{+3.0}_{-2.9}$ & $1.21^{+0.04}_{-0.05}$  & $1.24^{+0.06}_{-0.06}$\\
\\[-1em]
\hline
\\[-1em]

HARPS-N 2 & $-0.52^{+0.10}_{-0.11}$ & $15.6^{+3.7}_{-3.9}$& $101.5^{+33.7}_{-33.5}$& $-3.9^{+2.1}_{-2.4}$& $1.08^{+0.03}_{-0.03}$ &  $1.17^{+0.06}_{-0.07}$\\
\\[-1em]
\hline
\\[-1em]

HARPS-N 3 & $-0.69^{+0.10}_{-0.11}$ & $16.4^{+2.6}_{-2.4}$& $196.0^{+30.1}_{-32.0}$& $-2.8^{+2.2}_{-2.2}$& $1.15^{+0.04}_{-0.04}$ &  $1.22^{+0.06}_{-0.06}$ \\
\\[-1em]
\hline
\\[-1em]

CARMENES 1 & $-0.85^{+0.03}_{-0.03}$ & $22.6^{+0.9}_{-0.9}$ & $166.2^{+7.3}_{-7.4}$ & $-4.5^{+0.5}_{-0.5}$&  $1.21^{+0.01}_{-0.01}$ &  $1.27^{+0.02}_{-0.02}$\\ 

\\[-1em]
\hline
\\[-1em]

HARPS-N$x3$ & $-0.68^{+0.06}_{-0.06}$ & $19.0^{+1.7}_{-1.6}$ & $165.6^{+16.7}_{-16.6}$ & $-3.0^{+1.2}_{-1.2}$&  $1.13^{+0.02}_{-0.02}$ & $1.22^{+0.04}_{-0.04}$\\
\\[-1em]
\hline\hline
\end{tabular}\\
\begin{tablenotes}
\item Notes. $^($\footnotemark[1]$^)$ Effective radius value obtained from the best fit model of the CLV and RME effects. $^($\footnotemark[4]$^)$ Effective radius calculated considering the absorption value, $h$, and assuming a continuum level of $(R_p/R_{\star})^2 = 1.382\%$.
\end{tablenotes}
\label{Tab:resHa}
\end{table*}
\renewcommand{\thefootnote}{\arabic{footnote}}

In Figure~\ref{fig:Ha_res} we can see that the three nights observed with the HARPS-N are in general more noisy than the one with CARMENES, with residuals showing larger standard deviation. No obvious reason for this fact can be derived from checking the weather or seeing logs, only the ADC problem in HARPS-N Night 2 and 3 are noticeable. During the transit, the RME is observed as a red trail (positive relative flux) moving in the $\pm116~\mathrm{km/s}$ ($v\sin i$) range of velocities and the absorption as a blue trail (negative relative flux) around $\pm24~\mathrm{km/s}$, approximately.

\begin{figure}[h]
\centering
\includegraphics[width=0.45\textwidth]{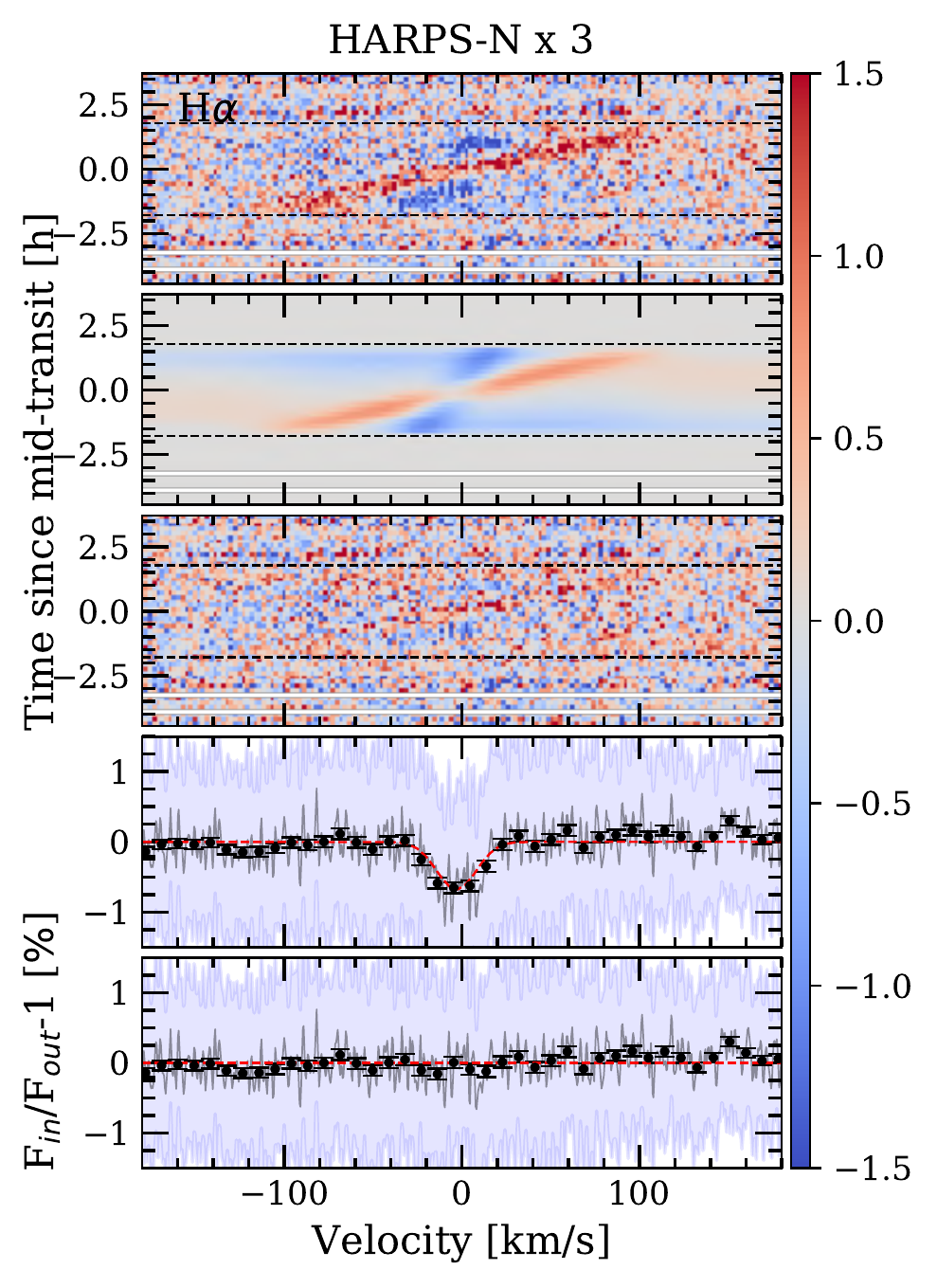}
\caption{Same as Figure~\ref{fig:Ha_res} but combining the three nights observed with HARPS-N.}
\label{fig:Ha_res_HARPS}
\end{figure}

\begin{figure}[h]
\centering
\includegraphics[width=0.48\textwidth]{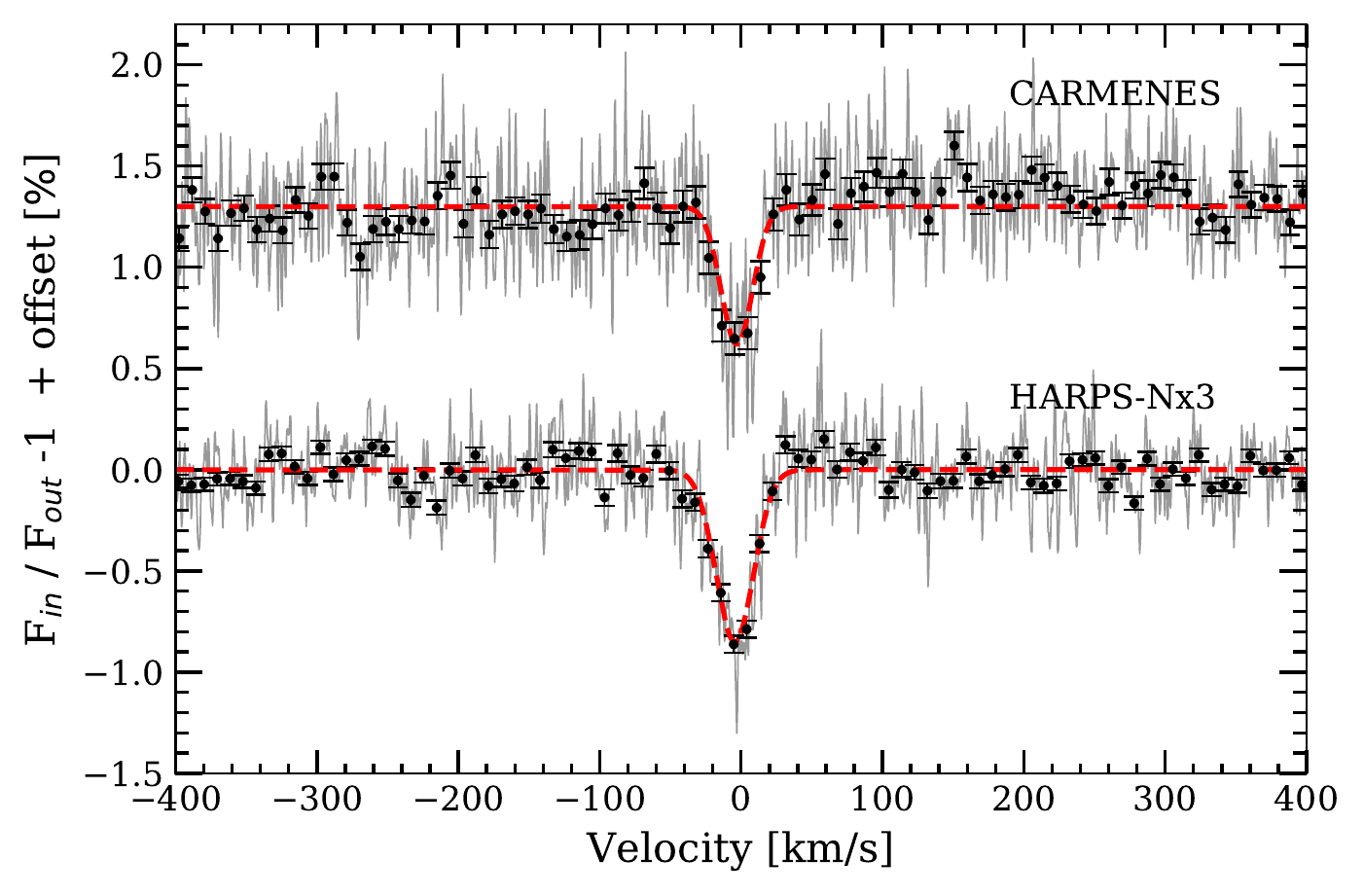}
\caption{Transmission spectra around H$\alpha$ line obtained with all nights of HARPS-N combined and CARMENES data (Night 4), with an offset for a better visualisation. The best-fit Gaussian profiles are shown in red. In grey we show the original transmission spectrum and the black dots show the data binned in $0.2~\mathrm{\AA}$ intervals.}
\label{fig:Ha_res_ts}
\end{figure}

Due to the noise, specially for Night 2 and Night 3 for which it increases during the transit, we are not able to totally fit the planetary absorption for the HARPS-N nights, which becomes underestimated by the model, and can not be recovered in the region were it intersects with the RME around mid-transit. However, for Night 4 where the SNR is higher, the absorption during the whole transit is recovered. In particular, this underestimation of the absorption can be observed in the residuals of the Gaussian profiles derived from the best fit model, which seem to present smaller contrast ($h$) than the absorption observed in the transmission spectrum of HARPS-N (see fourth and fifth row of Figure~\ref{fig:Ha_res}). When we fit all HARPS-N data together (see Figure~\ref{fig:Ha_res_HARPS}), the central absorption still remains within the noise level and the absorption is slightly underestimated. This is possibly caused by the decrease in SNR during the transit. We also analyzed the HARPS-N data considering only Night 1 and Night 3, with results closer to those obtained for CARMENES in Night 4 but the mid-transit region still remains within the noise level. The final transmission spectra of CARMENES~1 and the analysis combining all three HARPS-N nights, together with the Gaussian profiles derived from the best-fit parameters, can be observed in Figure~\ref{fig:Ha_res_ts}.

\begin{figure*}[t]
\centering
\includegraphics[width=0.98\textwidth]{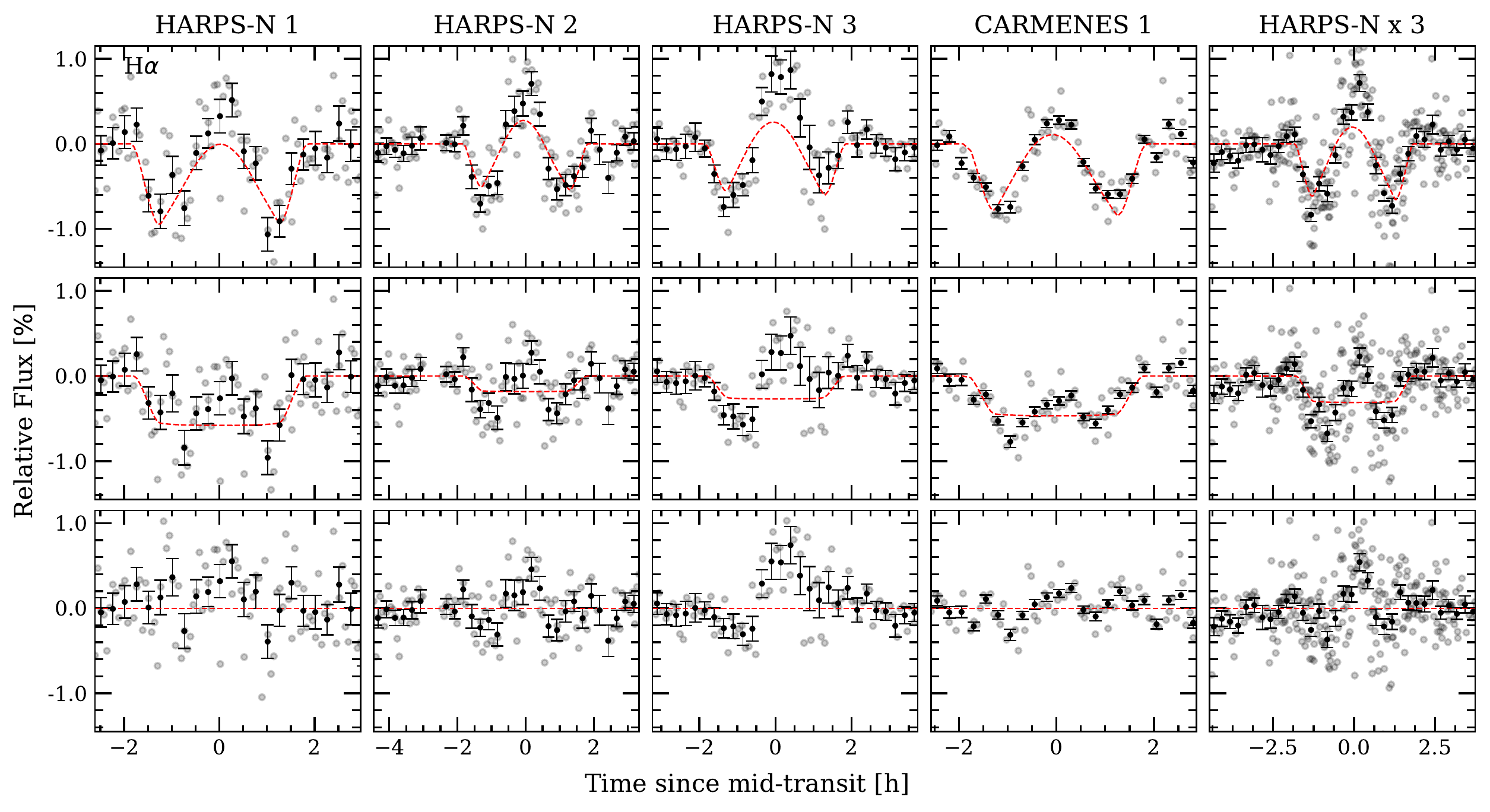}
\caption{H$\alpha$ transmission light curves for $1.5~\mathrm{\AA}$ bandwidth. Each column correspond to one night, and the last column (right) to the analysis of all HARPS-N nights together. First row (top): observed transmission light curves. These light curves contain the contribution of the CLV, RME and absorption of the planet. The red dashed line is the best-fit model derived from the fitting procedure of the residuals from Figure~\ref{fig:Ha_res}. Second row: H$\alpha$ transmission light curve after removing the CLV and RME. In red we show the best fit absorption model. Third row: residuals between the model and the data. In all cases the gray dots are the original data points, in black we show the data binned by $0.003$ orbital phase. We note the different x-scale of the different columns, as different nights present different time extension. }
\label{fig:TLC_Ha}
\end{figure*}

The observed transmission light curves before and after correcting the CLV and RME effects can be observed in Figure~\ref{fig:TLC_Ha}. The models are also obtained with the best fit parameters from the general fitting shown in Figure~\ref{fig:Ha_res}. In this light curves the RME effect and the time regions where the noise is higher can be clearly observed. The error bars are obtained by the error propagation from the photon noise. In order to include the ingress and egress shape of the transmission model we use the {\tt PyLDTk} \citep{Parviainen2015LDTK} and the {\tt PyTransit} \citep{Parviainen2015Pytransit} Python Packages.  The {\tt PyLDTk} estimates the limb-darkening coefficients for a given filter and stellar properties using the library of {\tt PHOENIX} stellar atmospheres \citep{Husser2013}. Then, {\tt PyTransit} allows us to obtain the transit shape with the same absorption ($h$) measured during the fitting procedure and the limb-darkening coefficients assumed from {\tt PyLDTk}. 

The absorption that we measure from the best fit model in the line core is $0.85\pm0.03~\%$ for CARMENES data and $0.68\pm0.06~\%$ for HARPS-N data combined, meaning an effective radius ($R_{\lambda} (h)$) of $\sim1.20~R_p$ as presented in \citet{Casasayas2018} where only HARPS-N Night~1 was available. We measure a mean standard deviation of $1.3$ and $2.3~\%$ in the residuals and a SNR of ${\sim 350}$ and ${\sim300}$ around the H$\alpha$ position ($\pm180~\mathrm{km/s}$) for CARMENES and HARPS-Nx3 analysis, respectively. We note that the results from both instruments are not totally consistent, as explained above. HARPS-N Night 1 and CARMENES observations are the most stable nights, with consistent values, while HARPS-N Night 2 and Night 3 present SNR variations during the transit that are then propagated to the final result when combining all three HARPS-N nights. These two nights, specially Night~3, present strong SNR variations during the observation due to the ADC problem. Although this effect is mainly removed during the analysis, in this process we apply additional steps that could introduce small over-corrections and cause variation of the final absorption level in these two nights. On the other hand, for these nights, the RME is not totally removed (see transmission light curves in Figure~\ref{fig:TLC_Ha}), and when combining the mid-transit signals we are averaging regions with RME residuals that decrease the absorption depth. 

The differences observed in the effective radius values obtained when fitting the CLV and RME models ($R_{\lambda}$) to the 2D residual maps and those derived from the absorption in the core of the lines ($R_{\lambda} ~(h)$) are possibly caused by intrinsic errors of the models. In particular, in case of hydrogen atoms, we are considering Solar abundance and LTE, but for MASCARA2 stellar-type we expect the non-LTE effects to become significant \citep{2004AtypemodelsNLTE}. There is also the low and variable SNR difficulty added when fitting the data. We can observe that for CARMENES data, which have higher SNR, the RME measurement is better.

We measure, in all cases, a shift of the absorption line with respect the theoretical position of H$\alpha$, which could be indicative of planetary winds. However, as discussed in Section~\ref{ssec:spectro}, we do not use the radial-velocity information from the instrument pipeline, and consequently we are not considering possible instrumental offsets in radial-velocity. For this reason, the winds here measured can be influenced by this effect. In all cases, before combining the residuals to obtain the transmission spectrum, we shift these residuals to the planet rest frame using the best fit $K_p$ parameter of each night. This value has large error bars for the noisiest nights, thus, we checked the differences when combining the spectra with different $K_p$ values within these $1-\sigma$ error bars, but only differences inside the noise level are noticeable. From \citet{2017Lund} we estimate that $K_p$ should be about $170~\pm7~\mathrm{km/s}$. We expect all nights and lines present consistent values (within the error bars) close to this estimation, however, caused by irregularities in the maps due to small residuals and the low S/N that makes difficult for the model to distinguish between CLV, RME and absorption regions, the measured $K_p$ values can be far from the expectation for a single night but become consistent after the combination of more nights. Small deviations of this measurements are introduced by the not corrected reflex motion from the host star.

The absorption depths measured for $1.5~\mathrm{\AA}$ passband in the final transmission spectra and light curves are presented in Table~\ref{tab:resAll}. We can observe that the results of CARMENES data and the combined HARPS-N nights are consistent, while individual HARPS-N nights give more dispersed values. We also note the slightly different results obtained in the analysis presented here and in \citet{Casasayas2018} for HARPS-N Night 1. The difference comes because here we are only considering the spectra fully in-transit (the ingress and egress files are not considered when computing the in-transit sample). Additionally, we are using different methods of telluric correction, $K_p$ values used when shifting the spectra to the planet rest-frame, and slightly improved corrections of the CLV and RME effects.

\subsection{H$\beta$ and H$\gamma$}
\label{subsec:HbHg}

\begin{figure}[h]
\centering
\includegraphics[width=0.5\textwidth]{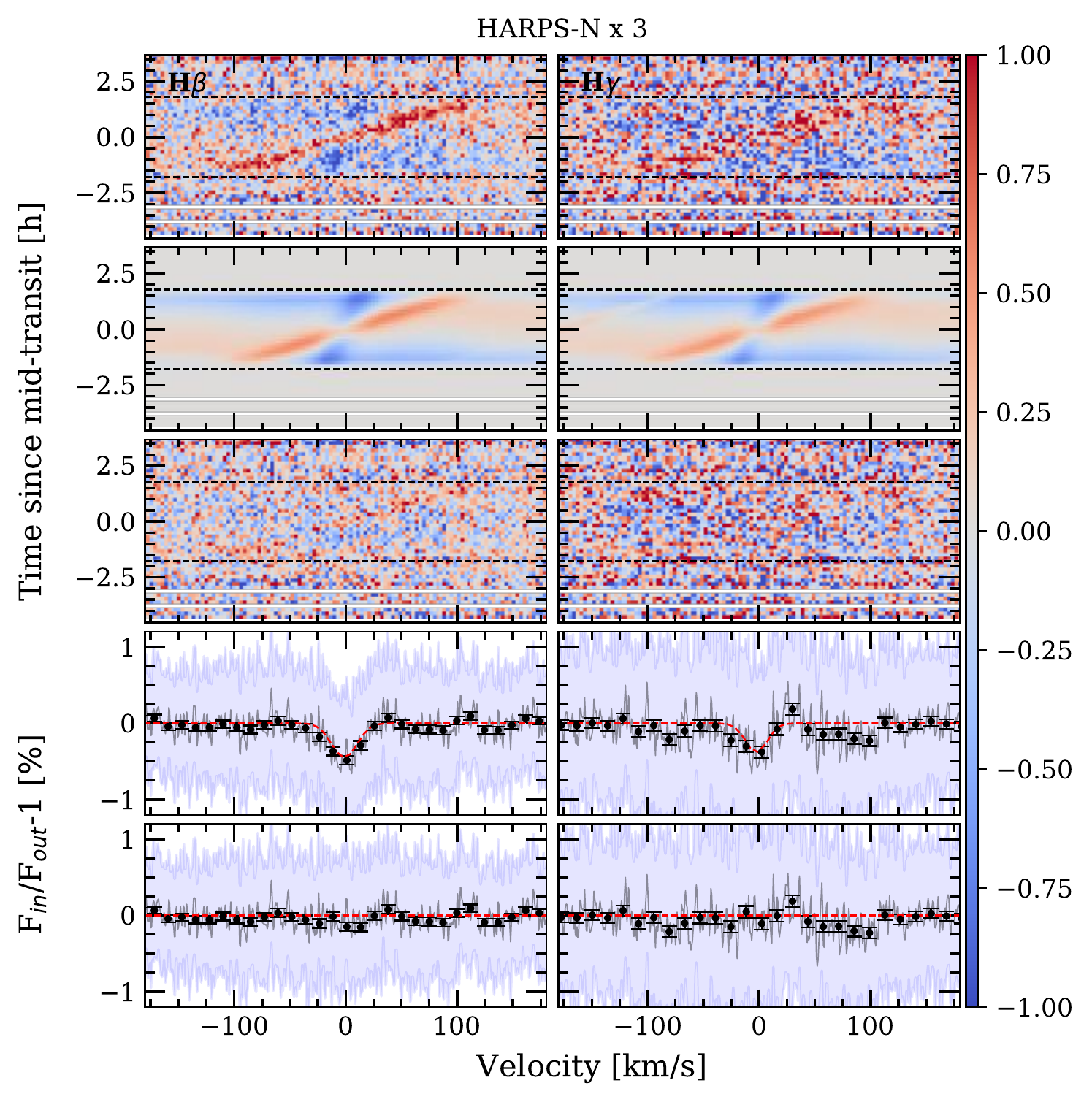}
\caption{Same as Figure~\ref{fig:Ha_res}, but for H$\beta$ and H$\gamma$ lines. The colour bar indicates the flux relative to the continuum (F$_{in}$/F$_{out}$ -1) in $\%$. }
\label{fig:resHbHg}
\end{figure}

The method applied to study the H$\beta$ ($4861.28~\mathrm{\AA}$) and H$\gamma$ ($4340.46~\mathrm{\AA}$) lines is exactly the same as for H$\alpha$. In Figure~\ref{fig:resHbHg} we present the results for the joint analysis of all nights observed with HARPS-N. The best-fit values are shown in Table~\ref{tab:resAll}. Note that CARMENES data do not cover this wavelength range, and thus only HARPS-N data is presented. Same as the H$\alpha$ line, the models were computed for a grid of $[0.7-2.5]~R_p$ with $0.1~R_p$ step.

In \citet{Casasayas2018} we observed noisy features in the transmission spectrum around the H$\beta$ line analysing only Night 1. Here, with three transits and after correcting the strong CLV and RME, we observe absorption centred at the H$\beta$ laboratory value. We note that some CLV and RME residuals remain after removing the best fit model (see the residuals in Figure~\ref{fig:resHbHg}). The absorption is observable in the analysis of the individual nights, except for Night 1 where the MCMC fitting is not able to distinguish between the noise and the absorption i.e. the parameters related to the absorption profile ($h$, $K_p$ and $FWHM$) are not well determined. However, when combining all three transits the absorption is well recovered. The absorption measured in this case corresponds to $-0.45\pm0.05~\%$ in the line core, which is consistent with the value measured in the transmission light curves during the transit (see Table~\ref{tab:resAll}). It is noticeable that the $R_{\lambda}$ values that we obtain in the fitting procedure are lower than $1~R_p$ in some cases. 

On the other hand, we find a H$\gamma$ absorption of $-0.4\pm0.1~\%$ in the line core when combining all nights. However, the in-transit trail in the 2D maps is not clear, possibly because the signal is less intense and remains partially hidden in the noise. This absorption is only observed when combining all data and in Night 2 when analysing each individual night. Additionally, the final transmission spectrum presents some residuals near the line.

In Table~\ref{tab:resAll} we give the absorption depth values measured in the spectra and the transmission light curves of H$\beta$ and H$\gamma$ for $1.5~\mathrm{\AA}$ passband. The results of each individual night and light curves are presented in Appendix~\ref{ap:individualTS} and ~\ref{ap:individualTLC}.

\subsection{The Ca$\mathrm{II}$ triplet}

The CaII triplet lines are located at $8498.02~\mathrm{\AA}$, $8542.09~\mathrm{\AA}$ and $8662.14~\mathrm{\AA}$. Only the transit observed with CARMENES (Night 4) is used to analyze these lines, as HARPS-N data do not cover this wavelength range. The telluric contamination observed in this region is low. The second and third lines (ordered in wavelength) are in the limit of two orders i.e. both lines are covered by two different CARMENES orders. The analysis shown here is performed using the orders where these lines are located in a most central region of the order. Although the detection of the lines, at lower SNR, is also possible in the other orders.

In Figure~\ref{fig:resCa} we show the 2D maps of these lines, where the planetary absorption trail is clear, specially for the second line ($\lambda8542~\mathrm{\AA}$) which is the strongest of the triplet. We measure absorption of $0.52\pm0.05~\%$, $0.60\pm0.04~\%$ and $0.55\pm0.06~\%$ in the lines core, respectively, which correspond to an effective radius of ${\sim1.2~R_{P}}$. The transmission spectrum around $\lambda8662~\mathrm{\AA}$ show an smaller second absorption peak in the right side of the main absorption feature. Even the telluric contamination is low in this region, it can be caused by a small telluric residual. The transmission light curves of each individual CaII line and the transmission light curve after combining of all three lines (see Appendix~\ref{ap:individualTLC}) present consistent absorption depth values in comparison with those measured in the transmission spectra (see Table~\ref{tab:resAll}). The transmission light curves and MCMC correlation diagrams can be observed in Appendix~\ref{ap:individualTLC} and \ref{ap:corners}.

As absorption residuals are observed at the CaII triplet lines, we also checked the CaII H \& K lines, which are only covered with HARPS-N observations. Unfortunately, these lines are located at the beginning of the first echelle order of the CCD and the spectra are too noisy in that region to be of any value. These lines are chromospherically sensitive lines, which could present variations caused by the contrast between different regions of the stellar disk \citep{Cauley2018}. However, as expected for this stellar-type, we do not notice variations or emissions in the line cores that could be produced by stellar activity.

\subsection{Na $\mathrm{I}$ doublet}
\label{subsec:Na}

Due to the rotational broadening, the NaI stellar lines expand over a large wavelength range, and a large number of telluric lines fall into the NaI lines, as we showed in Figure~\ref{fig:tell}. Additionally, interstellar medium (ISM) NaI absorption at the stellar lines core is observed for all  nights, affecting the region where we would expect the planetary signal. We used the LISM Kinematic Calculator \citep{2008ISMtool} in order to determine the origin of the interstellar NaI lines, which gives us information about the interstellar medium clouds that our line-of-sight traverses while observing a target. Although we expect this ISM NaI absorption at the same spectral position and to be stable during the night, any small variation of this absorption during the transit could mimic a planetary atmospheric absorption. As first approximation, assuming that the ISM remains totally stable, and not correcting by the stellar reflex motion around the Barycenter of the system, the ISM effects would be removed when dividing each spectrum by the master-out spectrum. 

Because of the large telluric contamination of the NaI stellar lines, we include an additional correction. After computing the ratio between each spectrum and the out-of-transit master spectrum, similarly to \citet{Snellen2010}, we remove the remaining telluric residuals by fitting the time variation in each pixel by, in this case, linear splines using the {\tt UnivariateSpline} class from {\tt scipy} to refuse outliers. The resulting fit is compared to the data, and those points differing more than $5~\%$ from the fit values are set to $1.0$ (we note that the residuals spectra are normalized to the unit). This would not affect the possible planetary signals because they are expected to be smaller than $5~\%$, as observed in \citet{Casasayas2018}.

\begin{figure*}[h!]
\centering
\includegraphics[width=1\textwidth]{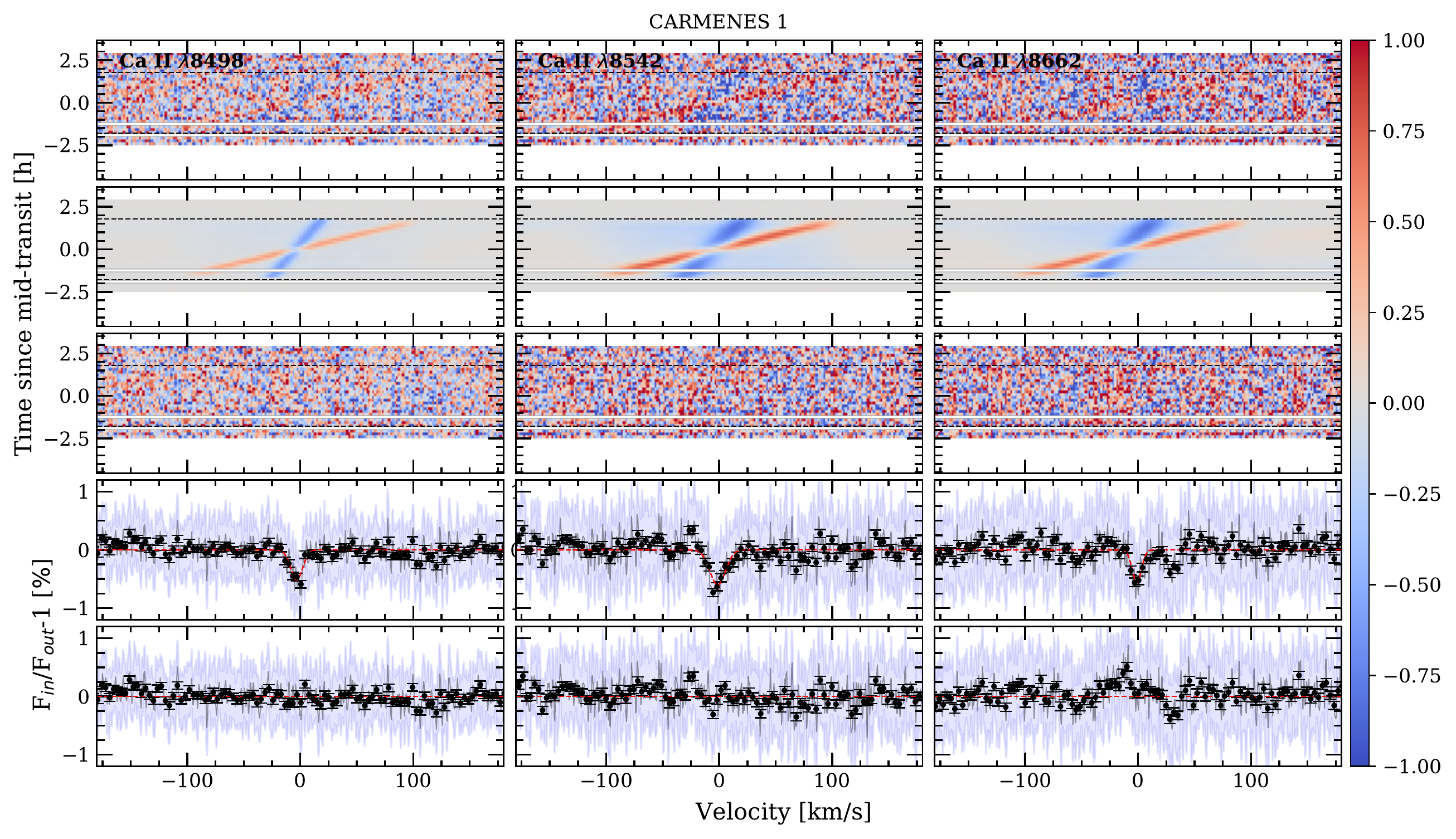}
\includegraphics[ width=1\textwidth]{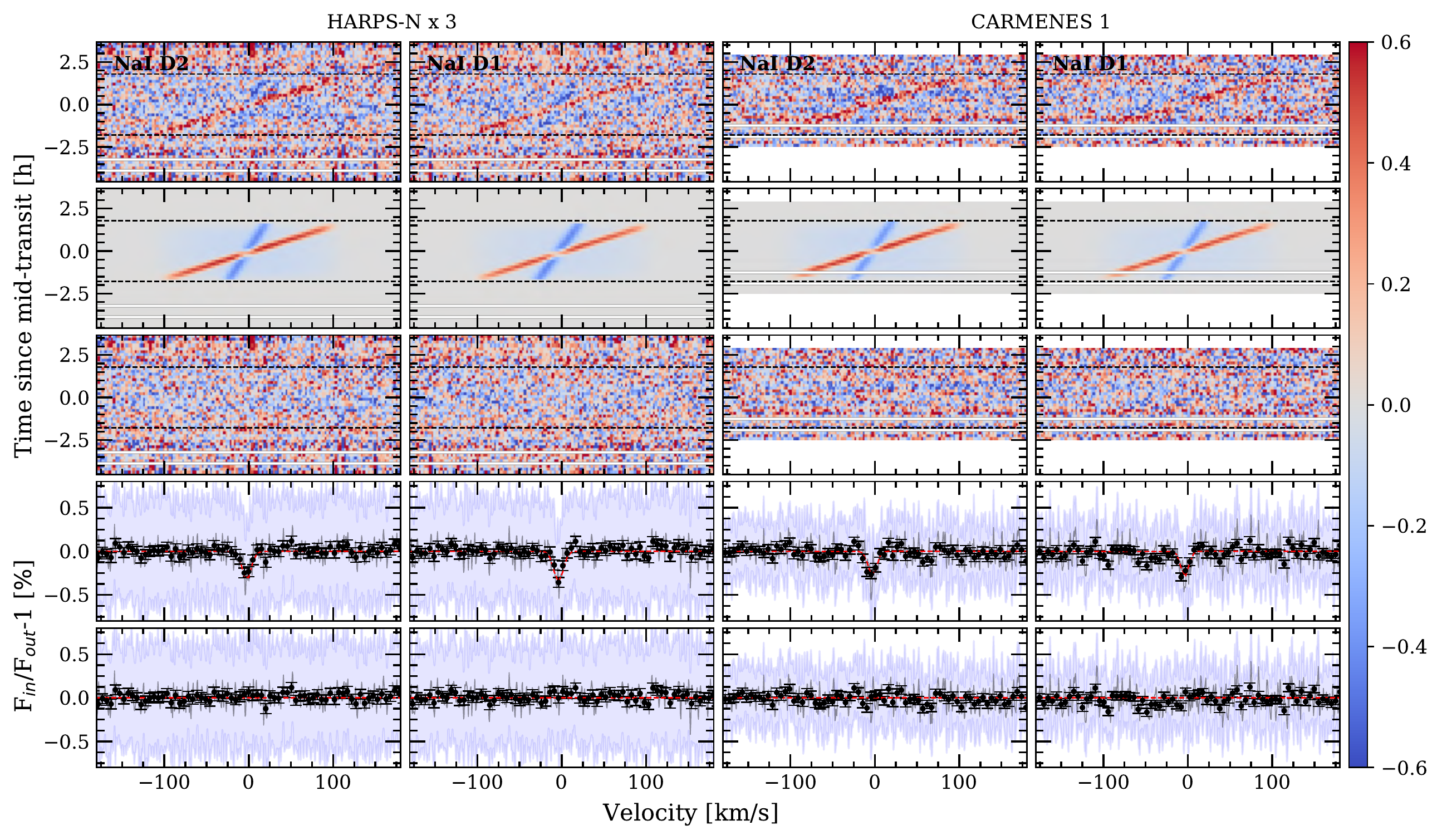}
\caption{Same as Figure~\ref{fig:Ha_res}, but for CaII (top) and NaI lines (bottom). The name of the line is indicated in each top-left panel and the titles indicate the instrument used to retrieve the data. The transmission spectra are binned by $0.1~\mathrm{\AA}$ (black dots). The colour bar indicates the flux relative to the continuum (F$_{in}$/F$_{out}$ -1) in $\%$.}
\label{fig:resCa}
\end{figure*}

As explained in Section~\ref{sec:modeling}, the stellar models computed with {\tt SME} in the NaI lines region assuming solar abundance were observed to be very different compared with the observational data in the same region. For this reason, we fit the abundance, and the final models were computed using the resulting NaI abundance value ([Na/H]$=0.98$). We then fit the residuals after dividing each spectrum by the master out spectrum as we did for the H$\alpha$ in section \ref{subsec:Ha}. The MCMC fitting procedure is applied to each NaI D2 and D1 lines separately (for which we take reference positions at $5889.95~\mathrm{\AA}$ and $5895.92~\mathrm{\AA}$, respectively) and the grid of models is computed for $[0.7-2.5]~R_p$ with steps of $0.1~R_p$. The results for CARMENES and combined HARPS-N data analysis are shown in Figure~\ref{fig:resCa}, and the best-fit parameters are given in Table~\ref{tab:resAll}.

In the 2D maps, the in-transit trail seems to be observable for both D$_2$ and D$_1$ lines and both CARMENES and HARPS-N analysis (see Figure~\ref{fig:resCa}), being clearer in HARPS-N. The MCMC analysis is able to distinguish and characterise the absorption, finding a consistent value of $K_p$ with the one theoretically predicted. When we analyse the nights separately, the D$_2$ line absorption can be reproduced every night except for HARPS-N Night 2, and the D$_1$ line can be reproduced in Night 1, CARMENES data, and when all HARPS-N data are combined. The measured absorption in the line core ($h$) is consistent for both analysis, with $\sim-0.30~\%$, which corresponds to an effective radius of $\sim1.1~R_p$ of absorption in both lines of the NaI doublet. These results are also consistent with the results obtained for Night 1 in \citet{Casasayas2018} applying a different analysis. 

 The light curves and transmission spectra present consistent absorption depth (see values in Table~\ref{tab:resAll}). See 2D maps of each night and light curves in Appendix~\ref{ap:individualTS} and ~\ref{ap:individualTLC}. The transmission light curves are computed for each line of the doublet separately and are then combined using the weighted mean.

\subsection{FeII}

Three other transmission signals are observed when scanning the residuals in the wavelength range covered by both spectrographs. The position of these lines are $5018.38\pm0.01~\mathrm{\AA}$, $5168.96\pm0.01$, and $5316.57\pm0.01~\mathrm{\AA}$, which we attribute to FeII $\lambda5018.43$,  $\lambda5169.03$ and $\lambda5316.61$, respectively. However, we note that due to the large rotation velocity of the star, it is difficult to distinguish a single line within the blended stellar lines. The FeII $\lambda5316~\mathrm{\AA}$ line is also covered by CARMENES at very low S/N (the mean S/N of this order is around $10$). The transmission spectrum obtained shows absorption in the same place as HARPS-N ($5316.61\pm0.02~\mathrm{\AA}$), but we also observe additional features which are not observed in HARPS-N results, and whose origin is not clear, possibly caused by the low S/N (see this results in Appendix~\ref{ap:individualTS}).

\begin{figure*}[h]
\centering
\includegraphics[ width=0.97\textwidth]{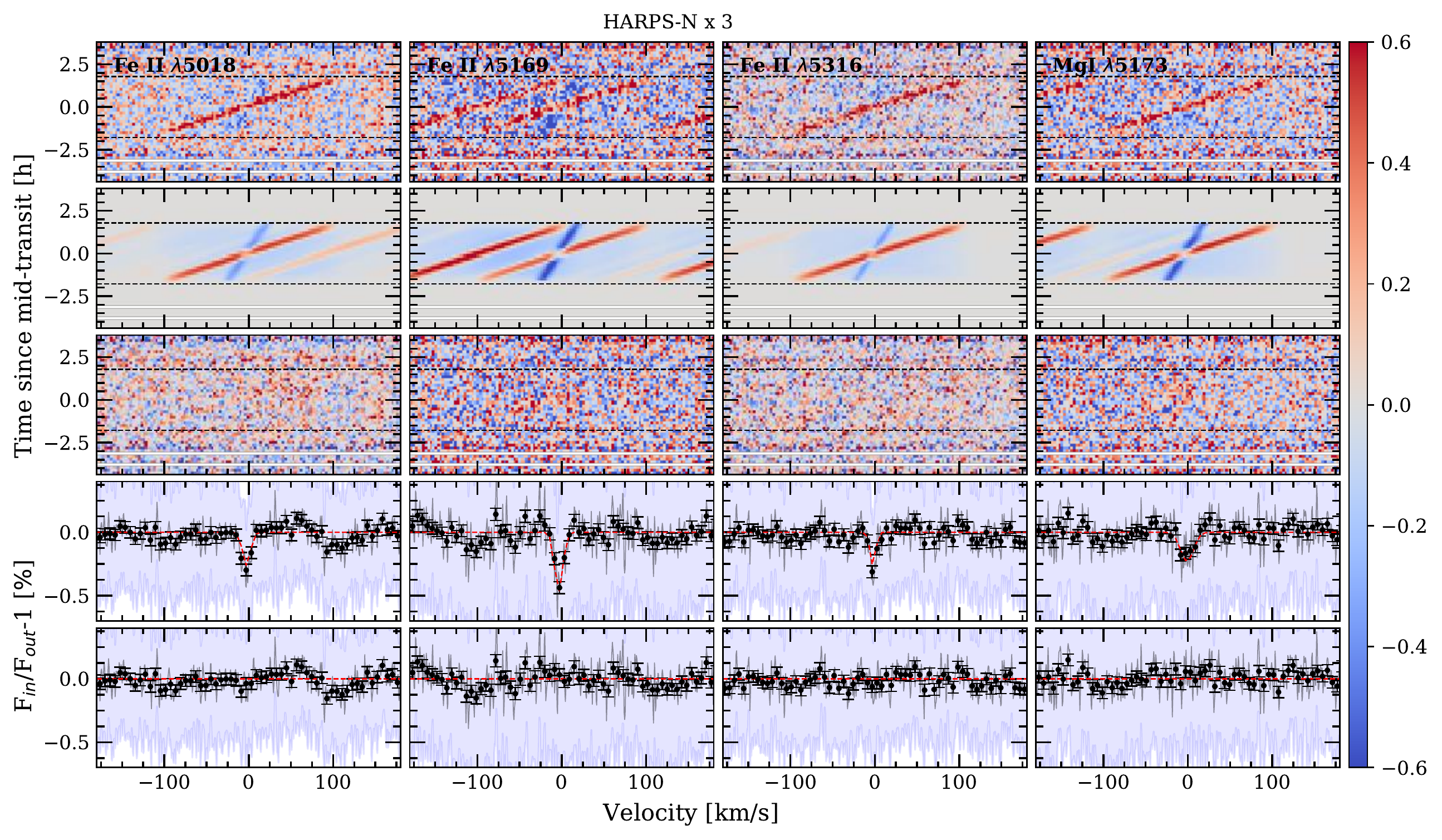}
\caption{ame as Figure~\ref{fig:Ha_res}, but for FeII and MgI lines. The name of the line is indicated in each top-left panel. The transmission spectra are binned by $0.1~\mathrm{\AA}$ (black dots). The colour bar indicates the flux relative to the continuum (F$_{in}$/F$_{out}$ -1) in $\%$.}
\label{fig:resFe}
\end{figure*}

In the same way as for the other lines, we remove the RM and CLV, and find the best-fit parameters (see Figure~\ref{fig:resFe} and Table~\ref{tab:resAll}). A faint trail during the transit is observed for each line 2D map, but stronger for FeII $\lambda5169$ line, which present an absorption of $0.47\pm0.05$ in the core. We note that near this line, we observe a second and fainter residual at $5172.66\pm0.02$. The unique specie that overlaps this position is MgI at $5172.68$, but the significance of this absorption signal is very low ($1.7\sigma$) with an absorption depth of $0.07\pm0.04\%$ in the transmission spectrum for a $0.75\mathrm{\AA}$ passband, and ${\sim -0.2\%}$ of absorption in the line core (see the results for this MgI residual in Figure~\ref{fig:resFe}). 

We compute the transmission light curves of each FeII and MgI lines, and the FeII combined transmission light curve (see Appendix~\ref{ap:individualTLC}). We then measure the absorption depth in the transmission spectra and light curves, with values presented in Table~\ref{tab:resAll}. The results from the transmission spectra and light curves are consistent.

\section{Systematic effects}
\label{sec:tests}

As presented in \citet{2008Redfield}, we can measure the stability of the individual observations by quantifying the systematic effects with the Empirical Monte Carlo (EMC) or bootstrapping diagnostic. The EMC analysis helps us to quantify the systematic errors, since the absorption depth measurements error bars come from the the propagation of the random photon noise from Poisson statistics through the process. This analysis involves the random selection of a number of individual exposures to build the "in-transit" sample and another random selection for the "out-of-transit" sample. With these two samples we then extract the transmission spectrum  and we measure the absorption depth of this result. This process is applied $10~000$ times with different random samples in order to have statistical significance. With this, we are able to check how likely it is that the signal we are observing has planetary origin, or is caused by a random combination of the data.

We assume three different scenarios, as presented in \citet{2008Redfield} and other recent studies  such as \citet{2015A&A...577A..62W} and \citet{2017A&A...602A..36W}. The first scenario is called "out-out" where we randomly split the out-of-transit observations in two samples with the same number of individual observations; one of these two samples will be used as the "in-transit" sample and the other as the "out-of-transit" sample. The second scenario is called "in-in" and constructed in the same way, but now splitting the in-transit observations. The final scenario is the "in-out" and represent the atmospheric absorption case. Here, the "in-transit" and "out-of-transit" samples are built with the in- and out-of-transit observations, respectively. For the "in-in" and "out-out" scenarios, we select the same number of exposures included in the "in-transit" and in the "out-of-transit" samples. This means that in the "in-in" scenario this value is half the total number of real in-transit observations and in the "out-out" is half the real out-of-transit observations. For the "in-out" scenario the number of exposures included in each sample changes randomly, never being less than half the total number of real in- or out-of-transit exposures and always accomplishing the number ratio of real in-transit to the out-of-transit observations. We note that each sample including in-transit data is corrected by the CLV and RME effects, computed considering the best fit parameters of each line presented in Section~\ref{sec:disc}.

This analysis is applied to each line here analysed. The distributions  for a $1.5~\mathrm{\AA}$ passband can be observed in Figure~\ref{fig:hist}. In all cases the "out-out" and "in-in" scenarios show distributions centered to $0~\%$ absorption while the "in-out" scenario is centered to a different position. The clearest case is the H$\alpha$ line, where the absorption scenario is centered to $0.36\pm0.14~\%$ and $0.40\pm0.14~\%$ (the error corresponds to the standard deviation of the distribution) for HARPS-N and CARMENES analysis, respectively, consistent with the absorption depth measured from the final transmission spectrum (see Table~\ref{tab:resAll}). In case of the CaII we observe the "in-out" samples centered on $0.10\pm0.08$, $0.12\pm0.11$ and $0.11\pm0.09$ absorption, for the $\lambda8498~\mathrm{\AA}$, $\lambda8542~\mathrm{\AA}$ and $\lambda8662~\mathrm{\AA}$ lines, respectively. 

For the NaI doublet case, the "in-out" scenario distributions clearly present different characteristics in comparison with the other scenarios for HARPS-N analysis. Both D$_2$ and D$_1$ distributions are centered to $0.07\pm0.05~\%$. In case of CARMENES, the three different scenarios give closer results; in this case, the distributions are centered on $0.05\pm0.09~\%$ for the D$_2$ line and $0.12\pm0.08\%$ for the D$_1$ line. The absorption scenario of H$\beta$ presents a distribution centered to $0.14\pm0.08~\%$. Finally, for H$\gamma$ the scenarios for all three distribution overlapped, being the "in-out" distribution centered on $0.04\pm0.09~\%$ absorption. 

The FeII lines obtained with HARPS-N data show "in-out" distributions centered to $0.03\pm0.03$, $0.04\pm0.04$ and $0.04\pm0.03$, for the $\lambda5018$, $\lambda5169$ and $\lambda5316$, respectively, close to the "in-in" and "out-out" scenarios. Similarly to H$\gamma$, the "in-out" scenario distributions obtained for FeII $\lambda5316$ using CARMENES data and MgI $\lambda5173$ using HARPS-N data, show mainly indistinguishable distributions with respect "in-in" and "out-out" scenarios, centered to $0.09\pm0.15$ and $0.03\pm0.05$, respectively.

\section{Discussion and conclusions}
\label{sec:disc}

\citet{Casasayas2018} presented the analysis of one HARPS-N transit observation of MASCARA-2b, Night 1 in this paper, and reported the detection of H$\alpha$ and NaI in its atmosphere, and a hint of H$\beta$ detection. Here we have repeated the analysis after combining two more transit observations of MASCARA-2b with HARPS-N and one more with the CARMENES spectrograph. After analysing the 2D residual transit maps, the transmission spectra, the transmission light curves and EMC results, we conclude that we have detected a hint to the presence of a possible ionosphere around MASCARA2-b, including the detection of CaII, FeII, NaI, H$\alpha$, H$\beta$, and H$\gamma$ transitions. We note that, with respect to the previous study, we are using here different telluric correction and only the spectra taken in the fully in-transit time are combined when computing the transmission spectra (the ingress and egress data are excluded from this sample). We summarise the results obtained in Table~\ref{tab:resAll} and Figure~\ref{fig:resAll}.

\renewcommand{\thefootnote}{\fnsymbol{footnote}}
\begin{table*}[]
\small
\centering
\caption{Summary of the results obtained in the analysis presented here, for the different species and instruments. Hx3 corresponds to the analysis of three HARPS-N nights combined, and Cx1 to the analysis of the unique transit observed with CARMENES.}

\begin{tabular}{llcccccccc}
\\[-1em] \hline\hline  \\[-1em]
                           &     & $h$ {[}$\%${]} & $R_{\lambda}(h)$ {[}R$_p${]} & $v_{wind}$ {[}km/s{]} & FWHM {[}km/s{]} & $K_p$ {[}km/s{]} & AD$_{\mathrm{TS}}^{1.5\mathrm{\AA}}$  {[}$\%${]} $^($\footnotemark[1]$^)$ & AD$_{\mathrm{TLC}}^{1.5\mathrm{\AA}}$ {[}$\%${]} $^($\footnotemark[3]$^)$& AD$_{\mathrm{EMC}}^{1.5\mathrm{\AA}}$  {[}$\%${]} $^($\footnotemark[5]$^)$\\  \\[-1em] \hline  \\[-1em]
{H$\alpha$} & Hx3 &  $-0.68\pm0.06$   & $1.22\pm0.04$  &  $-3.0\pm1.2$   &  $19.0\pm1.6$  & $165.6\pm16.7 $ & $0.33\pm0.05$   &  $0.30\pm0.05$     &  $0.36\pm0.14$  \\ \\[-1em] \\[-1em]
                           & Cx1 & $-0.85\pm0.03$ &   $1.27\pm0.02$    &   $-4.5\pm0.5$    & $22.6\pm0.9$   &$166.2\pm7.4 $ & $0.44\pm0.04$                          &   $0.43\pm0.05$      & $0.40\pm0.14$    \\ \\[-1em] \hline \\[-1em]
H$\beta$                   & Hx3 &  $-0.45\pm0.05$     &   $1.15\pm0.03$    &    $-1.2\pm1.4$    &  $19.4\pm2.5$  & $136.2\pm18.6$ & $0.13\pm0.04$     &  $0.17\pm0.04$  &  $0.14\pm0.08$ \\  \\[-1em] \hline  \\[-1em]

H$\gamma$                  & Hx3 & $-0.38\pm0.08$   &  $1.13\pm0.05$     &    $-2.3\pm2.7$    &   $16.6\pm4.2$    & $135.0\pm34.8 $&  $0.09\pm0.05$                &      $0.07\pm0.09$      & $0.08\pm0.04$ \\   \\[-1em] \hline  \\[-1em]
CaII                      & Cx1 & $-0.56\pm0.05$  &  $1.18\pm0.03$  &  $-1.9\pm0.6$  & $9.2\pm1.0$ &  $157.7\pm8.2 $ & $0.16\pm0.04$       &   $0.14\pm0.04$  & $0.11\pm0.09$   \\ \\[-1em] \hline  \\[-1em]
{NaI}      & Hx3 & $-0.34\pm0.05$   & $1.11\pm0.03$  &  $-3.1\pm0.9$ & $9.2\pm2.0$   & $182.5\pm14.3 $  & $0.05\pm0.03$     &  $0.08\pm0.03$   &  $0.07\pm0.05$  \\ \\[-1em] \\[-1em]
                           & Cx1 & $-0.29\pm0.04$    & $1.10\pm0.03$ & $-3.2\pm0.7$ & $8.0\pm1.2$   & $176.6\pm11.7 $ & $0.04\pm0.03$   &     $0.03\pm0.04$   &  $0.08\pm0.08$ \\   \\[-1em] \hline  \\[-1em]
FeII                      & Hx3 &      $-0.33\pm0.05$  &  $1.11\pm0.03$  &  $-2.8\pm0.8$  & $7.2\pm1.2$ & $174.4\pm14.0 $ &  $0.04\pm0.03$        &     $0.01\pm0.03$      &   $0.04\pm0.04$ \\    \\[-1em] \hline\hline  \\[-1em]

\end{tabular} \\

\begin{tablenotes}
\item Notes. $^($\footnotemark[1]$^)$Absorption depth measured in the transmission spectrum for a $1.5~\mathrm{\AA}$ passband. $^($\footnotemark[3]$^)$Absorption depth measured in the transmission light curve for a $1.5~\mathrm{\AA}$ passband. $^($\footnotemark[5]$^)$Center of the EMC distribution computed for a $1.5~\mathrm{\AA}$ passband.
\end{tablenotes}
\label{tab:resAll}
\end{table*}
\renewcommand{\thefootnote}{\arabic{footnote}}

\begin{figure*}[h!]
\centering
\includegraphics[width=0.92\textwidth]{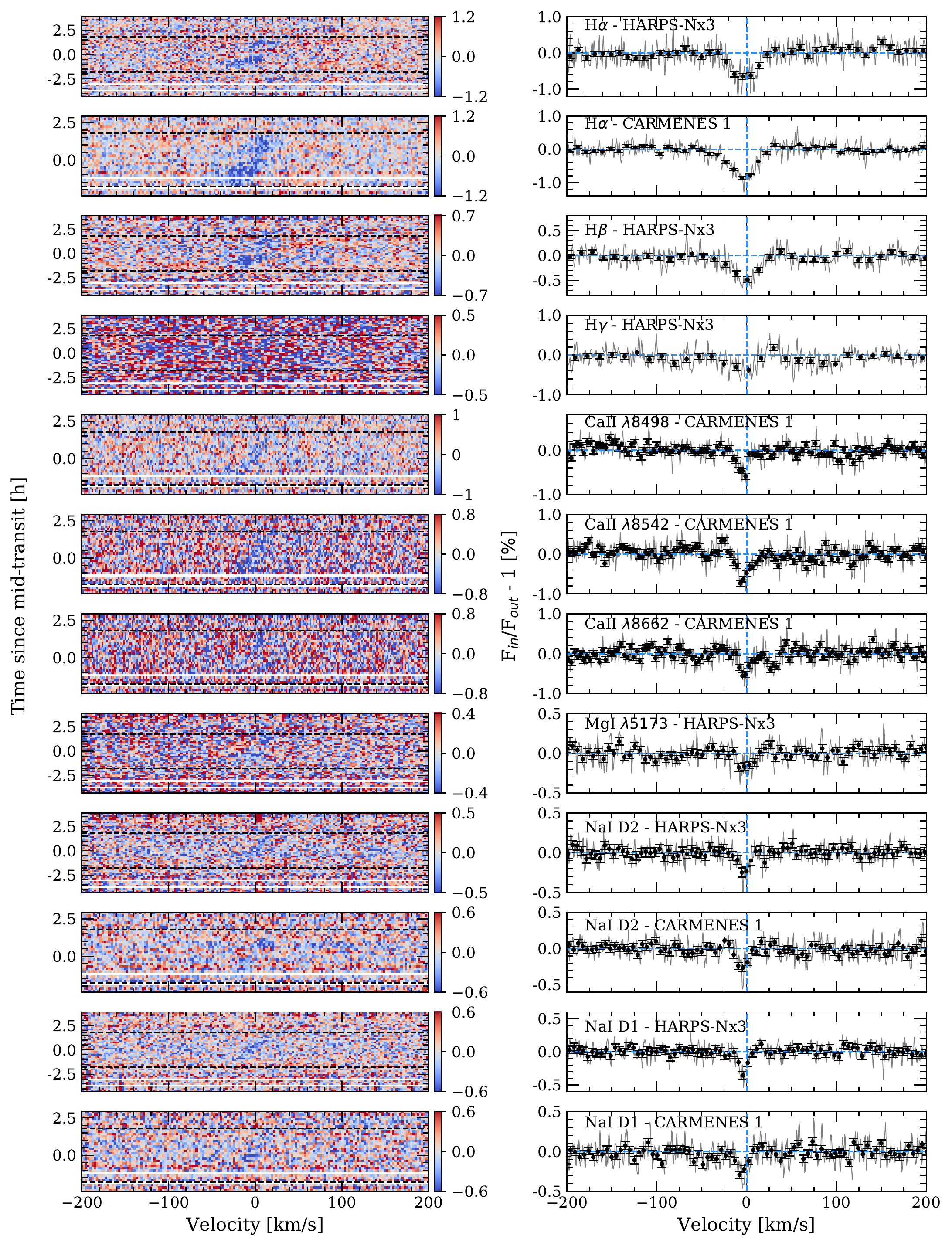}
\caption{Summary of the 2D maps with the remaining absorption after the CLV and RM correction (left column) and final transmission spectra (right column) of all species analysed. Each row corresponds to one line, whose name is indicated inside each transmission spectrum figure. The residual maps data is presented binned by $0.05~\mathrm{\AA}$ and $0.002$ in orbital phase in order to have better contrast. The transmission spectrum of Balmer lines are shown binned by $0.2~\mathrm{\AA}$ and the remaining lines by $0.1~\mathrm{\AA}$ (black dots). The colour bar indicates the flux relative to the continuum (F$_{in}$/F$_{out}$ -1) in $\%$. }
\label{fig:resAll}
\end{figure*}

\begin{figure*}[h!]
\ContinuedFloat
\centering
\includegraphics[width=0.92\textwidth]{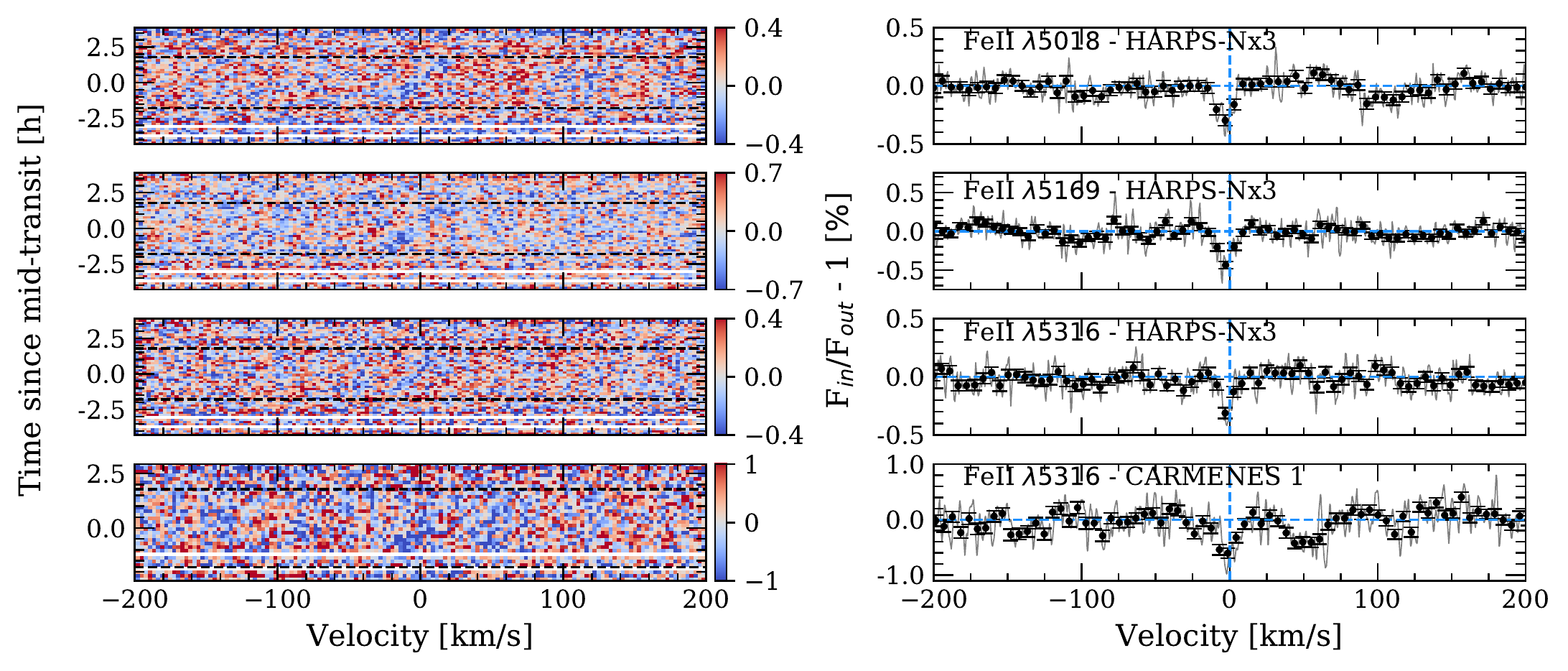}
\caption{Continued from previous page.}
\label{fig:resAll}
\end{figure*}

Focusing on the Balmer lines, the H$\alpha$ absorption by the planetary atmosphere can be observed by eye in the residuals of each single night, appearing during the transit and ranging different radial-velocities than the RME, which is also clearly observed. The planet orbital radial-velocity semi-amplitude ($K_p$) values measured are consistent (considering the error bars) with those from literature. For H$\beta$, on the other hand, the absorption by the planetary atmosphere is not clear if we do not combine the three transits observed with HARPS-N. But when doing so, the absorption in the transmission spectrum and the light curves is statistically significant ($5\sigma$ and $6\sigma$, respectively, for a $0.75~\mathrm{\AA}$ passband). Finally, the H$\gamma$ analysis of each individual night show mainly flat transmission spectra and light curves, but a faint H$\gamma$ signal moving at the planetary speed for MASCARA-2b results when combining all the HARPS-N data ($3\sigma$ for a $0.75~\mathrm{\AA}$ passband).

Recently, \citet{YanKELT9} detected H$\alpha$ in the atmosphere of KELT-9b, the ultra hot Jupiter with highest day-side temperature ($4600~\mathrm{K}$). They measure $1.15\pm0.05\%$ of absorption which corresponds to an hydrogen atmosphere around $1.64$ times the radius of the planet, close to the Roche Lobe ($1.91^{+0.22}_{-0.26}R_p$), and an estimated mass loss rate of ${\sim10}^{12}\mathrm{g~s^{-1}}$. In extreme cases, photoevaporation causes a transonic planetary wind and the mass loss rate could affect the planetary evolution \citep{Salz2015}. \citet{Cauley2019} reproduced the H$\alpha$ detection in KELT-9b and also reported significant absorption in H$\beta$. Here, for MASCARA-2b, the measurement of the H$\alpha$ absorption is ${\sim 0.7\%}$, which corresponds to an atmospheric extension of $1.2~R_p$. Considering that the lower-limit of the scale height ($H$) is $281~\mathrm{km}$ (calculated with $\mu = 2.3$ and the gravity upper-limit from \citet{2017Lund}), the atmosphere is extended by $545~H$. The atmosphere is expanded due to the gas heating caused by the irradiation received from the host star. For H$\beta$, on the other hand, the $0.45\%$ of absorption corresponds to $1.15~R_p$ ($525~H$). Considering the mass upper-limit from \citet{2017Lund}, we estimate that the Roche Lobe radius \citep{1983Eggleton} of MASCARA-2b is $<3.6~R_p$. If we now assume a mass of $2~$M$_J$ this radius decreases to $3~R_p$. The effective radius measured for the different atoms/ions do not reach the Roche Lobe in any of those cases. The measured FWHM of H$\alpha$ line is ${\sim20}$~km/s which is significantly lower than that of KELT-9b (${\sim51}$~km/s). The large FWHM of H$\alpha$ in KELT-9b is a result of high density of excited hydrogen atoms that causes the absorption to be optically thick at the line centre ($\tau\approx57$). According to our estimation using the analytic equation (3) in \citet{2017Huang}, the optical depth at the H$\alpha$ line centre of MASCARA-2b is small ($\tau\approx2$), indicating that the density of excited hydrogen is much lower than in KELT-9b. Such a lower density also explains why the effective radius of H$\alpha$ line does not reach a high altitude like in KELT-9b.
Thus, we conclude the Balmer lines absorption in MASCARA-2b are from relatively low altitude atmosphere which is far away from the Roche Lobe and therefore can not be used to estimate the mass loss rate. 

For CaII, the individual lines of the triplet are clearly detected in the CARMENES night that covers this wavelength range. The trail of the strongest line ($8\sigma$ of absorption significance for $0.75~\mathrm{\AA}$) at $\lambda8542~\mathrm{\AA}$ can be observed in the 2D maps tracing the radial-velocity change of the planet during the transit, while the other two lines of the triplet are also significant but fainter (around $5\sigma$). We note that this results from the analysis of only one transit. Using the combination of three HARPS-N nights, we also observe the presence of FeII ions in MASCARA-2b atmosphere at the $2\sigma$ level measured in the absorption depth for $0.75~\mathrm{\AA}$ passband, and an averaged significance around $7\sigma$ in the lines core. Finally, the traces of NaI planetary absorption are observable in each individual night, although the true absorption values retrieved here could be affected by the strong telluric and interstellar contamination in this region of the spectrum. The significance of these lines is around $7\sigma$ in the lines core and $2-3\sigma$ when averaging the absorption in a $0.75~\mathrm{\AA}$ passband, for both instruments.

Absorption excess in neutral Ca lines was firstly reported by \citet{2013A&A...557A..56A} in HD~209458b atmosphere, and the Ca$^+$ triplet lines variation during a transit was studied for the first time by \citet{Khalafinejad2018} as indicator of stellar activity. \citet{Hoeijmakers2018} detected Fe, Fe$^+$ and Ti$^+$ atoms in the KELT-9b atmosphere, using the cross-correlation technique. For MASCARA-2b, we are able to resolved individual lines of Ca$^+$ and Fe$^+$ in transmission. The individual Fe$^+$ lines and neutral Mg at $\lambda5173$ observed here were also observed in the atmosphere of KELT-9b by \citet{Cauley2019}. Fe$^+$, together with Ti$^+$, Ca$^+$, is expected to be more abundant than its neutral form \citep{Helling2019} due to thermal ionisation. In the high altitude atmosphere of MASCARA-2b, the temperature could be very high causing the Fe$^+$ more abundant than neutral Fe. Our detection of these species, in a planet with equilibrium temperature almost $2000$ K colder that KELT-9b, could indicate that either a large difference between actual and equilibrium temperatures or that there are strong mechanism of transport to move atmospheric species from day-side to the planet's terminator. Our detection of neutral Na falls in line also with the recent detection of this species in the atmosphere of WASP-76b, a hot planet with equilibrium temperature similar to MASCARA-2b \citep{Seidel2019arXiv190200001S}. 

\begin{figure}[h]
\centering
\includegraphics[width=0.43\textwidth]{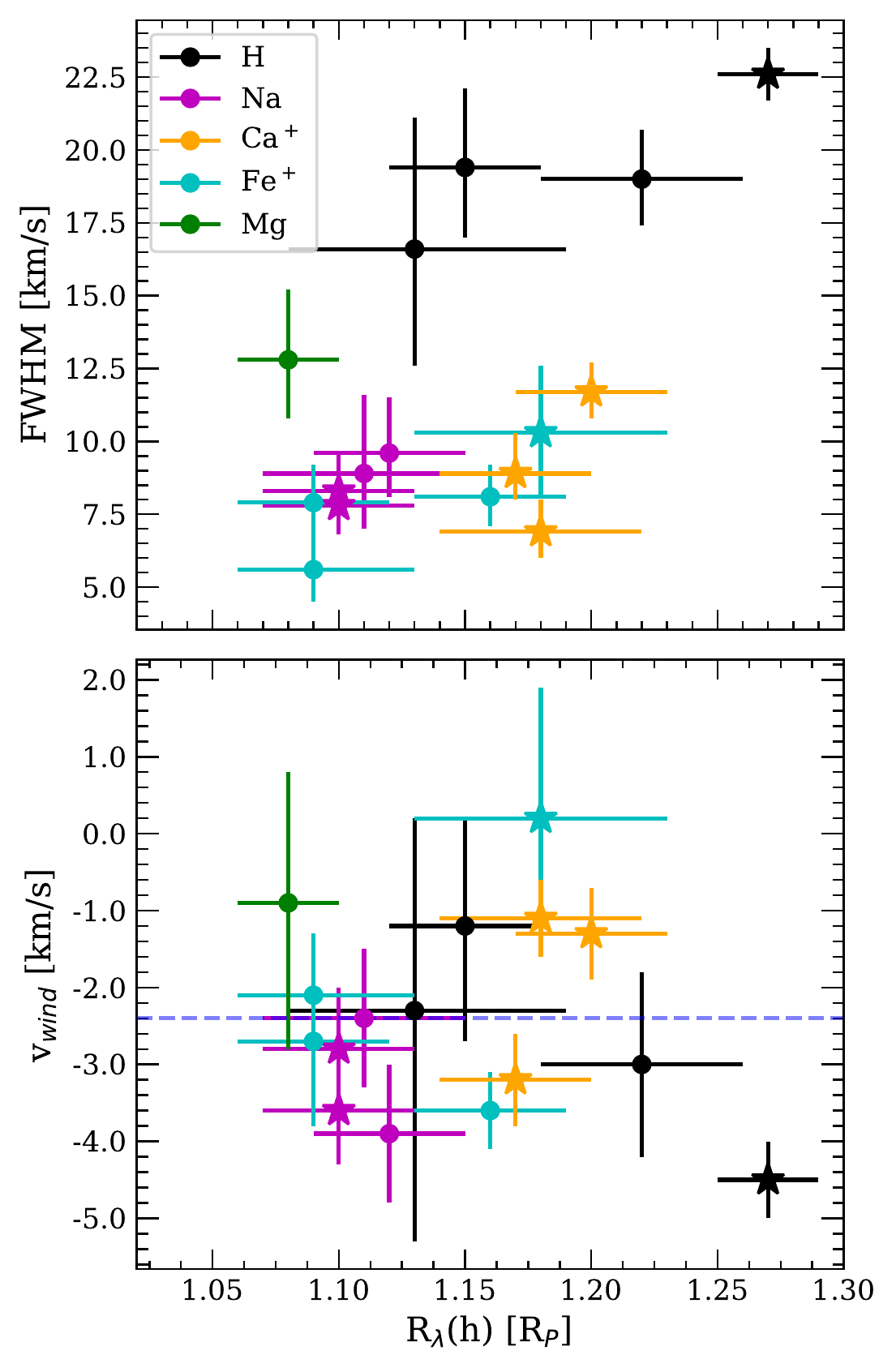}
\caption{Top: FWHM values versus $R_{\lambda}(h)$. Bottom: $v_{wind}$ values versus $R_{\lambda}(h)$. The measurements obtained with HARPS-N data are shown in dots while those obtained with CARMENES data are marked with stars. The horizontal blue dashed line show the mean $v_{wind}$ value at $-2.4\pm1.0~\mathrm{km/s}$. All values are extracted from Table~\ref{tab:resAll}.}
\label{fig:sumparams}
\end{figure}

For all species, the resulting absorption signatures are blueshifted with respect their reference values, with velocities larger than $1~\mathrm{km/s}$, while the instrumental drifts for the spectrographs used here are expected to be around $10~\mathrm{m/s}$. These blueshifts can be associated to atmospheric winds. For H$\alpha$ and NaI, the blueshift measured in CARMENES and HARPS-N data are consistent. For the H$\beta$ and H$\gamma$ lines, the significance is low ($<0.9\sigma$) to determine if the measured blueshift is real. On the other hand, the FWHMs measured for the different lines are consistent for those measured with HARPS-N and CARMENES. The Balmer lines are measured to be broader than CaII, NaI and FeII lines. The line spread function (LSF) of HARPS-N is $2.7~\mathrm{km/s}$ while for CARMENES is $3.5~\mathrm{km/s}$. In all cases, the FWHM measurements are larger than the resolution element of those instruments, and are probably associated to kinematic and/or thermal broadening. In Figure~\ref{fig:sumparams} we relate the velocity shifts ($v_{wind}$) and FWHM measurements with respect the absorption ($R_{\lambda}(h)$). In the $v_{wind}$ versus absorption relation no obvious correlation is observed, with a mean $v_{wind}$ value around $-2.4\pm1.0~\mathrm{km/s}$. For FWHM versus $R_{\lambda}(h)$, however, each element relation could be explained by a linear regression. \citet{Koll2018} studied the magnitude of the wind speeds expected in hot Jupiters, as a result of the incident flux that they receive and tidally locked spin states. Considering this study, the average wind measurement of $-2.4\pm1.0~\mathrm{km/s}$ is in agreement with the estimated values, for a planet with the equilibrium temperature of MASCARA-2b, assuming that the dissipation is caused by shear instabilities (equation 15).

Ultra hot Jupiters lie in the temperature transition region between gas-giants and stars, some of them having  very similar temperatures as the coldest stars. As such they are useful case tests to study the atmospheric chemistry, understand planetary mass loss and provide constrains for planetary theories of formation and evolution. As presented in several studies (\citealt{Parmentier2018}, \citealt{Arcangeli2018} and \citealt{Bell2018}), we expect these extremely hot planets to present close to stellar-like atmospheric temperatures. The stellar radiation heats the atmospheric gas to temperatures considerably higher than $3000$K causing the day-side atmosphere to be composed of atoms rather than molecules in addition to an increased fraction of ions (see discussion in \citet{2019arXiv190304565H}. Such a high irradiation leads to large day-night temperature differences and therefore we expect these planets to present hot days and cold nights, in particular for those planets with short orbital periods. For WASP-18b, for example, a planet with an effective temperature very close to MASCARA-2b, detailed theoretical studies show that we expect two very different sides, a cloudy night-side depleted of elements (which are then left to form atoms, ions, molecules and will make up the atmosphere of the planet) and a cloud-free day-side that forms a thermal ionosphere \citep{Helling2019}. As the day-side is $\approx 2500$K hotter than the night-side on WASP-18b, hydrogen is present in its atomic form (HI) such that n(HI)>n(H$_2$)$\gg$n(H$_2$O) in the low-pressure, upper atmospheric regions. The same pattern holds for most atoms, including Na, K, Ca, Ti, Al, Fe, Mg, and Si. Only at higher pressures, molecule like H$_2$,  SiO, AlH  are the dominating species of their element. Na, K, Ca, Ti, Al, Fe, Mg, and Si are singly ionised, with Na$^+$, Ca$^+$, K$^+$, Al$^+$, and Ti$^+$ being more abundant than their neutral atomic/ionic form. On the night-side, the thermal ionisation is smaller and the elements remain in their neutral state of ionization or are bound in molecules. The detection of HI, NaI, Ca$^+$ (CaII) and Fe$^+$ (FeII) in MASCARA-2b using transmission spectroscopy data from the planet's terminator is consistent with this theoretical works.
We note, however, that transmission spectroscopy does not probe the day-side directly but rather indirectly through the observation of the terminator regions. The terminator regions are transition regions between the day- and the night-side where the temperature differences will not be as large as between the day- and night-side. The effect on the chemistry will therefore also be less pronounced. The observation of HI, NaI, Ca$^+$  and Fe$^+$ is therefore indicative for rather warm terminator regions. The blueshifted HI, NaI, Ca$^+$  and Fe$^+$ absorption suggestes that strong winds emerge from the day-side transporting hot and considerably ionised material into the respective terminator region, hence, providing us with information about the chemistry on the day-side, too. We would therefore expect the evening terminator to be geometrically more extended than the cooler morning terminator of MASCARA-2b. We note, however, that with the observations presented here we can not conclude if the heat is coming from the day-side or we are observing at very high altitudes where the atmosphere is hotter.

Thus, the theoretical picture of UHJs possessing extended ionospheres and two differentiated day-side/night-side chemistry's is consistent with our MASCARA-2b results. However, more observational evidence is desirable to underpin this hypothesis such as CCF studies. For this planet the detection of several species at the same time has become possible with 3-4 meter class telescopes. UHJs present an exciting opportunity to study their detailed composition, with state-of-the-art spectrographs such as ESPRESSO, which will be able to characterize the H$\gamma$ features observed in MASCARA-2b and possibly other fainter elements predicted by the theoretical models. In the near future high-resolution spectrographs, such as HIRES for the ELT, will be able to disentangle the complex picture of UHJ atmospheres through the simultaneous detection of multiple species in their atmospheres.


\begin{acknowledgements}
The authors would like to thank Dr Sven Buder for his useful comments and discussion about the Na abundances. Based on observations made with the Italian Telescopio Nazionale Galileo (TNG) operated on the island of La Palma by the Fundación Galileo Galilei of the INAF (Istituto Nazionale di Astrofisica) at the Spanish Observatorio del Roque de los Muchachos of the Instituto de Astrofisica de Canarias. CARMENES is an instrument for the Centro Astron\'omico Hispano-Alem\'an de Calar Alto (CAHA, Almer\'ia, Spain) funded by the German Max-Planck-Gesellschaft (MPG), the Spanish Consejo Superior de Investigaciones Cient\'ificas (CSIC), the European Union through FEDER/ERF~FICTS-2011-02 funds, and the members of the CARMENES Consortium. This work is partly financed by the Spanish Ministry of Economics and Competitiveness through project ESP2016-80435-C2-2-R. G.C. acknowledges the support by the National Natural Science Foundation of China (Grant No. 11503088, 11573073, 11573075) and by the project ``Technology of Space Telescope Detecting Exoplanet and Life'' from National Defense Science and Engineering Bureau civil spaceflight advanced research project (D030201). F.Y. acknowledges the support of the DFG priority program SPP 1992 "Exploring the Diversity of Extrasolar Planets (RE 1664/16-1). A.F. acknowledges the support from JSPS KAKENHI Grant Number JP17H04574. This work is partly supported by JSPS KAKENHI Grant Numbers JP18H01265 and 18H05439, and JST PRESTO Grant Number JPMJPR1775. This article is based on observations made with the MuSCAT2 instrument, developed by ABC, at Telescopio Carlos Sánchez operated on the island of Tenerife by the IAC in the Spanish Observatorio del Teide. This work made use of PyAstronomy and of the VALD database, operated at Uppsala University, the Institute of Astronomy RAS in Moscow, and the University of Vienna.

\end{acknowledgements}

%
%
\bibliographystyle{bibtex/aa.bst} 
\bibliography{bibtex/aa.bib} 





   
  



%

%
\onecolumn
\begin{appendix} 

\section{Individual transmission spectra}
\label{ap:individualTS}
We present here the transmission spectra obtained for each individual night.

\subsection{H$\beta$}

\begin{figure}[h]
\centering
\includegraphics[width=0.98\textwidth]{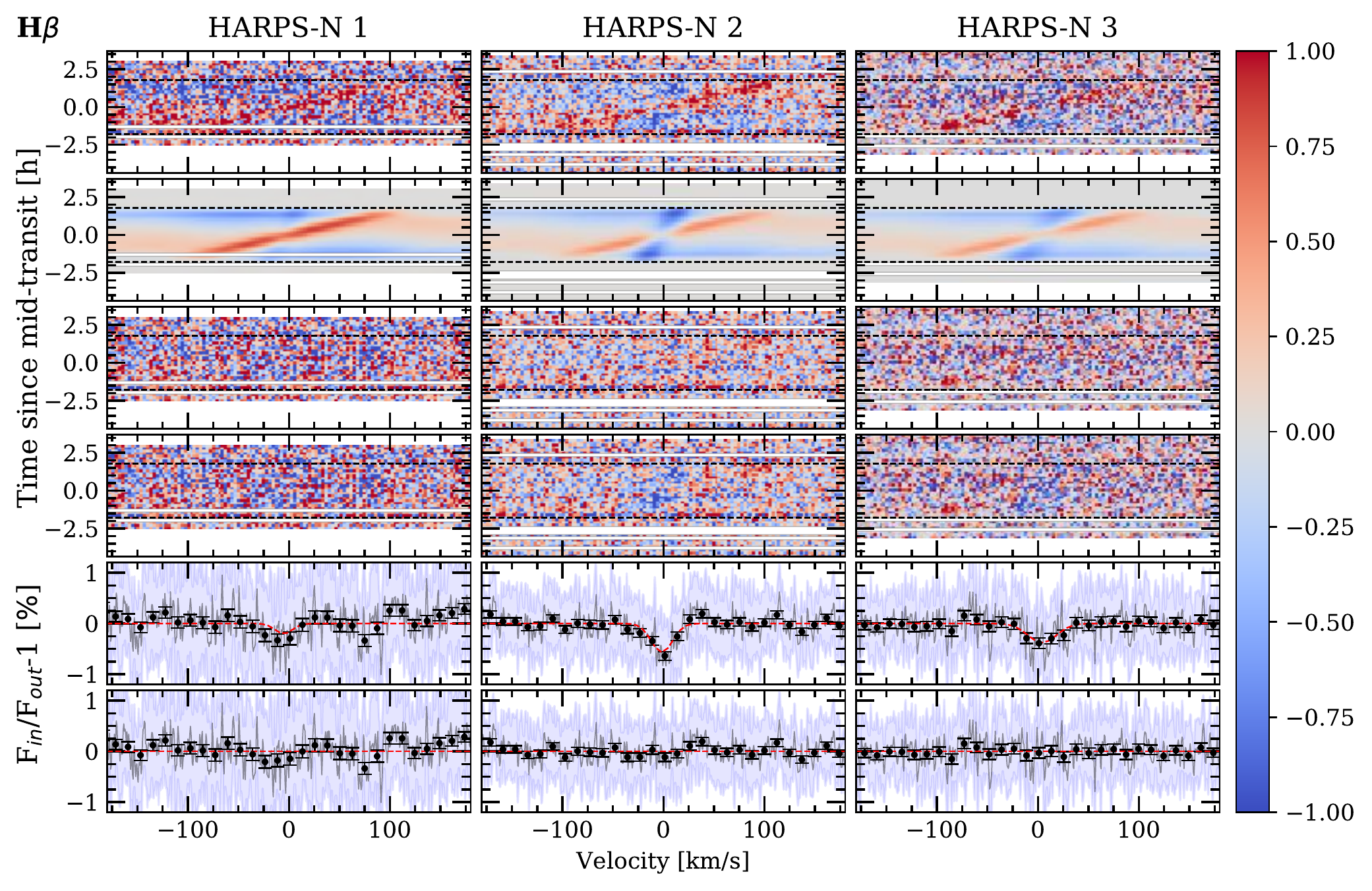}
\caption{Results around the H$\beta$ of Night 1 (left column), Night 2 (middle column), and Night 3 (right column). First row (top): results after dividing each spectrum by the out-of-transit master spectrum. Second row: best fit model of the residuals. The RME is observed in red color (see color bar) and the absorption and CLV in blue (negative relative flux). Third row: residuals when subtracting the best fit model (second row) to the data (first row). Fourth row: absorption signature remaining after correcting the CLV and RME effects from first row residuals. Fifth row: transmission spectrum. Sixth row: residuals of removing the Gaussian profile model to the data in fourth row. In black dots we show the result binned by $0.2~\mathrm{\AA}$, the light blue region is the standard deviation of the residuals and the red line is the Gaussian computed with the best-fit parameters. In all panels, except for the last two rows, the data is shown with $0.05~\mathrm{\AA}$ bins in wavelength and $0.002$ in orbital phase. }
\label{fig:ts_Hb_indiv}
\end{figure}

\newpage
\subsection{H$\gamma$}

\begin{figure}[h]
\centering
\includegraphics[width=0.98\textwidth]{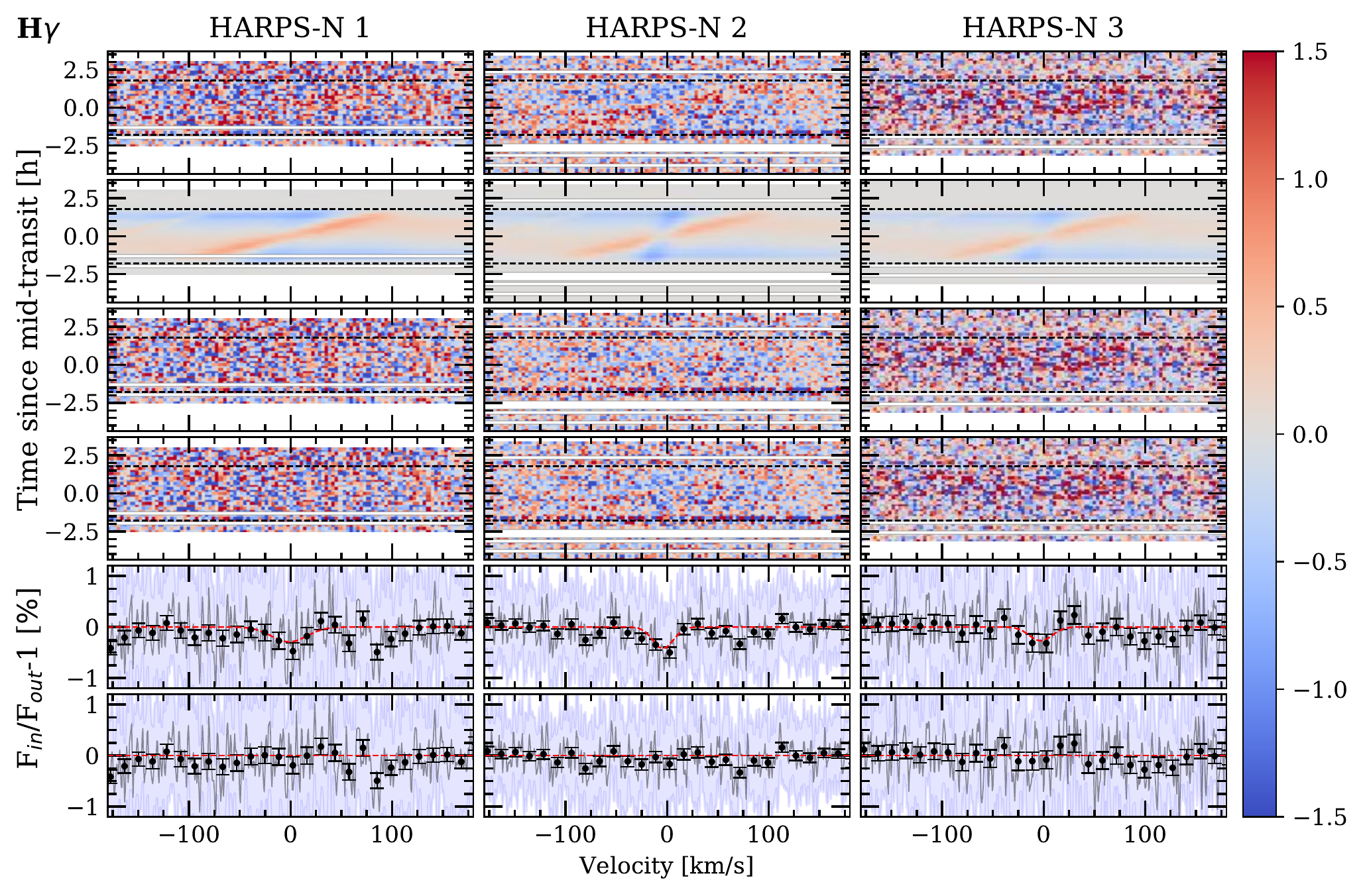}
\caption{Same as Figure~\ref{fig:ts_Hb_indiv} but for H$\gamma$ line.}
\label{fig:ts_Hg_indiv}
\end{figure}

\newpage

\subsection{NaI D$_2$}

\begin{figure}[h]
\centering
\includegraphics[width=0.98\textwidth]{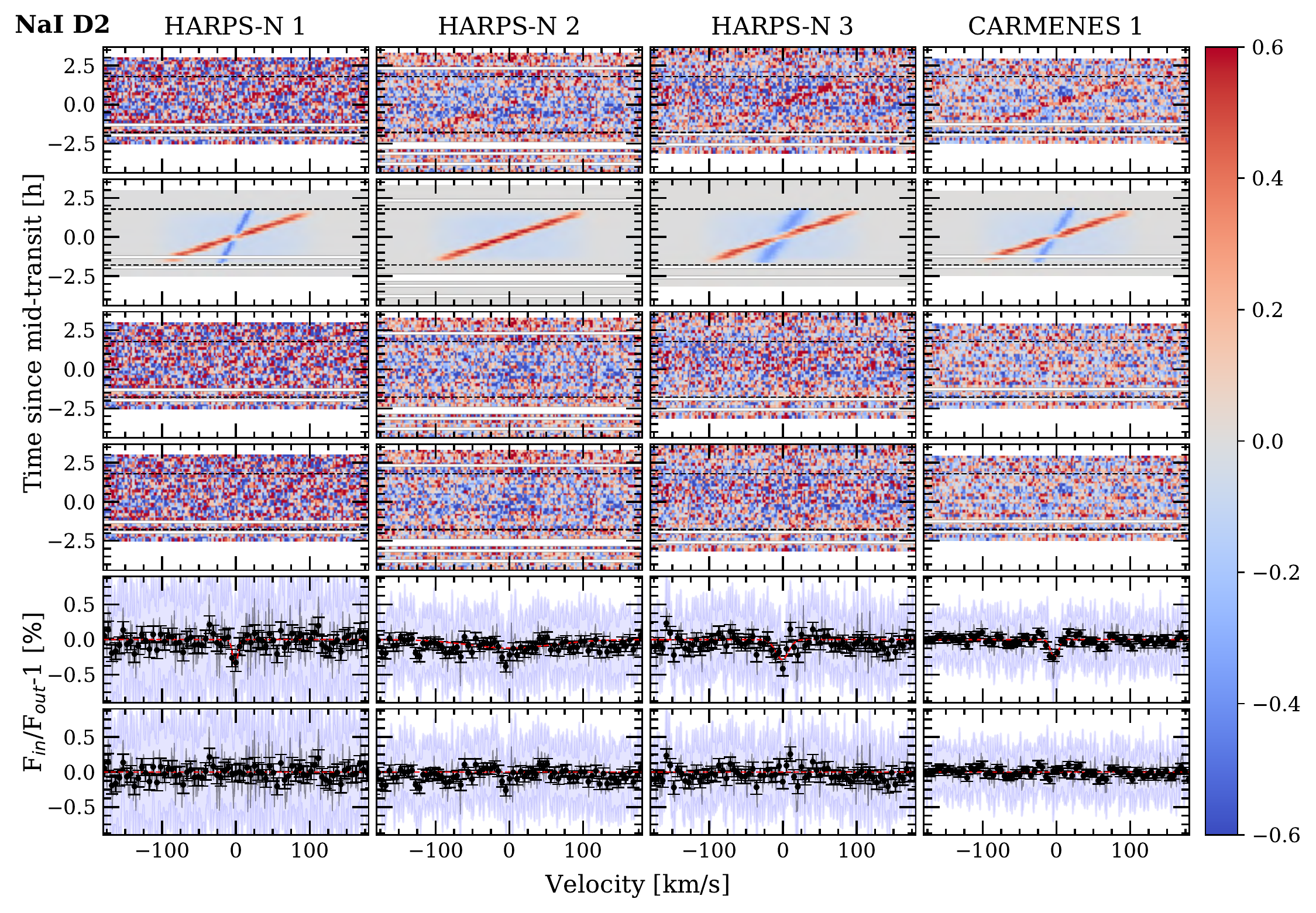}
\caption{Results around the NaI D$_{2}$ of HARPS-N Night 1 (first column starting from left), Night 2 (second column), Night 3 (third column) and CARMENES data (fourth column). The different panels are explained in Figure~\ref{fig:ts_Hb_indiv}. In the last two rows (starting from top) in black dots we show the result binned by $0.1~\mathrm{\AA}$.}
\label{fig:ts_NaD2_indiv}
\end{figure}

\newpage

\subsection{NaI D$_1$}
\begin{figure}[h]
\centering
\includegraphics[width=0.98\textwidth]{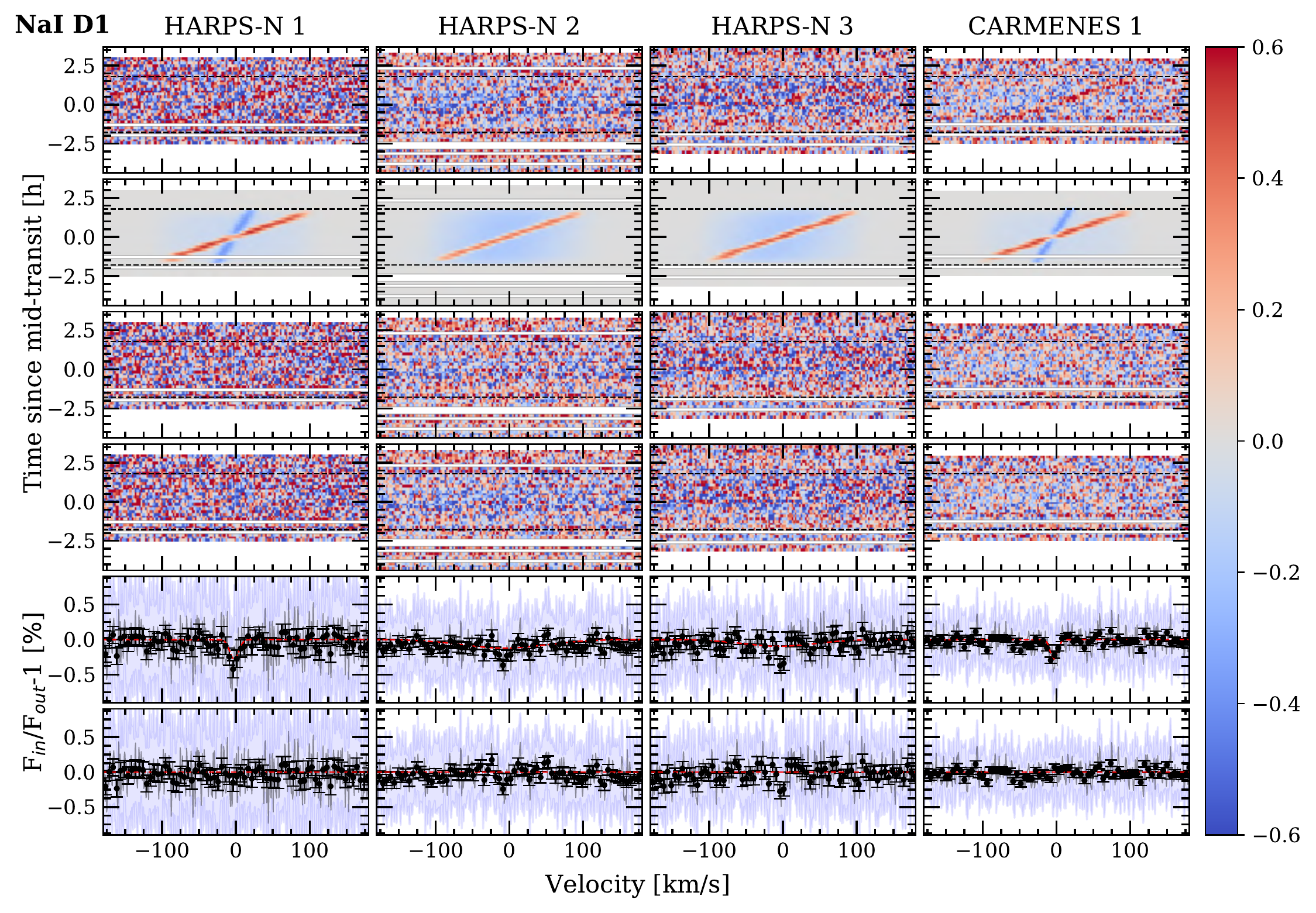}
\caption{Same as Figure~\ref{fig:ts_NaD2_indiv} but for Na D$_1$ line.}
\label{fig:ts_NaD1_indiv}
\end{figure}

\newpage

\subsection{FeII}
\begin{figure}[h]
\centering
\includegraphics[width=0.4\textwidth]{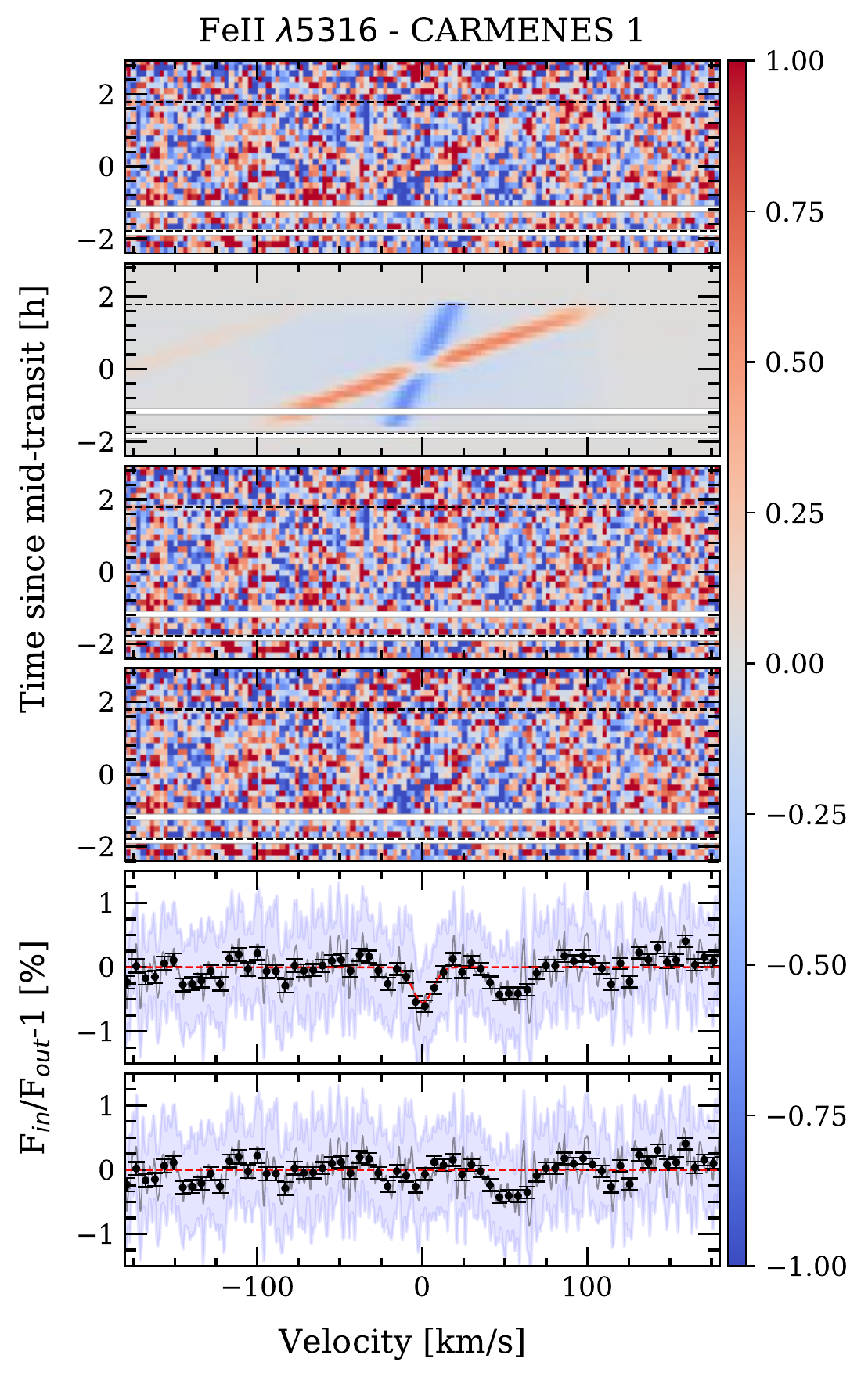}
\caption{Same as Figure~\ref{fig:ts_NaD2_indiv} but for FeII $\lambda5316~\mathrm{\AA}$ analysis obtained with the night observed with CARMENES. We note the strong residual observed close to the FeII line position. }
\label{fig:ts_FeMg_indiv}
\end{figure}

\newpage
\section{Best fit parameters and absorption depth values}

We present here the best-fit parameters obtained with the MCMC fitting procedure and the absorption depth values measured in the transmission spectra and light curves.

\subsection{MCMC best-fit parameters}

\renewcommand{\thefootnote}{\fnsymbol{footnote}}
\begin{table*}[h]
\centering
\caption{Best fit parameters and 1-$\sigma$ error bars from the MCMC analysis.}
\begin{tabular}{llcccccc}
\hline\hline
\\[-1em]
 &  & h & FWHM& K$_p$& v$_{wind}$ &  R$_{\lambda}$ $^($\footnotemark[1]$^)$ &  R$_{\lambda}~(h)$ $^($\footnotemark[4]$^)$\\ 
 \\[-1em]
 & & [$\%$]& [km/s] &[km/s] & [km/s] &[R$_p$] &[R$_p$] \\ \hline
\\[-1em]
\\[-1em]
\hline
\\[-1em]
H$\beta$ & HARPS-N 1 &  $-0.20^{+0.16}_{-0.19}$ & $14.7^{+16.0}_{-9.4}$ & $110.6^{+85.5}_{-71.7}$& $-5.1^{+8.0}_{-6.6}$&  $1.12^{+0.04}_{-0.03}$ & $1.07^{+0.11}_{-0.13}$\\
\\[-1em]
\\[-1em]
& HARPS-N 2 &  $-0.57^{+0.07}_{-0.08}$ & $17.9^{+2.8}_{-2.6}$ & $131.8^{+19.3}_{-18.4}$ & $-1.5^{+1.6}_{-1.7}$&  $0.89^{+0.03}_{-0.03}$ & $1.19^{+0.04}_{-0.05}$ \\
\\[-1em]
\\[-1em]
& HARPS-N 3 & $-0.38^{+0.08}_{-0.09}$ & $24.2^{+7.0}_{-6.0}$ & $166.3^{+55.6}_{-43.7}$ & $+4.2^{+3.6}_{-3.5}$&  $0.86^{+0.03}_{-0.03}$ & $1.13^{+0.05}_{-0.06}$ \\
\\[-1em]
\\[-1em]
&HARPS-Nx3 &  $-0.45^{+0.05}_{-0.05}$ & $19.4^{+2.7}_{-2.4}$ & $136.2^{+18.7}_{-18.5}$ & $-1.2^{+1.4}_{-1.5}$ & $0.91^{+0.02}_{-0.02}$ & $1.15^{+0.03}_{-0.03}$\\ 
\\[-1em]
\hline\hline
\\[-1em]
H$\gamma$ & HARPS-N 1 &  $-0.29^{+0.13}_{-0.26}$ & $29.5^{+56.3}_{-15.8}$ & $189.7^{+43.7}_{-105.8}$& $-1.4^{+7.2}_{-6.3}$&  $1.01^{+0.05}_{-0.05}$ & $1.10^{+0.08}_{-0.17}$\\
\\[-1em]
& HARPS-N 2 &  $-0.43^{+0.10}_{-0.12}$ & $16.9^{+5.6}_{-4.6}$ & $101.5^{+43.7}_{-105.8}$ & $-4.0^{+2.8}_{-3.0}$&  $0.86^{+0.03}_{-0.03}$& $1.15^{+0.07}_{-0.08}$\\
\\[-1em]
\\[-1em]
& HARPS-N 3 & $-0.28^{+0.17}_{-0.23}$ & $18.5^{+18.6}_{-11.2}$ & $104.6^{+81.0}_{-65.8}$ & $-2.8^{+8.5}_{-7.5}$&  $0.78^{+0.05}_{-0.05}$ & $1.13^{+0.05}_{-0.06}$\\
\\[-1em]
\\[-1em]
& HARPS-Nx3 &  $-0.38^{+0.08}_{-0.09}$ & $16.6^{+4.5}_{-4.0}$  & $135.0^{+34.1}_{-35.6}$  & $-2.3^{+2.5}_{-3.0}$  & $0.85^{+0.02}_{-0.03}$ & $1.13^{+0.05}_{-0.06}$\\
\\[-1em]
\hline\hline
\\[-1em]
NaI D$_2$ & HARPS-N 1 & $-0.37^{+0.14}_{-0.16}$ & $5.93^{+4.0}_{-2.2}$ & $151.4^{+35.4}_{-36.4}$ & $ -0.9^{+2.5}_{-2.6}$ &  $0.92^{+0.05}_{-0.05}$ &$1.13^{+0.09}_{-0.10}$ \\
\\[-1em]
\\[-1em]
&HARPS-N 2 & $-0.12^{+0.02}_{-0.02}$ & $96.4^{+2.7}_{-6.0}$ & $137.1^{+78.3}_{-91.1}$ & $1.4^{+9.0}_{-9.6}$ &  $0.97^{+0.04}_{-0.04}$ & $1.04^{+0.01}_{-0.01}$\\
\\[-1em]
\\[-1em]
&HARPS-N 3 & $-0.28^{+0.08}_{-0.10}$ & $14.9^{+8.5}_{-6.2}$ & $103.3^{+26.8}_{-26.4}$ &$-1.3^{+2.2}_{-2.4}$ & $0.94^{+0.04}_{-0.04}$  & $1.10^{+0.05}_{-0.07}$\\
\\[-1em]
\\[-1em]
&HARPS-Nx3 & $-0.33^{+0.06}_{-0.06}$ &  $8.9^{+2.7}_{-1.9}$ & $171.1^{+15.2}_{-12.8}$ & $-2.4^{+0.9}_{-0.9}$& $0.94^{+0.03}_{-0.03}$ & $1.11^{+0.04}_{-0.04}$\\ 
\\[-1em]
\\[-1em]
& CARMENES 1 & $-0.29^{+0.04}_{-0.04}$ & $8.3^{+1.3}_{-1.2}$  & $176.5^{+12.8}_{-12.3}$ & $-2.8^{+0.8}_{-0.8}$ & $0.91^{+0.02}_{-0.02}$ & $1.10^{+0.03}_{-0.03}$\\ 
\\[-1em]
\hline\hline
\\[-1em]  
NaI D$_1$ & HARPS-N 1 &  $-0.29^{+0.11}_{-0.11}$ & $10.2^{+5.1}_{-3.2}$ & $177.8^{+39.9}_{-41.4}$ & $-2.2^{+2.4}_{-2.4}$ & $0.97^{+0.06}_{-0.05}$ & $1.10^{+0.07}_{-0.07}$ \\
\\[-1em]
\\[-1em]
&HARPS-N 2 & $-0.12^{+0.01}_{-0.02}$  & $95.8^{+3.2}_{-7.5}$ & $143.3^{+75.2}_{-93.1}$ &  $ -7.4^{+8.6}_{-5.7}$& $0.90^{+0.04}_{-0.03}$  & $1.04^{+0.01}_{-0.01}$  \\
\\[-1em]
\\[-1em]
&HARPS-N 3 & $-0.09^{+0.02}_{-0.02}$ & $81.7^{+12.9}_{-21.8}$ & $196.5^{+40.9}_{-79.3}$& $-6.4^{+9.6}_{-6.1}$ & $0.97^{+0.04}_{-0.04}$ & $1.03^{+0.01}_{-0.02}$\\
\\[-1em]
\\[-1em]
 &HARPS-Nx3 & $-0.35^{+0.05}_{-0.05}$ & $9.6^{+1.9}_{-1.5}$ & $194.0^{+15.1}_{-14.4}$ & $-3.9^{+0.9}_{-0.9}$ & $0.91^{+0.03}_{-0.03}$ & $1.12^{+0.03}_{-0.03}$\\ 
 \\[-1em]
\\[-1em]
& CARMENES 1 & $-0.29^{+0.04}_{-0.04}$ & $7.8^{+1.2}_{-1.0}$  & $176.8^{+10.8}_{-10.6}$ & $-3.6^{+0.7}_{-0.7}$ & $0.90^{+0.02}_{-0.02}$ & $1.10^{+0.03}_{-0.03}$ \\ 
 \\[-1em]
\hline\hline
\\[-1em] 
 CaII $\lambda8498~\mathrm{\AA}$ & CARMENES 1 & $-0.52^{+0.05}_{-0.05}$ & $8.9^{+1.4}_{-0.9}$ & $173.2^{+7.8}_{-7.5}$ & $-3.2^{+0.6}_{-0.6}$ & $0.87^{+0.03}_{-0.03}$ &  $1.17^{+0.03}_{-0.03}$\\ 
\\[-1em]
\\[-1em]
CaII $\lambda8542~\mathrm{\AA}$ & CARMENES 1 & $-0.60^{+0.04}_{-0.04}$ & $11.7^{+1.0}_{-0.9}$ & $139.2^{+8.9}_{-9.2}$ & $-1.3^{+0.6}_{-0.6}$ & $1.09^{+0.02}_{-0.02}$ & $1.20^{+0.03}_{-0.03}$\\ 
\\[-1em]
\\[-1em]
CaII $\lambda8662~\mathrm{\AA}$ & CARMENES 1 & $-0.55^{+0.06}_{-0.06}$ & $6.9^{+1.1}_{-0.9}$ & $160.7^{+8.0}_{-7.8}$ & $-1.1^{+0.5}_{-0.5}$ &  $0.95^{+0.02}_{-0.03}$ & $1.18^{+0.04}_{-0.04}$\\ 
 \\[-1em]
\hline\hline
\\[-1em] 
 FeII $\lambda5018~\mathrm{\AA}$ & HARPS-Nx3 & $-0.26^{+0.04}_{-0.05}$ & $7.9^{+1.3}_{-1.3}$ & $181.9^{+18.3}_{-18.7}$ &$-2.7^{+1.1}_{-1.1}$ & $1.19^{+0.03}_{-0.03}$ &  $1.09^{+0.03}_{-0.03}$ \\ 
\\[-1em]
\\[-1em]
FeII $\lambda5169~\mathrm{\AA}$ & HARPS-Nx3 & $-0.47^{+0.05}_{-0.05}$ & $8.1^{+1.1}_{-1.0}$ & $170.6^{+8.3}_{-7.8}$ & $-3.6^{+0.5}_{-0.5}$ & $1.18^{+0.02}_{-0.02}$ & $1.16^{+0.03}_{-0.03}$\\ 
\\[-1em]
\\[-1em]
FeII $\lambda5316~\mathrm{\AA}$  & HARPS-Nx3 & $-0.26^{+0.05}_{-0.06}$ & $5.6^{+1.4}_{-1.1}$ & $170.8^{+14.4}_{-16.6}$ & $-2.1^{+0.8}_{-0.8}$ &  $1.18^{+0.03}_{-0.03}$  & $1.09^{+0.03}_{-0.04}$\\  
\\[-1em]
\\[-1em]
 & CARMENES 1 &$-0.55^{+0.08}_{-0.09}$ & $10.3^{+2.3}_{-2.0}$ & $139.6^{+32.3}_{-22.6}$ & $+0.2^{+1.7}_{-1.3}$ & $1.24^{+0.05}_{-0.06}$ & $1.18^{+0.05}_{-0.05}$\\ 
 \\[-1em]
\hline\hline
\\[-1em] 
MgI $\lambda5173$ & HARPS-Nx3 & $-0.23^{+0.03}_{-0.04}$  & $12.8^{+2.4}_{-2.0}$ & $159.6^{+30.9}_{-23.8}$ &  $ -0.9^{+1.7}_{-1.9}$& $1.23^{+0.02}_{-0.02}$  & $1.08^{+0.02}_{-0.02}$  \\
 \\[-1em]
\hline\hline
\end{tabular}\\
\begin{tablenotes}
\item Notes. $^($\footnotemark[1]$^)$ Effective radius value obtained from the best fit model of the CLV and RME effects. $^($\footnotemark[4]$^)$ Effective radius calculated considering the absorption value, $h$, from the best-fit model and assuming a continuum level of $(R_p/R_{\star})^2 = 1.382\%$
\end{tablenotes}
\label{Tab:res}
\end{table*}
\renewcommand{\thefootnote}{\arabic{footnote}}

\newpage

\subsection{Absorption depth values}
\renewcommand{\thefootnote}{\fnsymbol{footnote}}
\begin{table*}[h]
\centering

\caption{Absorption depth (in $\%$) measured for $0.75~\mathrm{\AA}$ and $1.5~\mathrm{\AA}$ bandwidths calculated in the final transmission spectra (TS) and transmission light curves (TLC).}
\begin{tabular}{llcccc}
\hline\hline
\\[-1em]
 &  & \multicolumn{2}{c}{TS} & \multicolumn{2}{c}{TLC}                 \\ \\[-1em]
\cline{3-6} 
\\[-1em]
&          & $0.75~\mathrm{\AA}$ & $1.5~\mathrm{\AA}$ & $0.75~\mathrm{\AA}$ & $1.5~\mathrm{\AA}$ \\ \\[-1em]
          \hline
          \\[-1em]
          
H$\alpha$ & HARPS-N 1   &    $0.69\pm0.10$  &   $0.44\pm0.07$    &  $0.69\pm0.11$                   &   $0.45\pm0.10$                 \\ \\[-1em]
& HARPS-N 2   &   $0.41\pm0.08$     &  $0.20\pm0.06$  &     $0.37\pm0.10$                &     $0.16\pm0.08$               \\ \\[-1em]
& HARPS-N 3   &   $0.61\pm0.10$     &    $0.24\pm0.07$                 &       $0.45\pm0.12$              &      $0.10\pm0.10$              \\ \\[-1em]
& HARPS-N$x3$   &  $0.59\pm0.07$   &  $0.33\pm0.05$     &     $0.58\pm0.06$                &     $0.30\pm0.05$               \\ 
& CARMENES 1   & $0.68\pm0.05$   &  $0.44\pm0.04$    &    $0.68\pm0.06$   &  $0.43\pm0.05$                  \\ \\[-1em]
\\[-1em]
\hline\hline
\\[-1em]
H$\beta$ & HARPS-N 1   &    $0.30\pm0.09$  &   $0.17\pm0.06$    &     $0.35\pm0.10$                &    $0.22\pm0.08$                \\ \\[-1em]
& HARPS-N 2   &   $0.35\pm0.07$     &  $0.14\pm0.05$  &     $0.37\pm0.08$                &     $0.17\pm0.07$               \\ \\[-1em]
& HARPS-N 3   &   $0.23\pm0.08$     &    $0.10\pm0.06$                 &        $0.28\pm0.10$             &      $0.15\pm0.08$              \\ \\[-1em]
& HARPS-N$x3$   &  $0.28\pm0.06$   &  $0.13\pm0.04$     &    $0.31\pm0.05$                 &     $0.17\pm0.04$               \\ 
\\[-1em]
\hline\hline
\\[-1em]
H$\gamma$  & HARPS-N 1   &    $0.19\pm0.10$  &   $0.09\pm0.07$    &    $0.18\pm0.11$                 &     $0.09\pm0.10$               \\ \\[-1em]
& HARPS-N 2   &   $0.28\pm0.08$     &  $0.14\pm0.06$  &      $0.29\pm0.09$               &    $0.15\pm0.08$                \\ \\[-1em]
& HARPS-N 3   &   $0.13\pm0.10$     &    $0.04\pm0.07$                 &        $0.08\pm0.13$             &       $0.00\pm0.11$             \\ \\[-1em]
& HARPS-N$x3$   &  $0.21\pm0.07$   &  $0.09\pm0.05$     &     $0.17\pm0.06$                &   $0.07\pm0.05$                 \\ 
\\[-1em]
\hline\hline
\\[-1em]
NaI D$_2$ & HARPS-N 1   &    $0.07\pm0.07$  &   $0.02\pm0.05$    &    $0.06\pm0.08$                 &   $0.01\pm0.07$                 \\ \\[-1em]
&HARPS-N 2   &   $0.12\pm0.06$     &  $0.05\pm0.04$  &       $0.15\pm0.07$              &   $0.09\pm0.06$                 \\ \\[-1em]
&HARPS-N 3   &   $0.10\pm0.06$     &    $0.06\pm0.05$                 &       $0.10\pm0.08$              &     $0.06\pm0.07$               \\ \\[-1em]
&HARPS-N$x3$   &  $0.09\pm0.05$   &  $0.05\pm0.03$     &     $0.15\pm0.04$                &   $0.07\pm0.04$                 \\ 
&CARMENES 1   & $0.10\pm0.05$   &  $0.03\pm0.03$    &    $0.08\pm0.05$   &  $0.01\pm0.05$                  \\ \\[-1em]
\\[-1em]
\hline\hline
\\[-1em]
NaI D$_1$ & HARPS-N 1   &    $0.16\pm0.07$  &   $0.10\pm0.05$    &    $0.14\pm0.04$                 &     $0.08\pm0.07$               \\ \\[-1em]
&HARPS-N 2   &   $0.09\pm0.06$     &  $0.04\pm0.04$  &  $0.05\pm0.07$                   &     $0.01\pm0.06$               \\ \\[-1em]
&HARPS-N 3   &   $0.11\pm0.06$     &    $0.05\pm0.05$                 &       $0.10\pm0.08$              &      $0.04\pm0.07$              \\ \\[-1em]
&HARPS-N$x3$   &  $0.10\pm0.05$   &  $0.06\pm0.03$     &   $0.16\pm0.04$                  &     $0.09\pm0.04$               \\ 
&CARMENES 1   & $0.11\pm0.04$   &  $0.06\pm0.03$    &    $0.09\pm0.05$   &  $0.04\pm0.04$                  \\ \\[-1em]
\\[-1em]
\hline\hline
\\[-1em]
Na D$_{21}$ $^($\footnotemark[1]$^)$ & HARPS-N 1   &    $0.11\pm0.07$  &   $0.06\pm0.05$    &    $0.10\pm0.07$                 &     $0.05\pm0.06$               \\ \\[-1em]
& HARPS-N 2   &   $0.10\pm0.06$     &  $0.04\pm0.04$  &  $0.10\pm0.06$                   &     $0.05\pm0.05$               \\ \\[-1em]
& HARPS-N 3   &   $0.10\pm0.06$     &    $0.05\pm0.05$                 &       $0.10\pm0.07$              &      $0.05\pm0.06$              \\ \\[-1em]
& HARPS-N$x3$   &  $0.09\pm0.05$   &  $0.05\pm0.03$     &   $0.15\pm0.04$                  &     $0.08\pm0.03$               \\ 
& CARMENES 1   & $0.11\pm0.04$   &  $0.04\pm0.03$    &    $0.09\pm0.04$   &  $0.03\pm0.04$                  \\ \\[-1em]
\\[-1em]
\hline\hline
\\[-1em]

CaII $\lambda8498~\mathrm{\AA}$  & CARMENES 1 &   $0.28\pm0.05$  &   $0.15\pm0.04$    &     $0.28\pm0.06$                &    $0.14\pm0.05$                \\ \\[-1em]
CaII $\lambda8542~\mathrm{\AA}$   & CARMENES 1&   $0.41\pm0.05$     &  $0.18\pm0.04$  &     $0.40\pm0.06$                &     $0.17\pm0.05$               \\ \\[-1em]
CaII $\lambda8662~\mathrm{\AA}$   & CARMENES 1 &   $0.27\pm0.06$     &    $0.14\pm0.04$                 &        $0.26\pm0.07$             &      $0.15\pm0.04$              \\ \\[-1em]
CaII Combined$^($\footnotemark[3]$^)$ & CARMENES 1  &   $0.32\pm0.05$     &    $0.16\pm0.04$                 &        $0.31\pm0.05$             &      $0.14\pm0.04$              \\
\\[-1em]
\hline\hline
\\[-1em]

FeII $\lambda5018~\mathrm{\AA}$  &  HARPS-Nx3 &  $0.09\pm0.04$  &   $0.04\pm0.03$    &     $0.03\pm0.04$                &    $0.02\pm0.03$                \\ \\[-1em]
FeII $\lambda5169~\mathrm{\AA}$  &  HARPS-Nx3 &   $0.09\pm0.04$     &  $0.04\pm0.03$  &     $0.12\pm0.04$                &     $0.05\pm0.03$               \\ \\[-1em]
FeII $\lambda5316~\mathrm{\AA}$  &  HARPS-Nx3 &   $0.07\pm0.04$     &    $0.04\pm0.03$                 &        $0.04\pm0.03$             &      $0.00\pm0.03$              \\ \\[-1em]
FeII Combined$^($\footnotemark[3]$^)$ &  HARPS-Nx3  &   $0.08\pm0.04$     &    $0.04\pm0.03$                 &        $0.06\pm0.03$             &      $0.01\pm0.03$              \\

\\[-1em]
\hline\hline
\\[-1em]
MgI $\lambda5173~\mathrm{\AA}$ HARPS-Nx3 &  HARPS-Nx3  &   $0.07\pm0.04$     &    $0.02\pm0.03$                 &        $0.09\pm0.04$             &      $0.04\pm0.03$              \\

\\[-1em]
\hline\hline
\\[-1em]
\end{tabular} \\
\begin{tablenotes}
\item Notes. $^($\footnotemark[1]$^)$ Combined absorption depth of both NaI D$_2$ and D$_1$ lines. $^($\footnotemark[3]$^)$ Combined absorption depth of the three FeII lines.
\end{tablenotes}
\label{tab:AD_vals}
\end{table*}
\renewcommand{\thefootnote}{\arabic{footnote}}

\newpage

\section{Individual transmission light curves}
\label{ap:individualTLC}
We present here the transmission light curves obtained for each specie analysed here.

\subsection{H$\beta$}
\label{ap:cornHb}

\begin{figure}[h]
\centering
\includegraphics[width=0.80\textwidth]{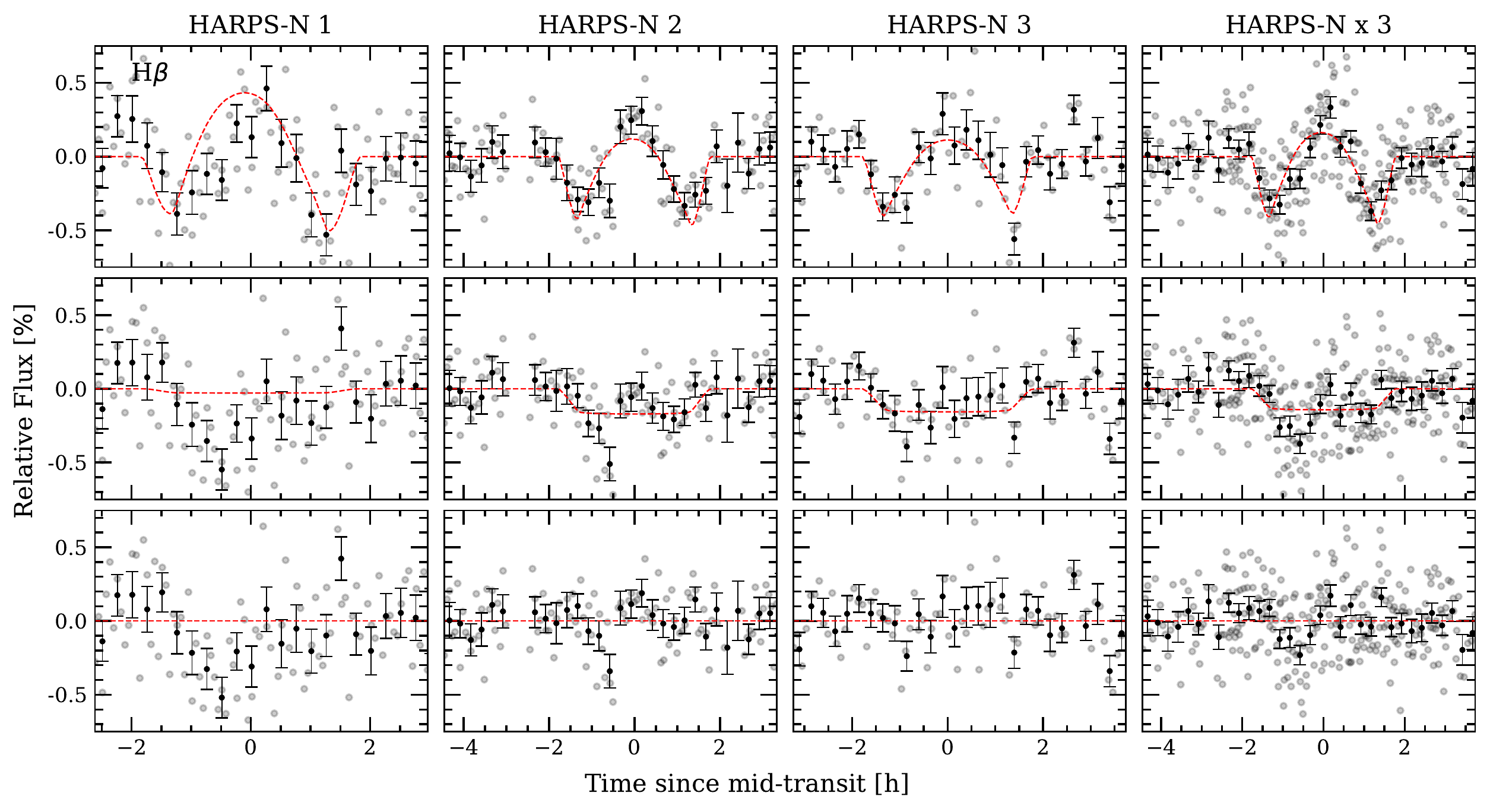}
\caption{H$\beta$ transmission light curves of HARPS-N Night 1 (first column, left), Night 2 (second column), Night 3 (third column) and join analysis of all HARPS-N observations (fourth column, right) computed for $1.5~\mathrm{\AA}$  passband. First row (top): observed transmission light curve. Second row: after correcting for the CLV and RME. Third row (bottom): residuals when subtracting the model to the data. The gray dots are the original data, the black dots is the data binned $0.003$ in orbital phase. The red-dashed line of the first row is the best-fit model containing the CLV, RME and absorption. The red-dashed line of the second row corresponds to the model containing only absorption. In the third row, the red-line is a reference showing the null flux level.}
\label{fig:TLC_Hb_indiv}
\end{figure}

\subsection{H$\gamma$}

\begin{figure}[h]
\centering
\includegraphics[width=0.80\textwidth]{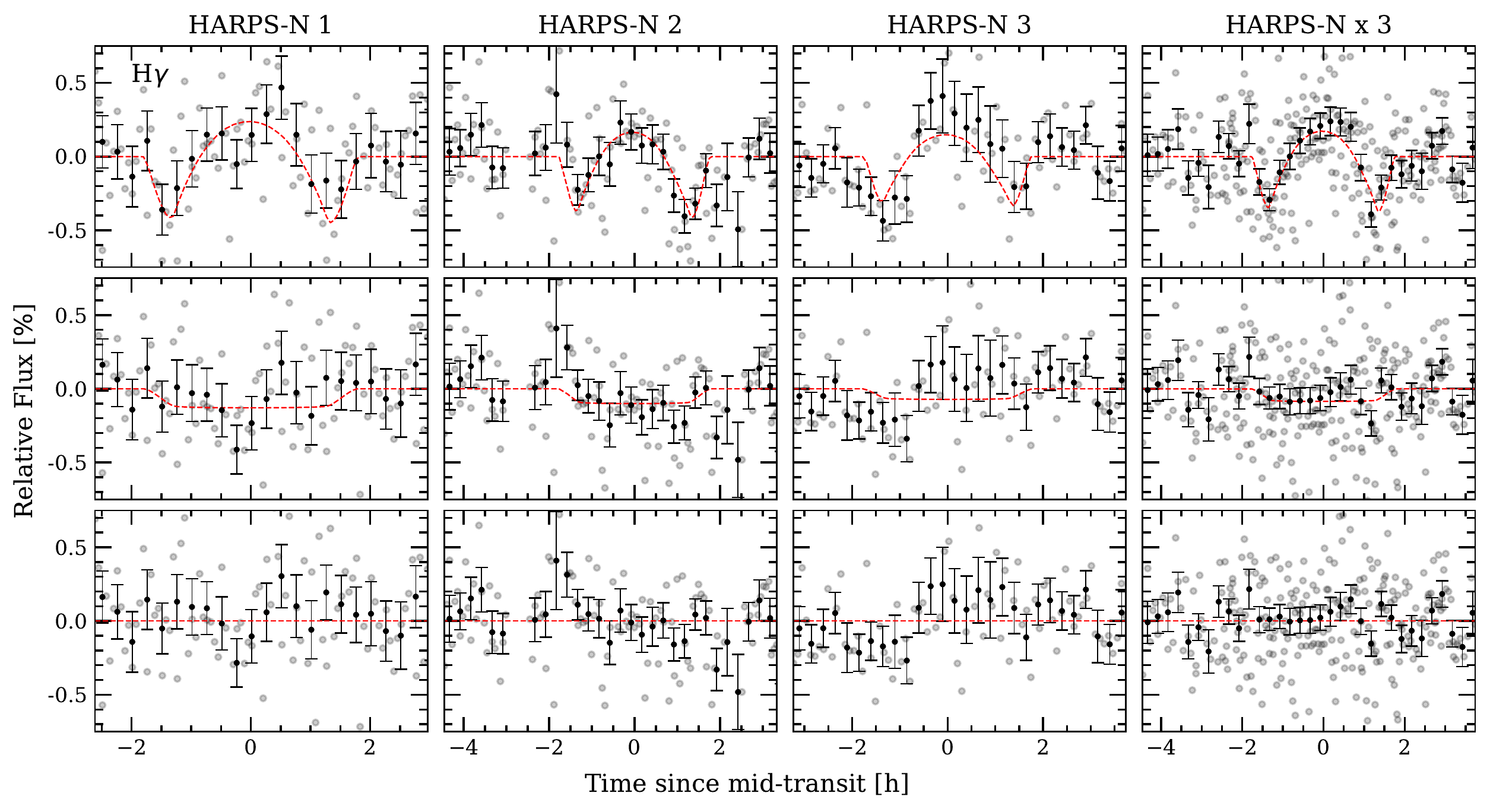}
\caption{Same as Figure~\ref{fig:TLC_Hb_indiv} but for H$\gamma$ line.}
\label{fig:TLC_Hg_indiv}
\end{figure}

\newpage

\subsection{CaII triplet}
\label{ap:cornHb}

\begin{figure}[h]
\centering
\includegraphics[width=0.6\textwidth]{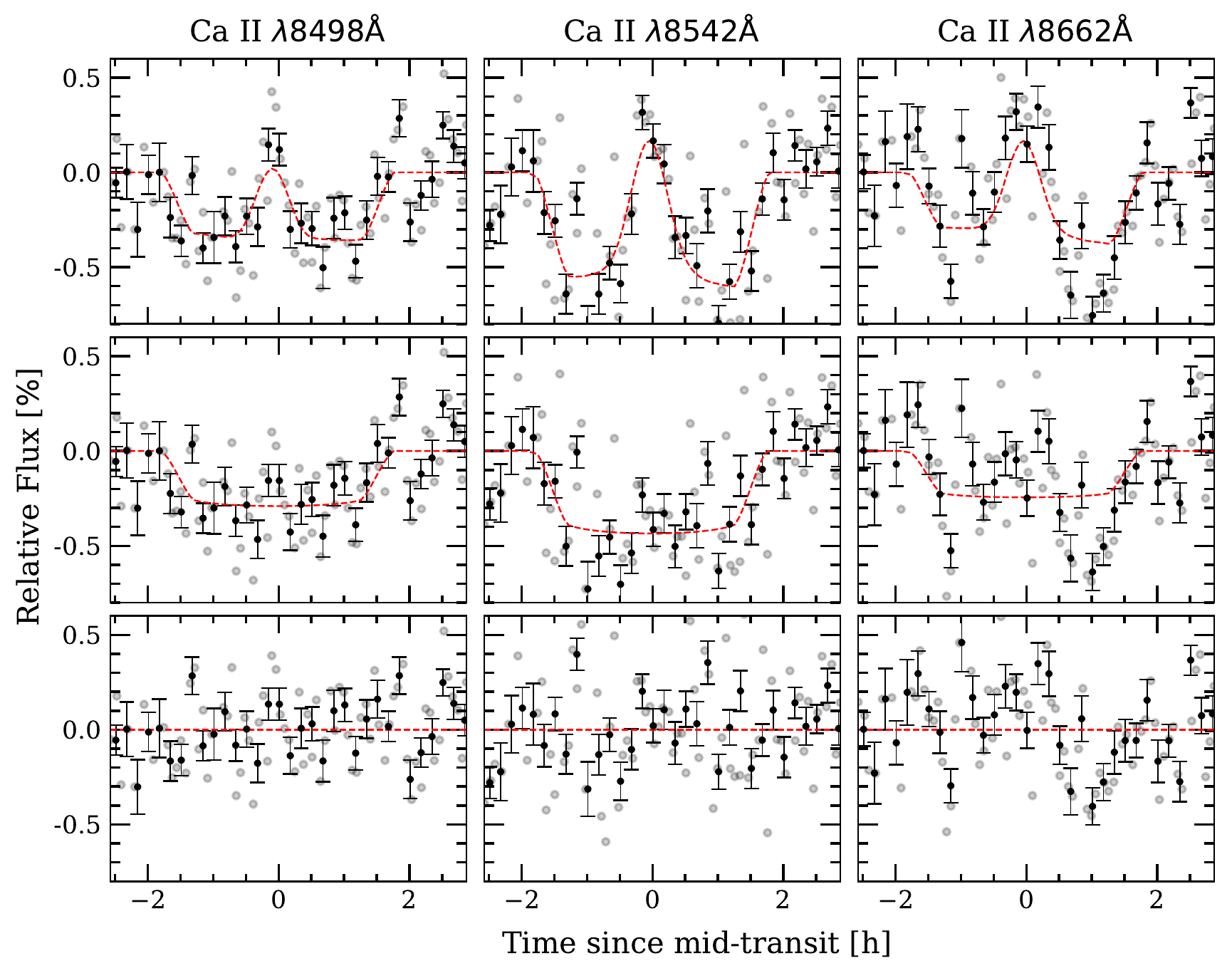}
\includegraphics[width=0.235\textwidth]{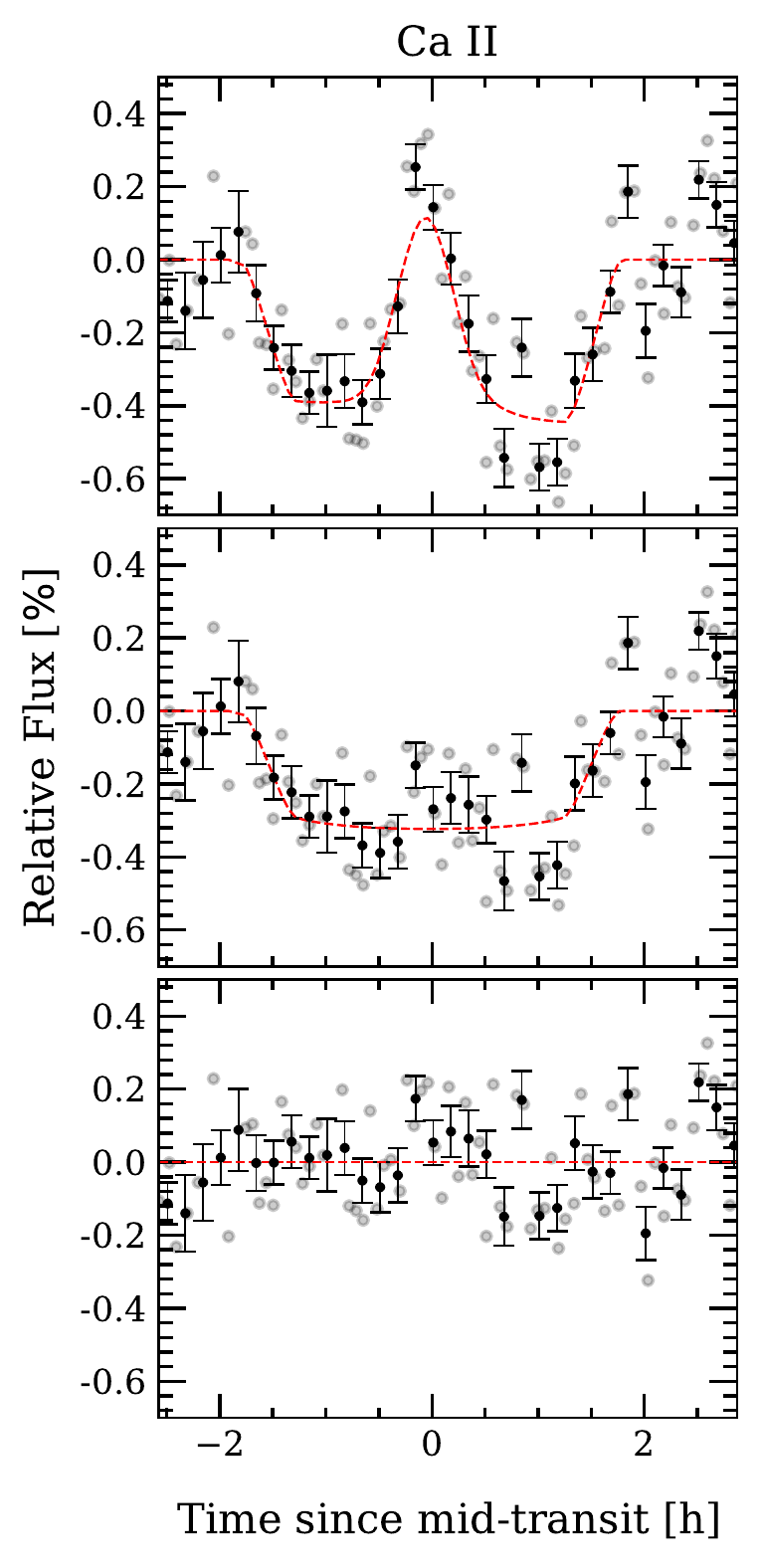}
\caption{CaII transmission light curves obtained with CARMENES data. Left: Same as Figure~\ref{fig:TLC_Hb_indiv} but for the CaII triplet lines. The analysis of each line is shown in one column. Right: Combined light curve (using the weighted mean) of the three individual light curves of CaII triplet lines observed in the left panel. These are the result for a $0.75~\mathrm{\AA}$ passband analysis. Data binned $0.002$ in orbital phase. }
\label{fig:TLC_Ca_indiv}
\end{figure}

\subsection{NaI doublet}

\begin{figure}[h]
\centering
\includegraphics[width=0.85\textwidth]{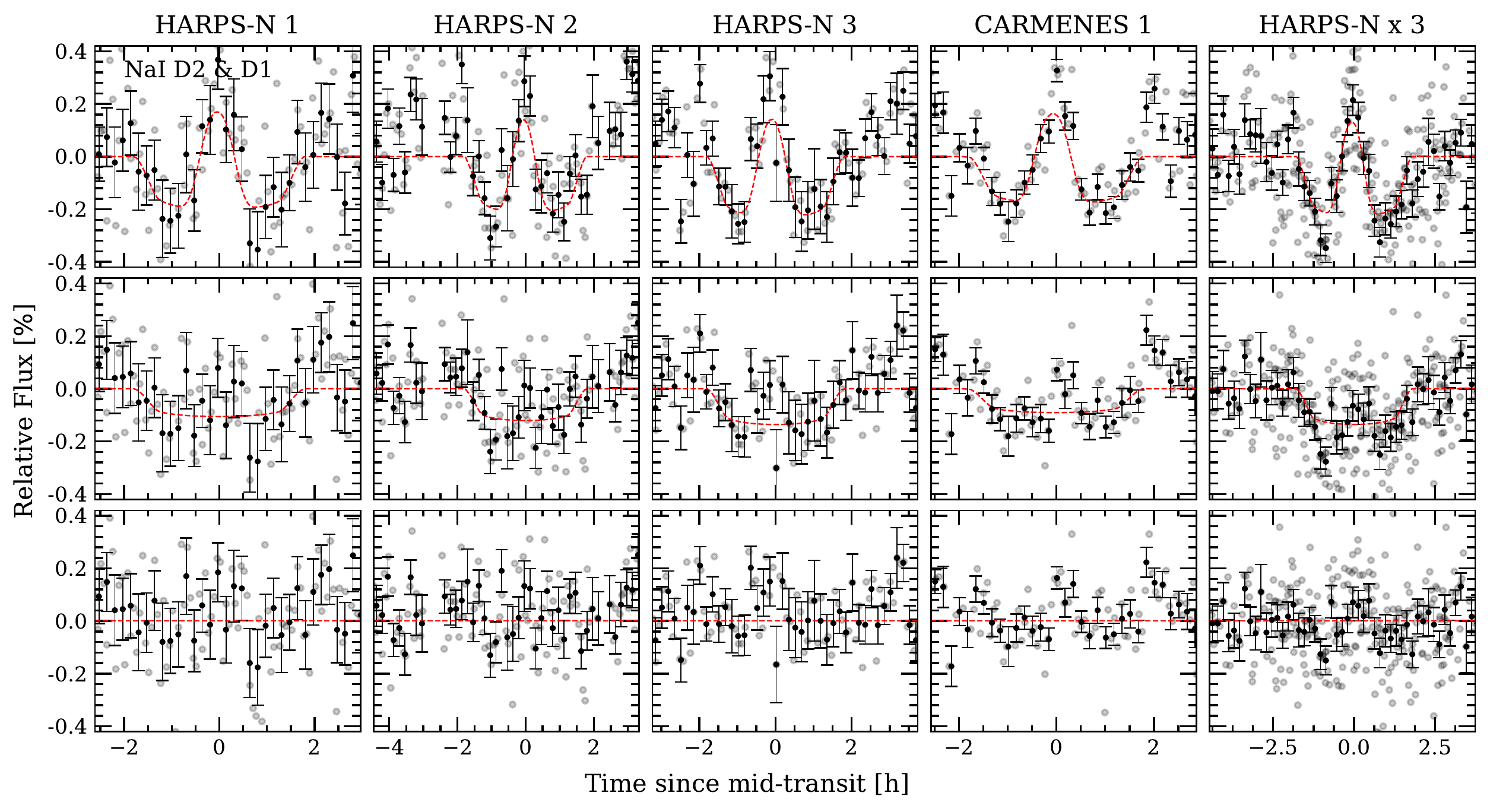}
\caption{Same as Figure~\ref{fig:TLC_Hb_indiv} but for both NaI D$_2$ and D$_1$ lines. This is the result of computing the weighted mean of both individual NaI D lines transmission light curves for a $0.75~\mathrm{\AA}$ passband. Data binned $0.002$ in orbital phase. }
\label{fig:TLC_NaD21_indiv}
\end{figure}

\subsection{FeII}
\begin{figure}[h]
\centering
\includegraphics[width=0.62\textwidth]{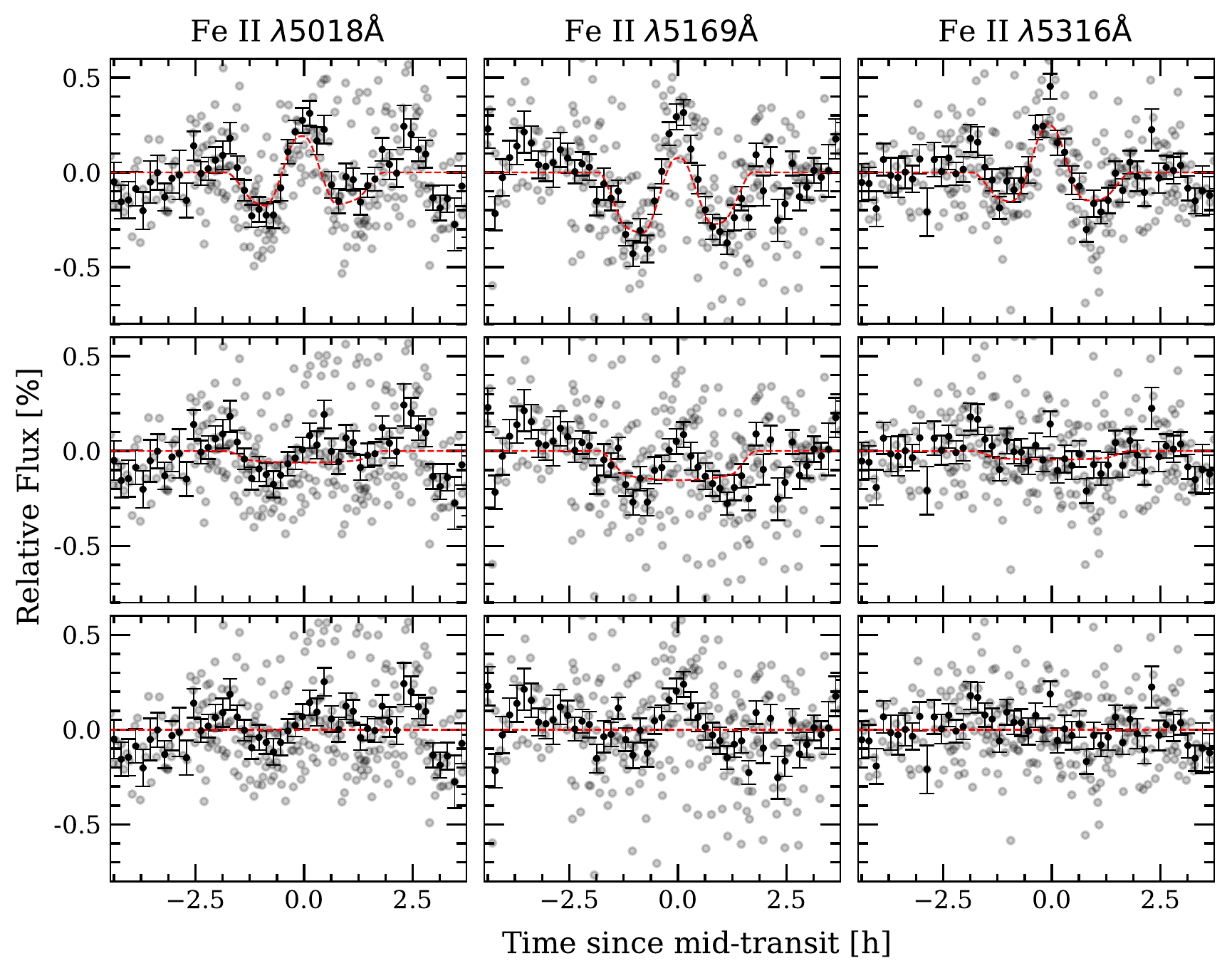}
\includegraphics[width=0.24\textwidth]{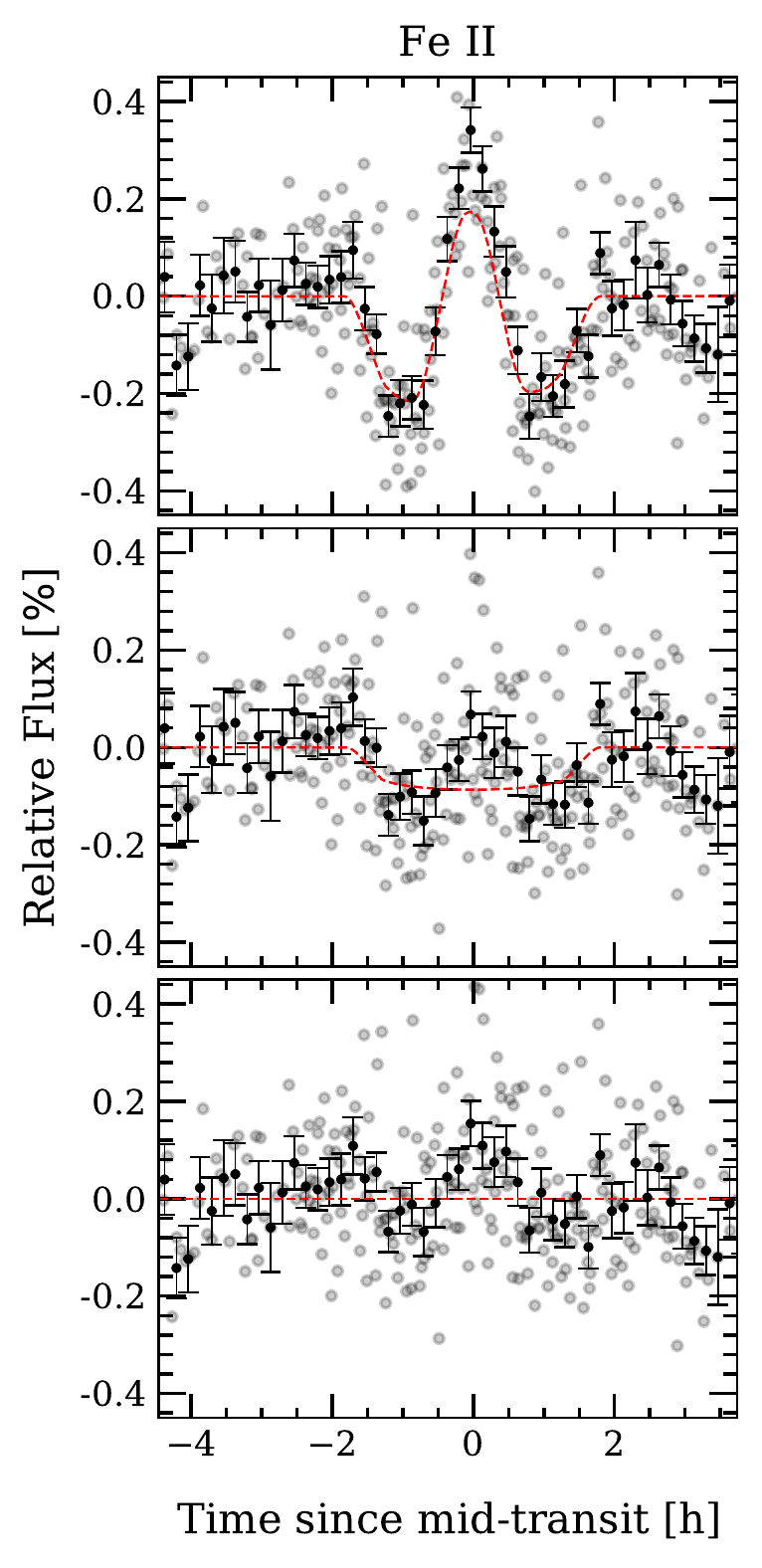}
\caption{FeII transmission light curves obtained combining the three nights of HARPS-N data. Left: Same as Figure~\ref{fig:TLC_Ca_indiv} but for three individual FeII lines and using a $0.75~\mathrm{\AA}$ passband. We note the asymmetry of the observed transmission light curve (first row) of the FeII$\lambda 5169$ line, caused by the RM effect from the closest line crossing the position of the FeII for times close to the egress. Right: transmission light curve after combining the three individual FeII lines. The data is binned by $0.002$ in orbital phase.}
\label{fig:TLC_Fe_indiv}
\end{figure}

\subsection{MgI}
\begin{figure}[h]
\centering
\includegraphics[width=0.24\textwidth]{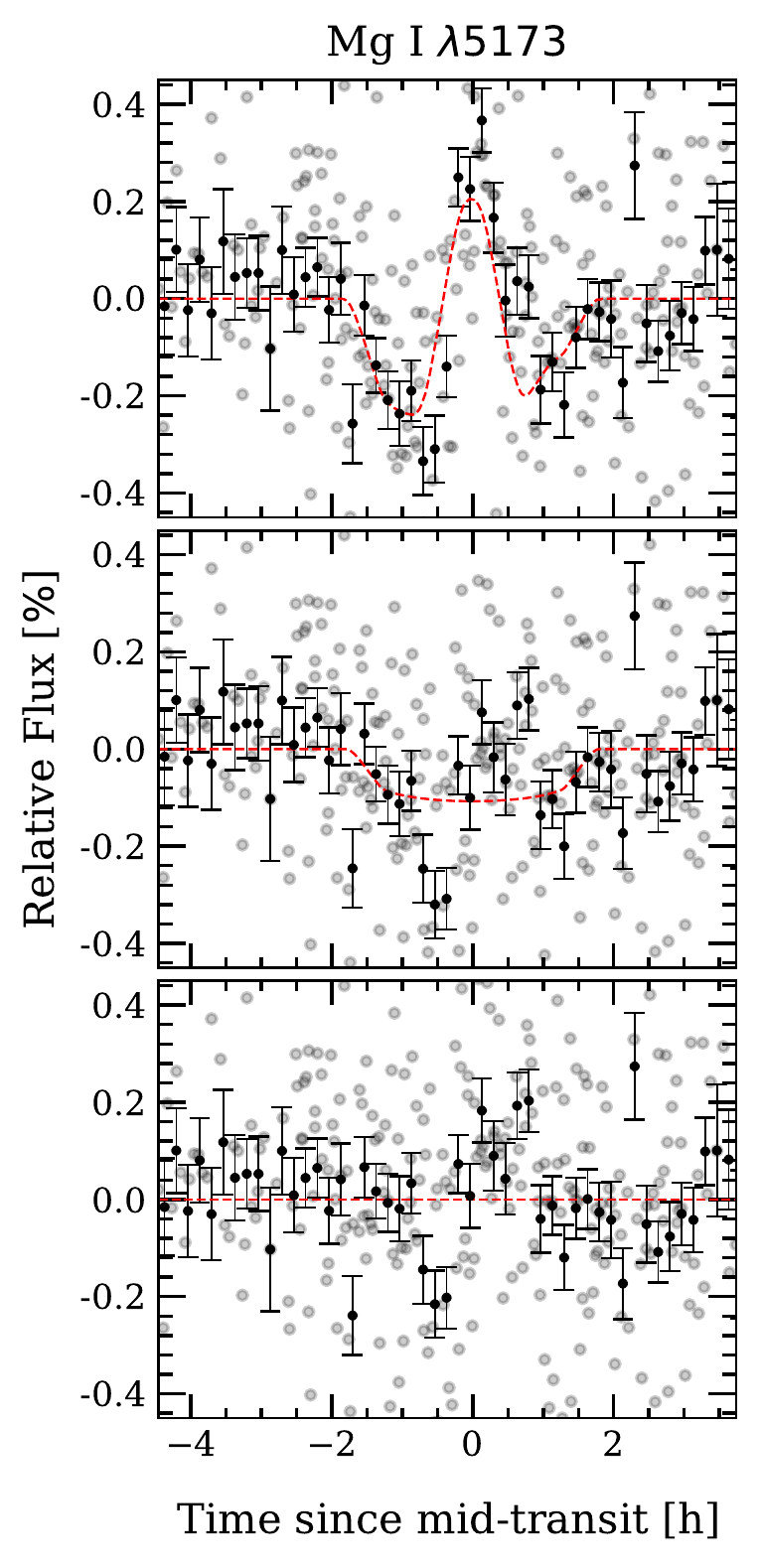}
\caption{Same as Figure~\ref{fig:TLC_Fe_indiv} right but for the MgI $\lambda5173~\mathrm{\AA}$. We note the asymmetry of the observed transmission light curve (first row) caused by a second RM effect from the closest line crossing the position (see Figure~\ref{fig:ts_FeMg_indiv}).}
\label{fig:TLC_Mg_indiv}
\end{figure}

\newpage
\section{Correlation diagrams of the best-fit models}
\label{ap:corners}
We present here the correlation diagrams of the best-fit CLV, RME and absorption model for the different nights and wavelength regions.

\subsection{H$\alpha$}
\label{ap:cornHa}

\begin{figure}[h]
\centering
\includegraphics[width=0.45\textwidth]{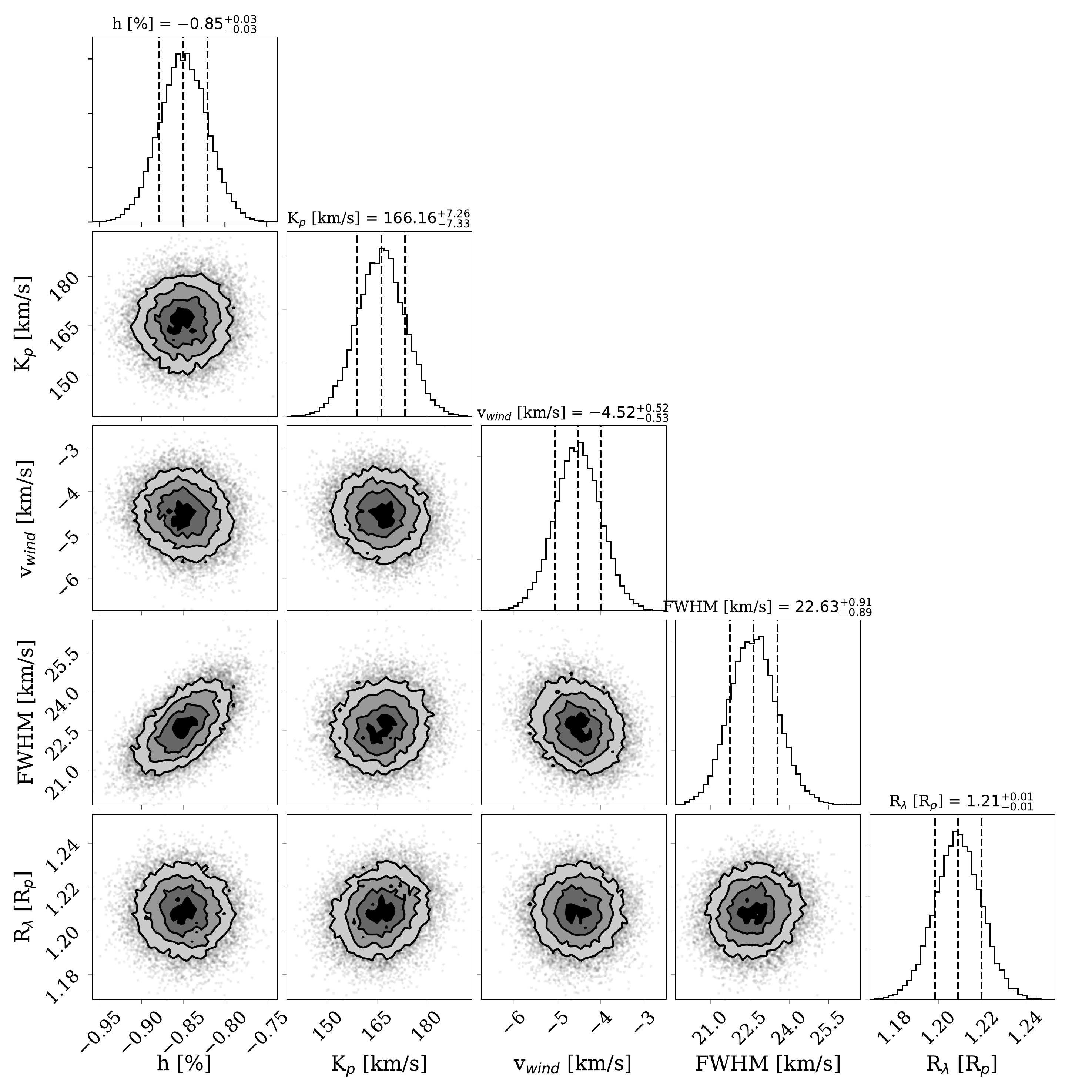}
\includegraphics[width=0.45\textwidth]{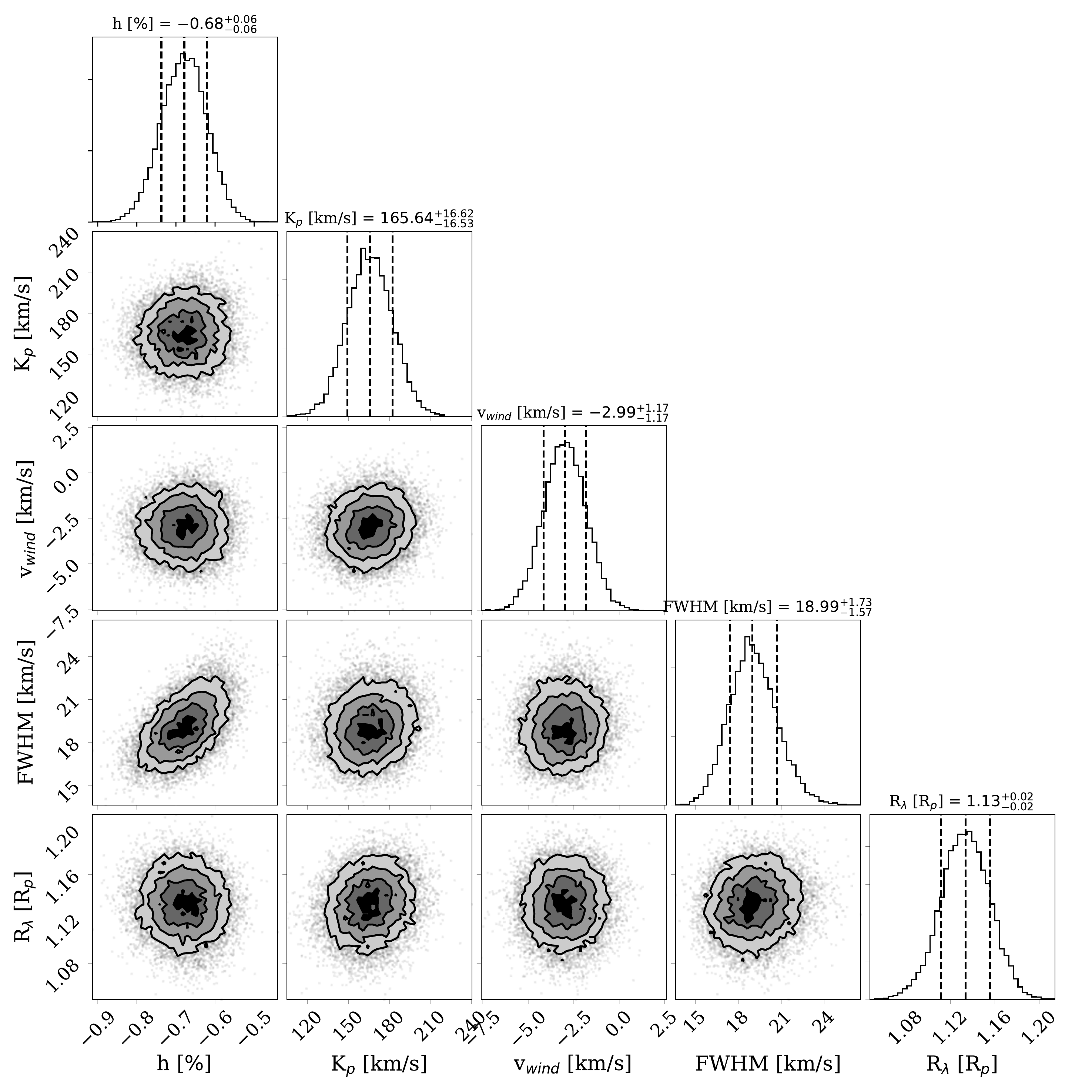}
\caption{Correlation diagrams of MCMC analysis of the H$\alpha$ line of CARMENES (left) and HARPS-N data analysis (right).}
\label{fig:mcmc_Ha_indiv}
\end{figure}

\subsection{H$\beta$ and H$\gamma$}
\label{ap:cornHb}
\begin{figure}[h]
\centering
\includegraphics[width=0.45\textwidth]{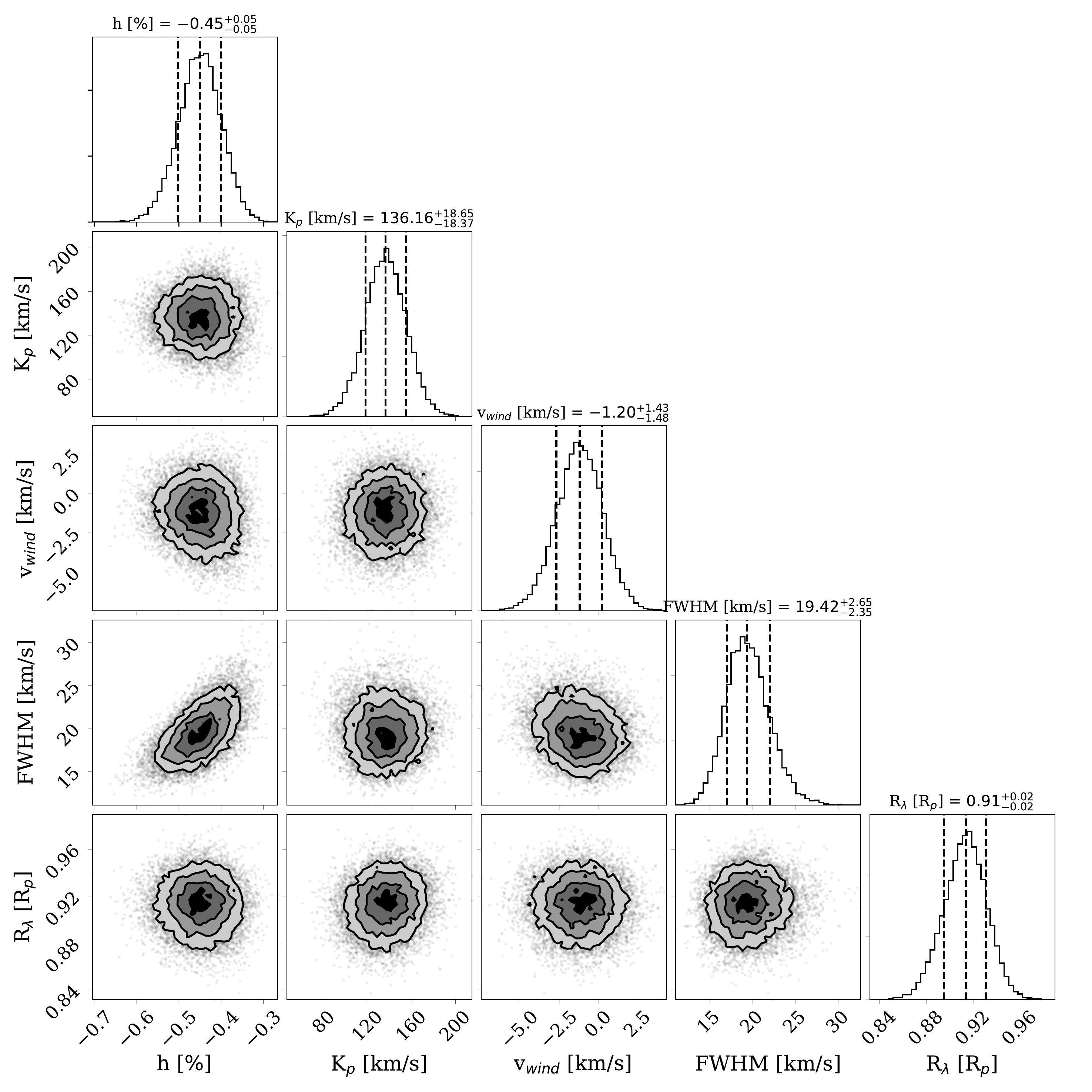}
\includegraphics[width=0.45\textwidth]{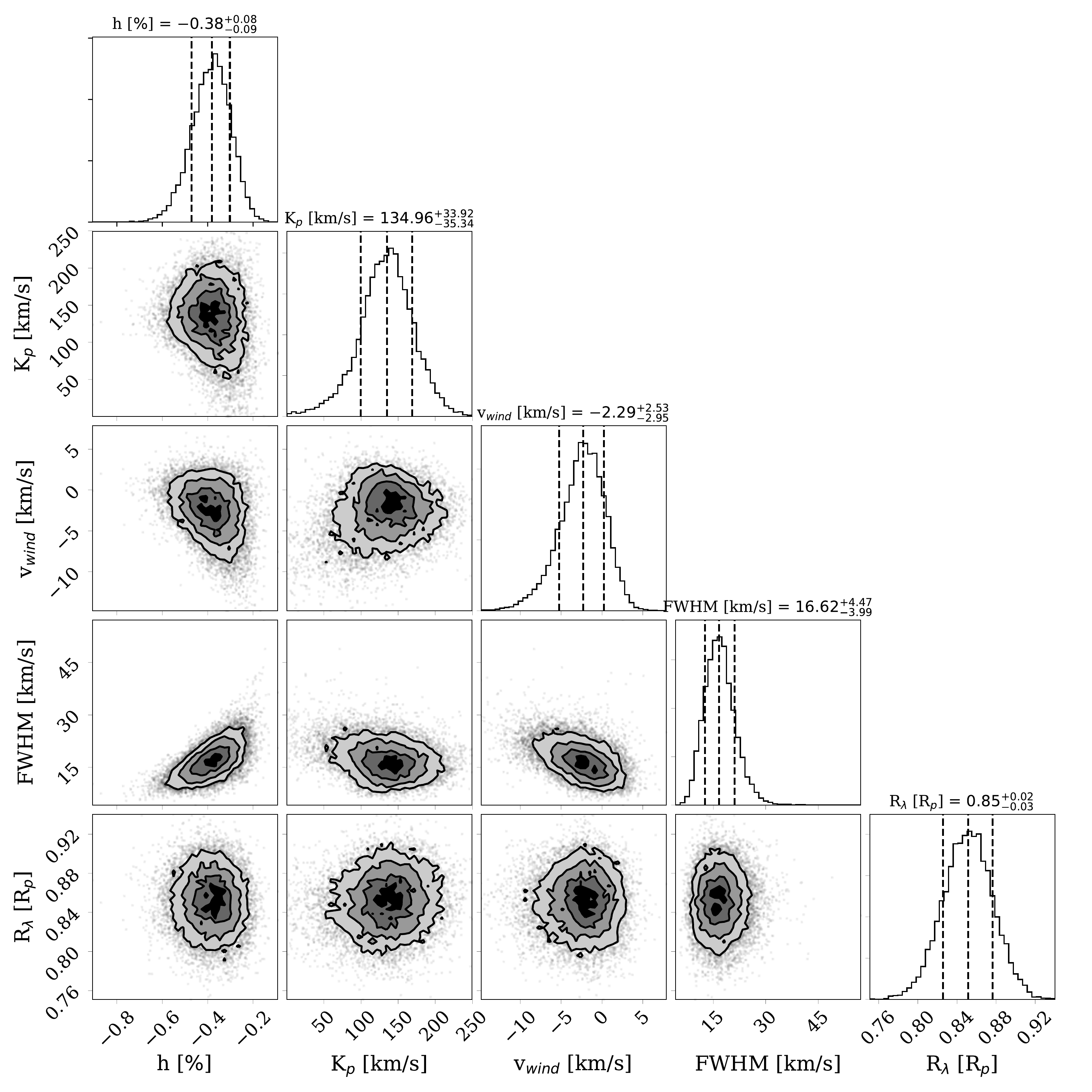}
\caption{Correlation diagrams of MCMC analysis of the H$\beta$ (left) and H$\gamma$ (right) lines obtained with the HARPS-N data analysis.}
\label{fig:mcmc_HbHg_indiv}
\end{figure}

\newpage
\subsection{CaII}
\label{ap:cornCa}

\begin{figure}[h]
\centering
\includegraphics[width=0.45\textwidth]{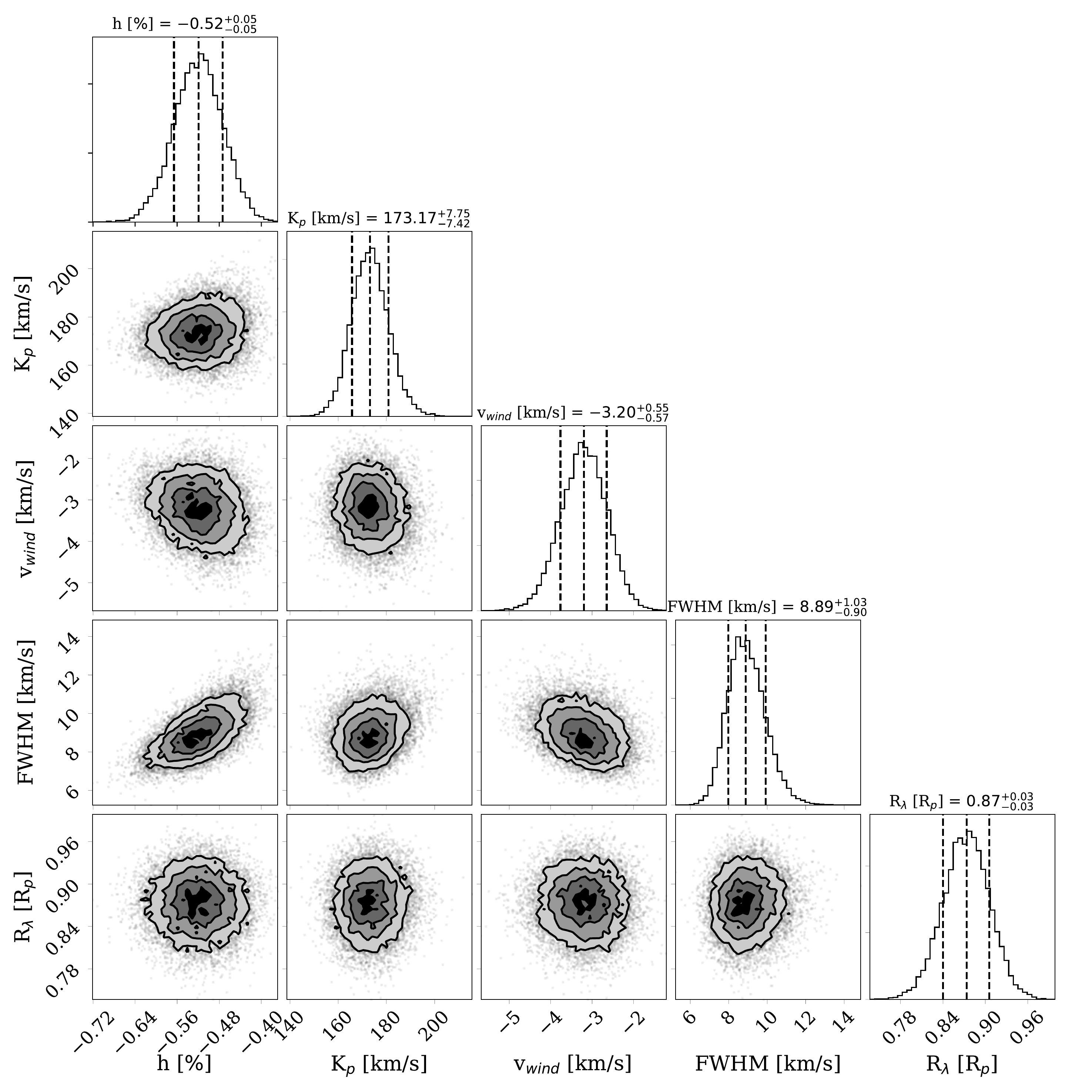}
\includegraphics[width=0.45\textwidth]{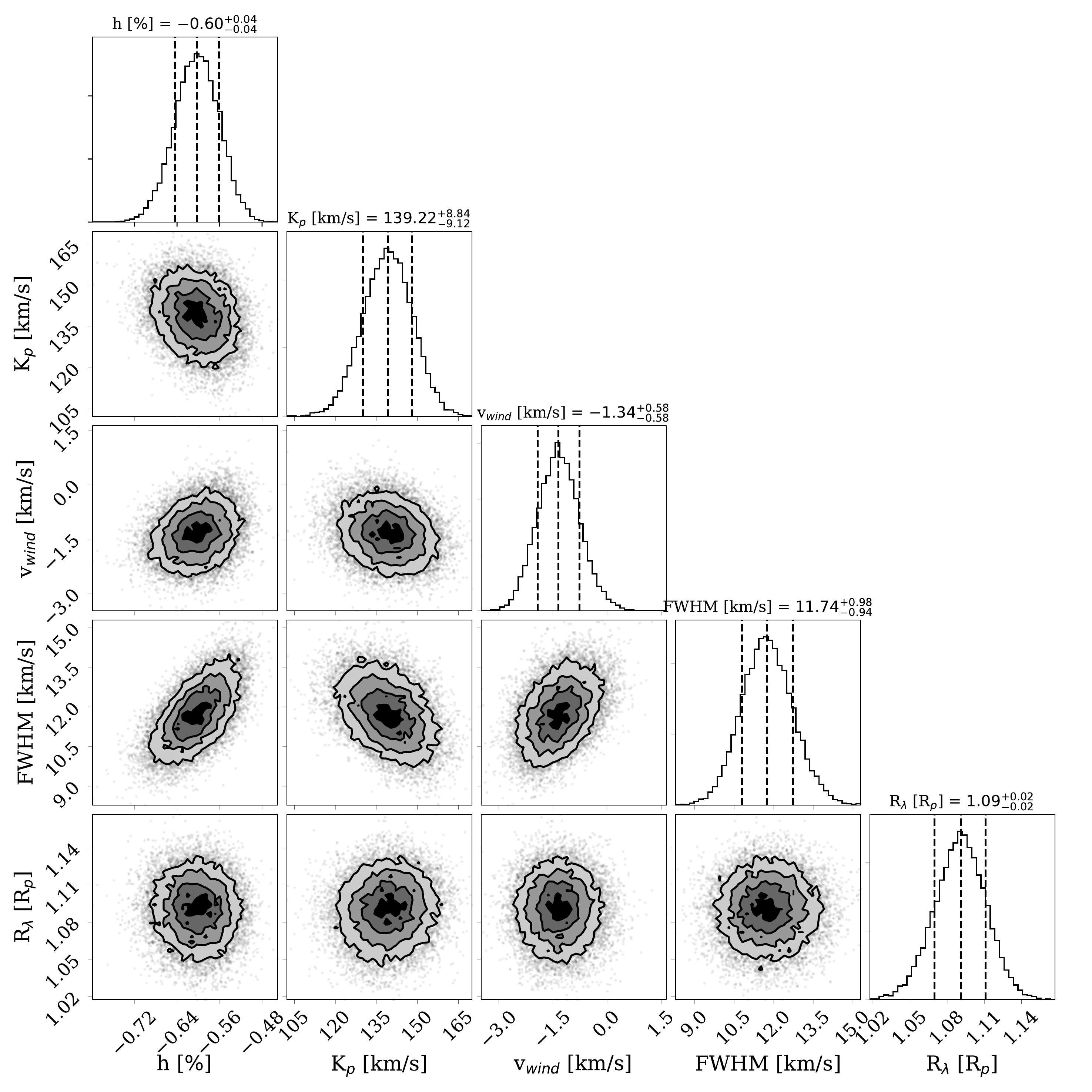}
\includegraphics[width=0.45\textwidth]{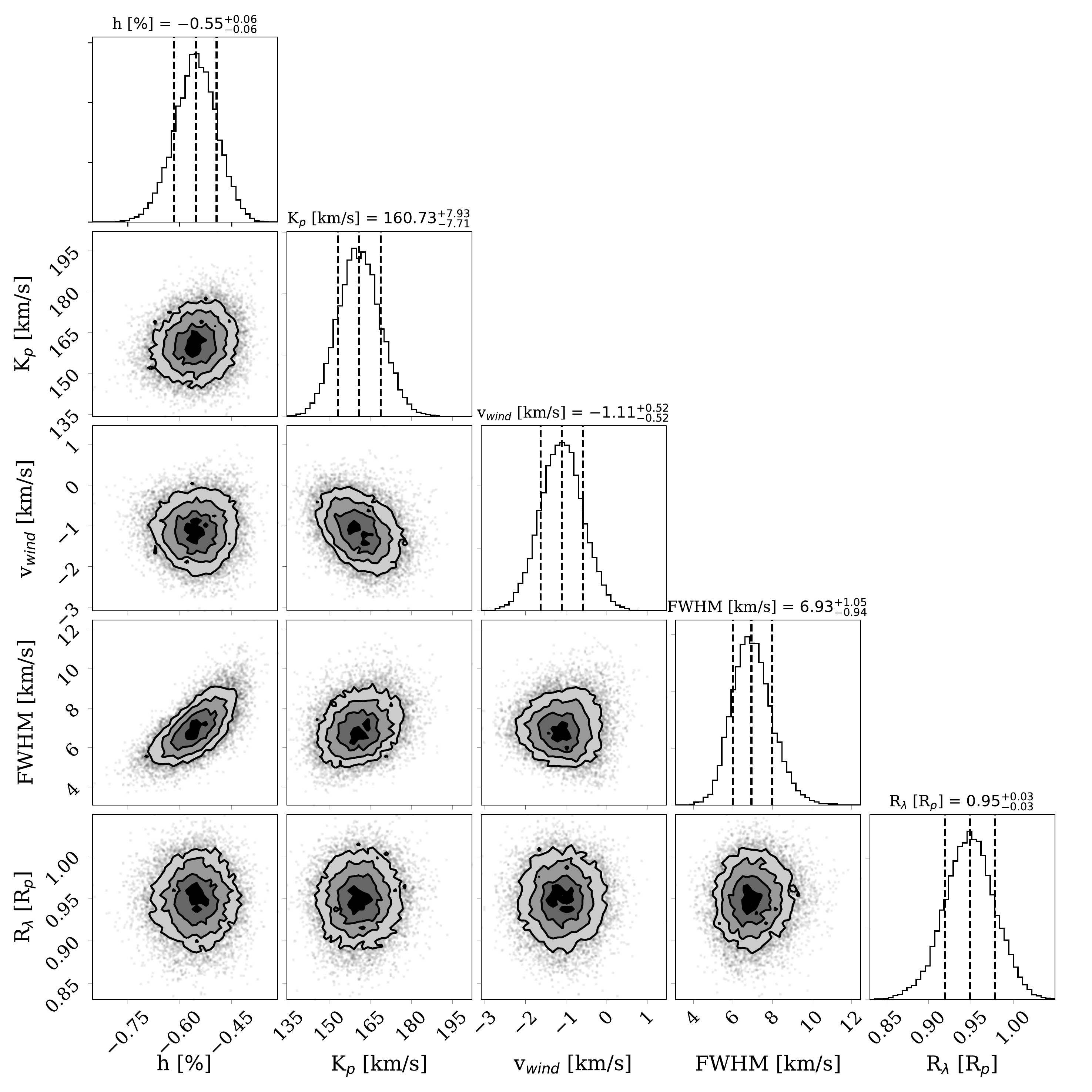}
\caption{Correlation diagrams of MCMC analysis of the CaII triplet lines of CARMENES. The CaII $\lambda8498~\mathrm{\AA}$ diagram is shown on top left, the CaII $\lambda8542~\mathrm{\AA}$ on top right, and the $\lambda8662~\mathrm{\AA}$ on bottom.}
\label{fig:mcmc_Ca_indiv}
\end{figure}

\newpage
\subsection{NaI}
\label{ap:cornNa}

\begin{figure}[h]
\centering
\includegraphics[width=0.45\textwidth]{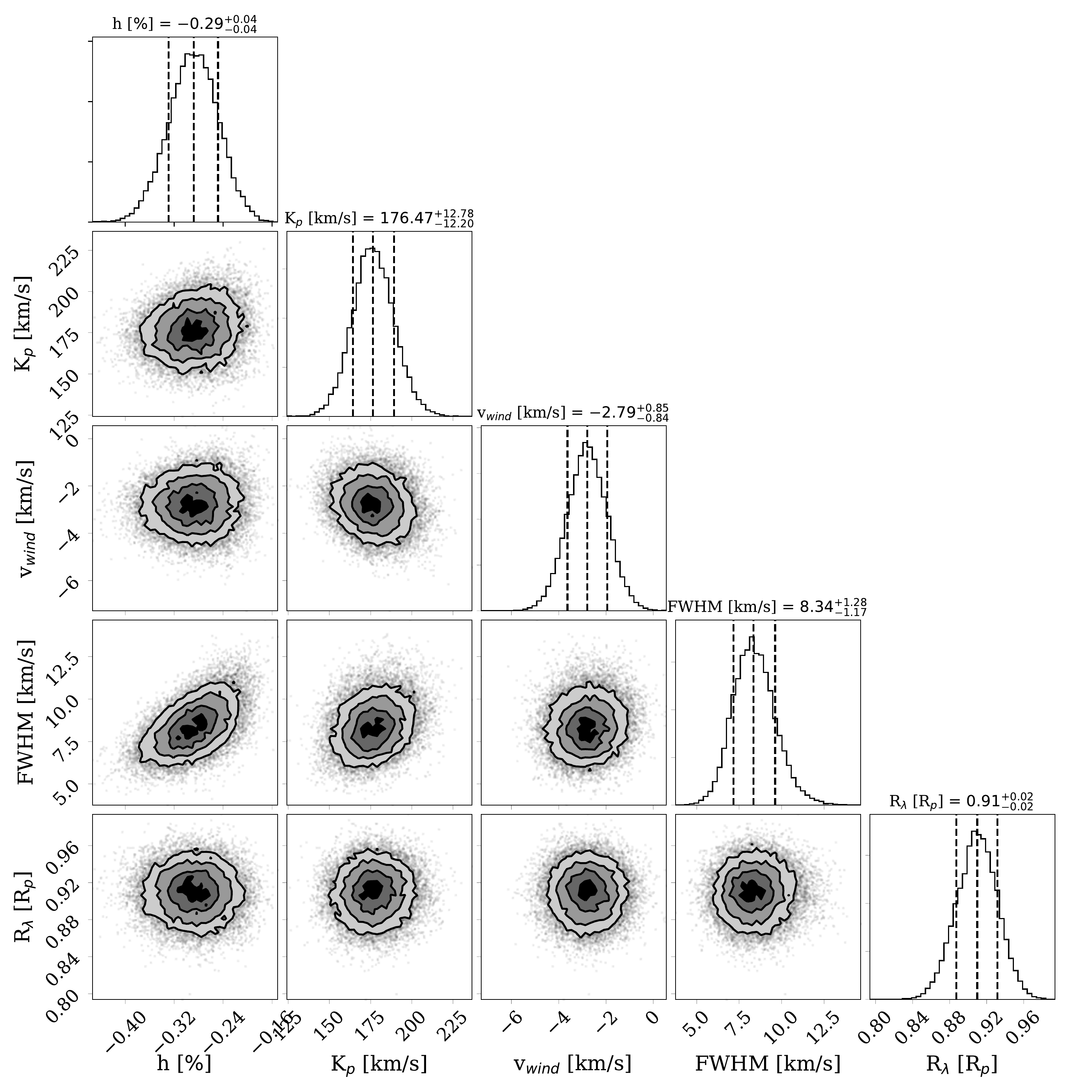}
\includegraphics[width=0.45\textwidth]{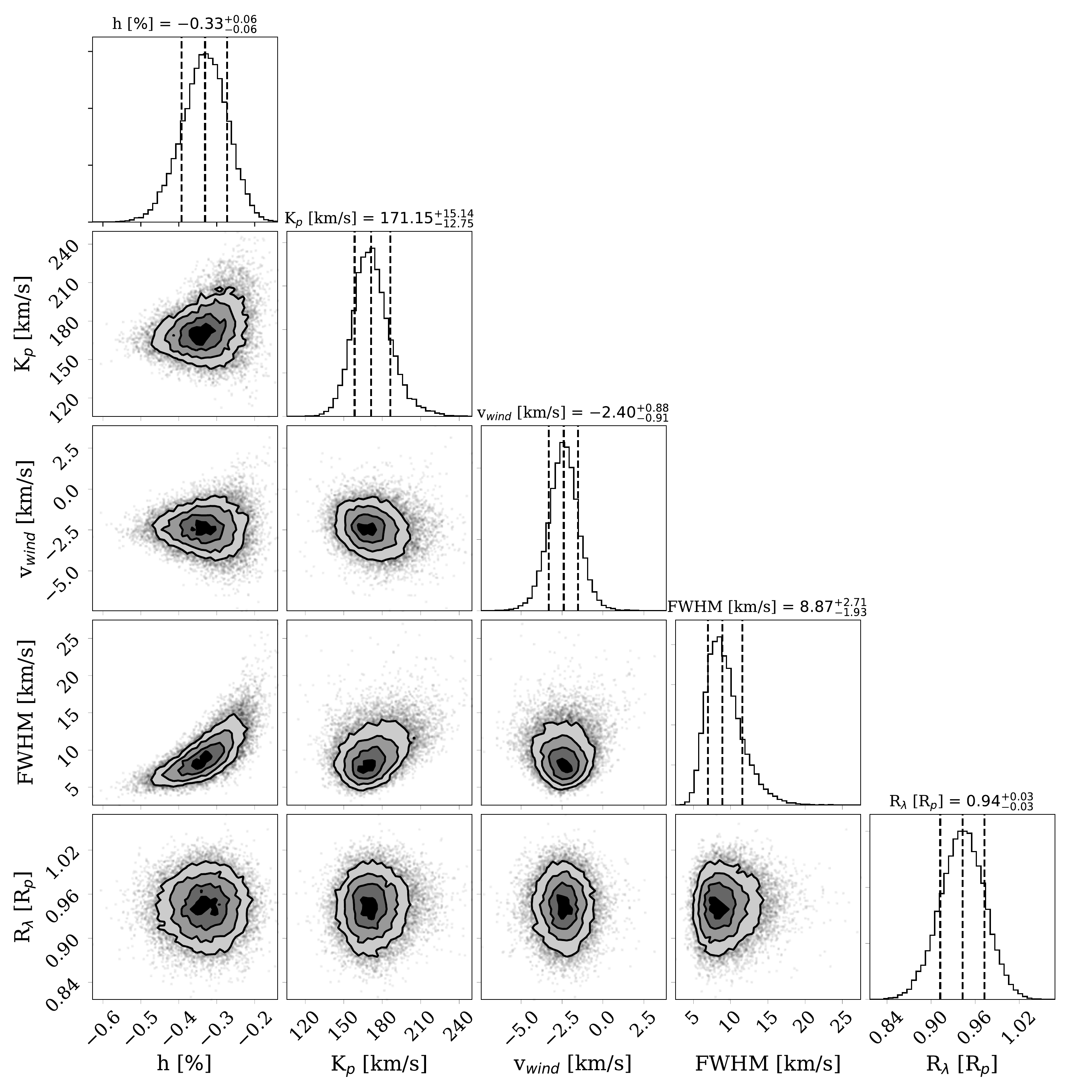}
\caption{Correlation diagrams of MCMC analysis of the NaI D$_2$ line of CARMENES (left) and HARPS-N data analysis (right).}
\label{fig:mcmc_NaD2_indiv}
\end{figure}

\begin{figure}[h]
\centering
\includegraphics[width=0.45\textwidth]{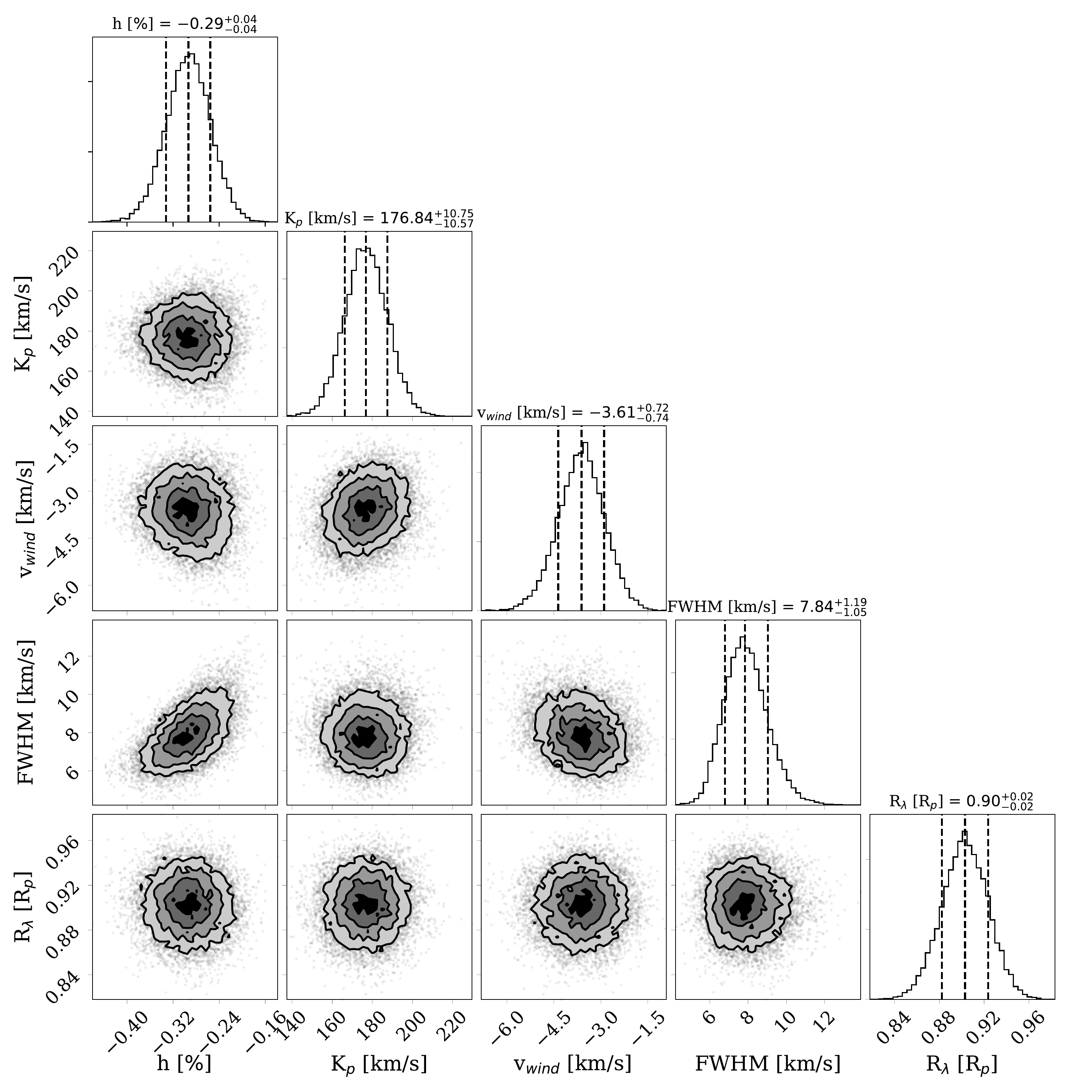}
\includegraphics[width=0.45\textwidth]{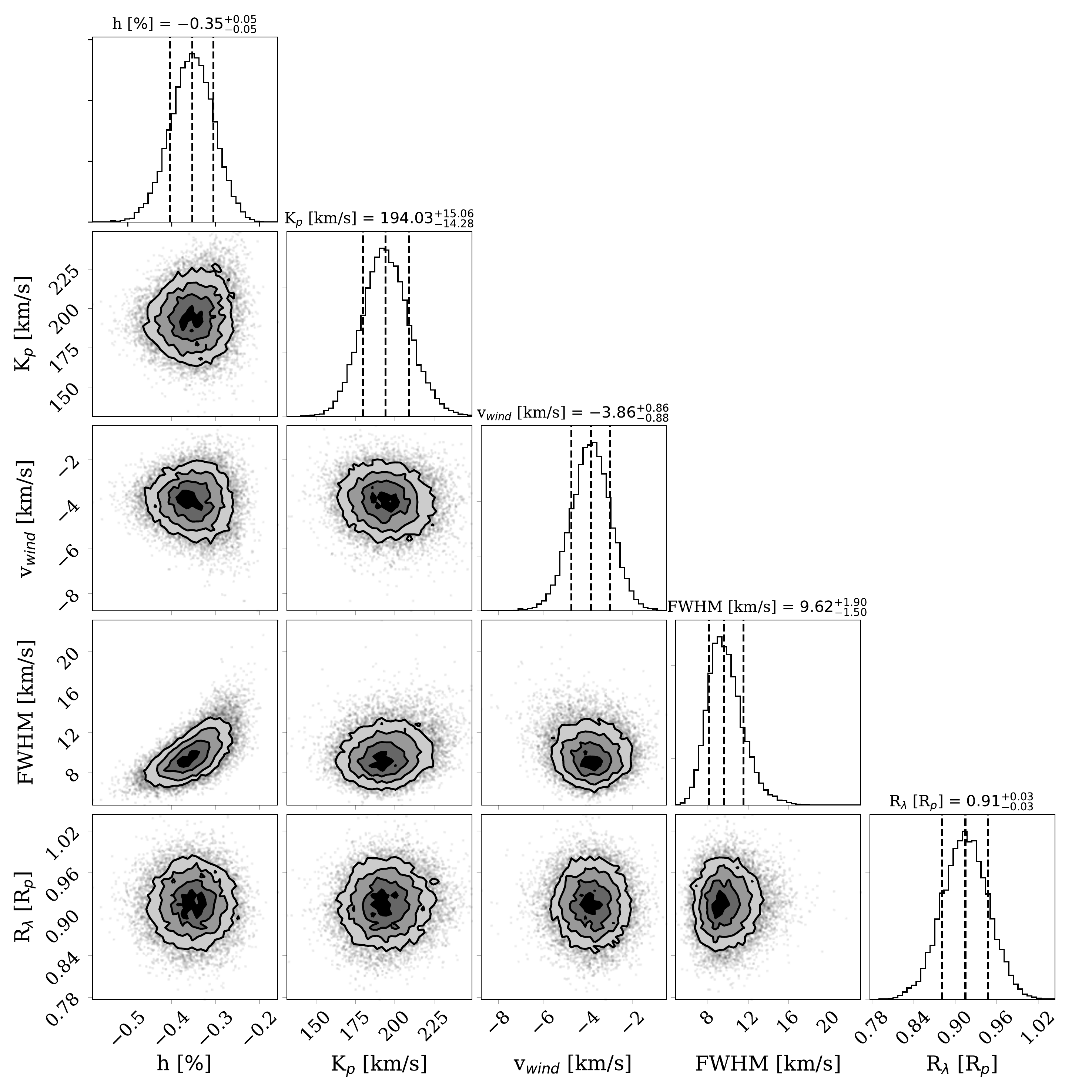}
\caption{Correlation diagrams of MCMC analysis of the NaI D$_1$ line of CARMENES (left) and HARPS-N data analysis (right).}
\label{fig:mcmc_NaD1_indiv}
\end{figure}

\newpage
\subsection{FeII}
\label{ap:cornCa}

\begin{figure}[h]
\centering
\includegraphics[width=0.45\textwidth]{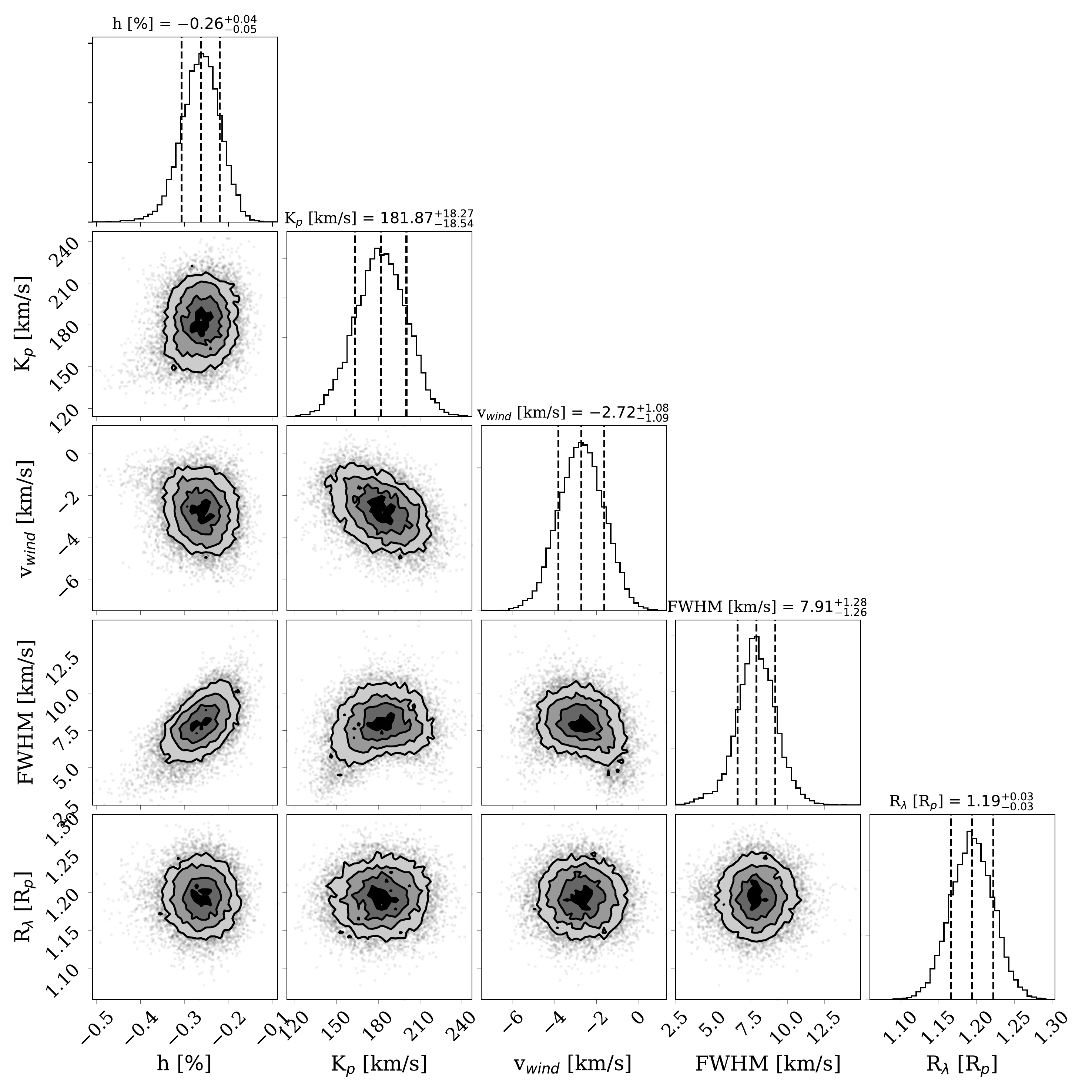}
\includegraphics[width=0.45\textwidth]{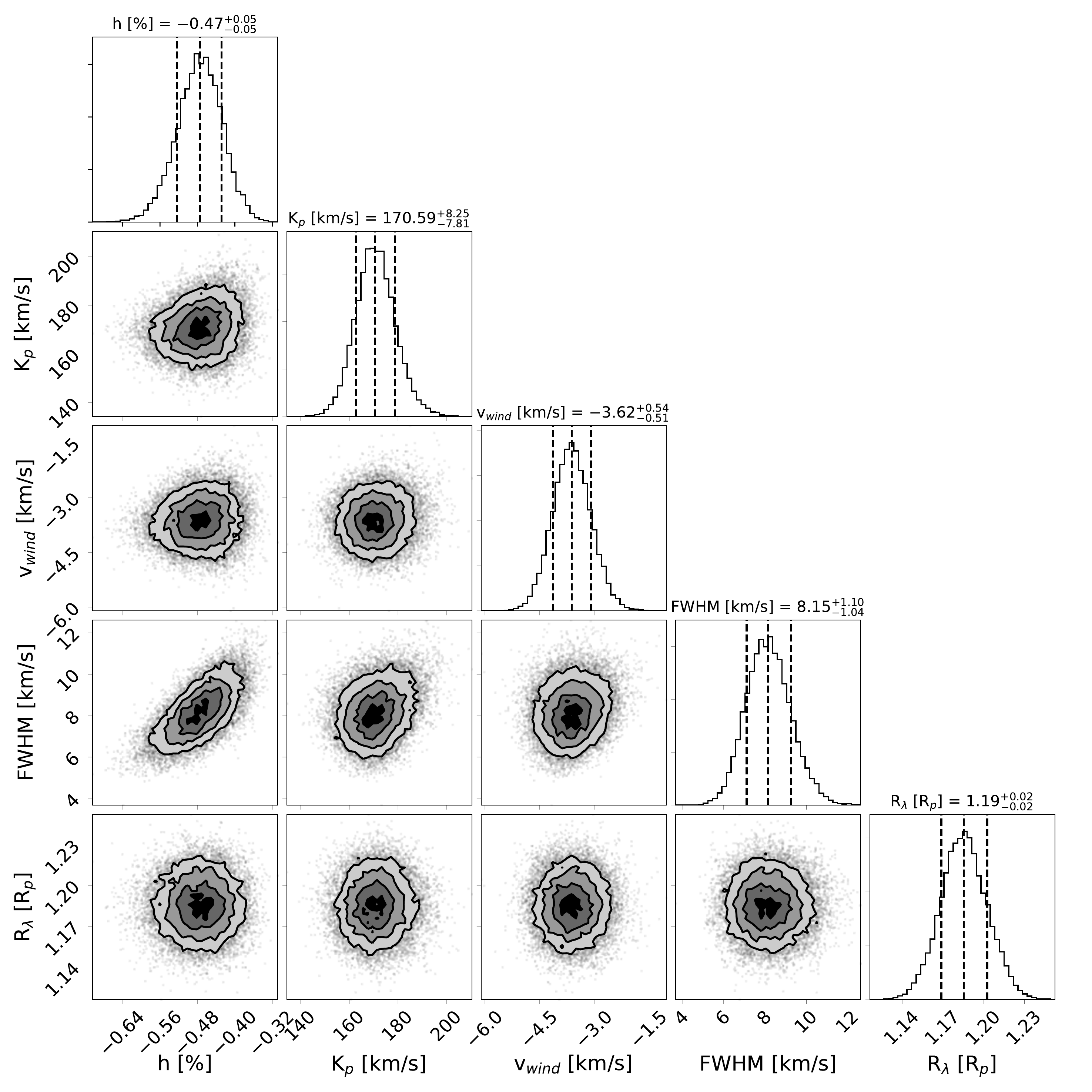}
\includegraphics[width=0.45\textwidth]{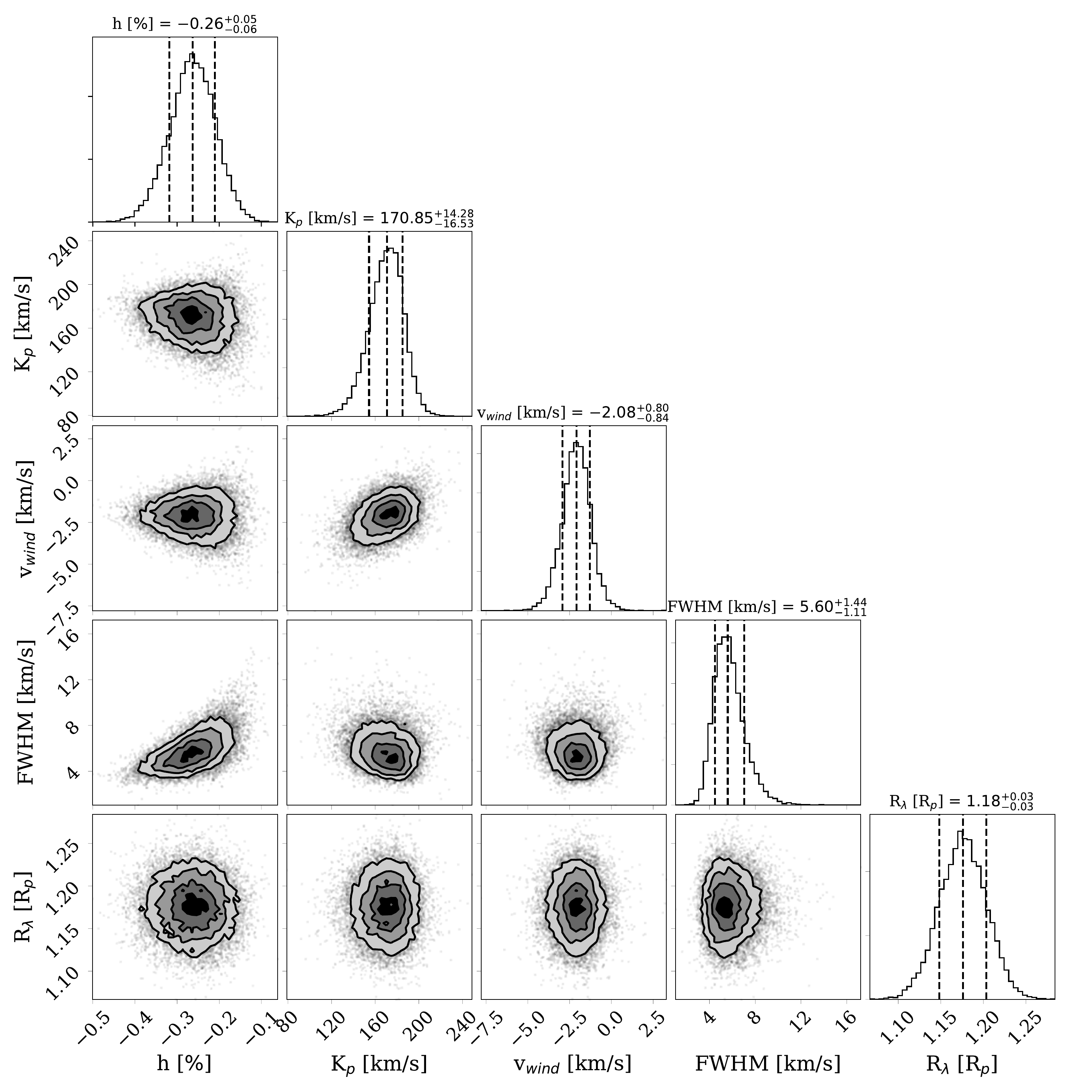}
\includegraphics[width=0.45\textwidth]{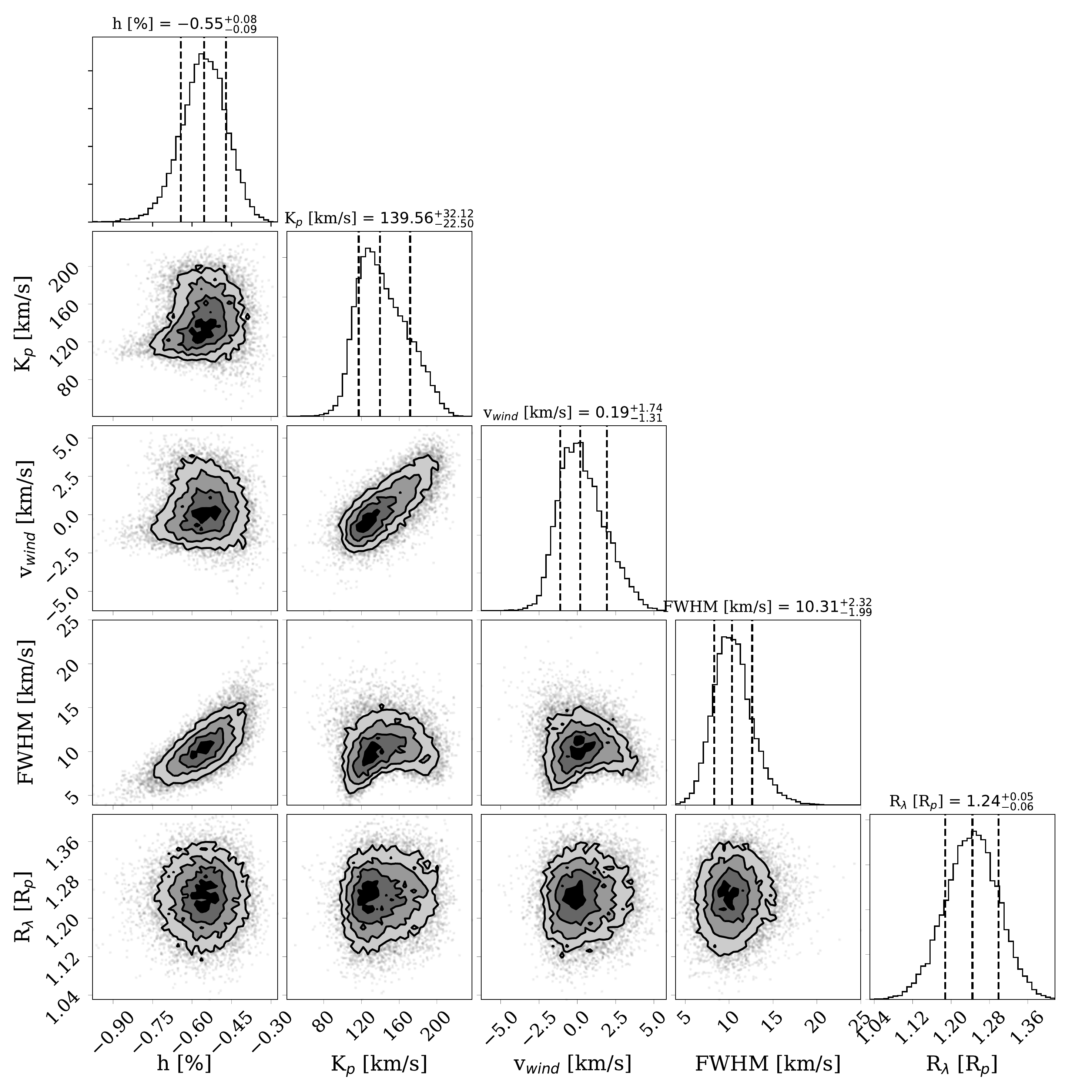}
\caption{Correlation diagrams of MCMC analysis of the FeII triplet lines. The FeII $\lambda5018~\mathrm{\AA}$ diagram is shown on top left and the FeII $\lambda5169~\mathrm{\AA}$ on top right, both result from the HARPS-N combined data analysis. The FeII $\lambda5316~\mathrm{\AA}$ diagrams are shown in the bottom panels: the HARPS-N analysis on the left and CARMENES on the right.}
\label{fig:mcmc_Ca_indiv}
\end{figure}

\newpage 
\subsection{MgI}
\label{ap:cornCa}

\begin{figure}[h]
\centering
\includegraphics[width=0.45\textwidth]{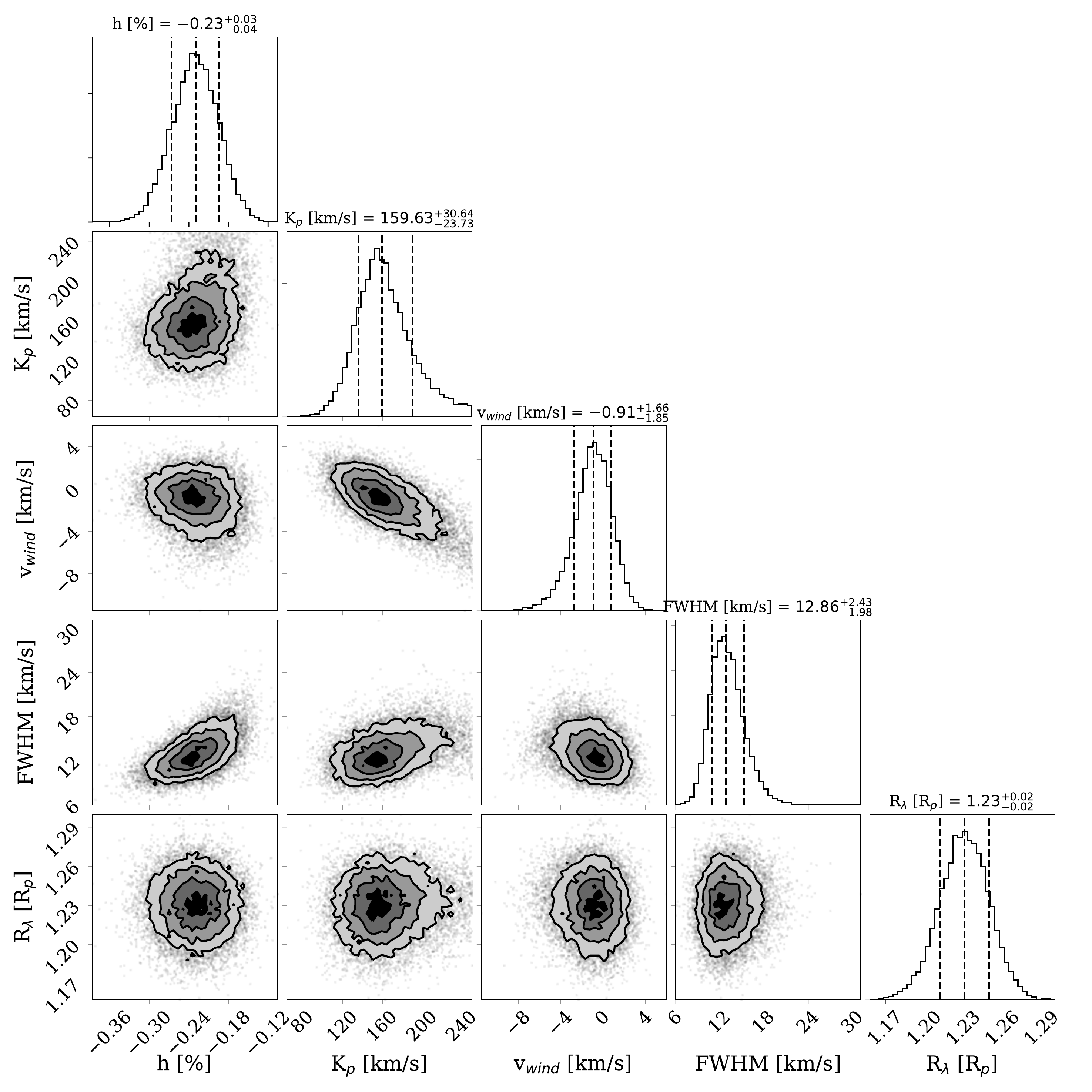}
\caption{Correlation diagrams of MCMC analysis of the MgI line at $5173~\mathrm{\AA}$.}
\label{fig:mcmc_Ca_indiv}
\end{figure}

\newpage 
\section{Empirical Monte-Carlo distributions}

\begin{figure*}[h]
\centering
\includegraphics[width=0.99\textwidth]{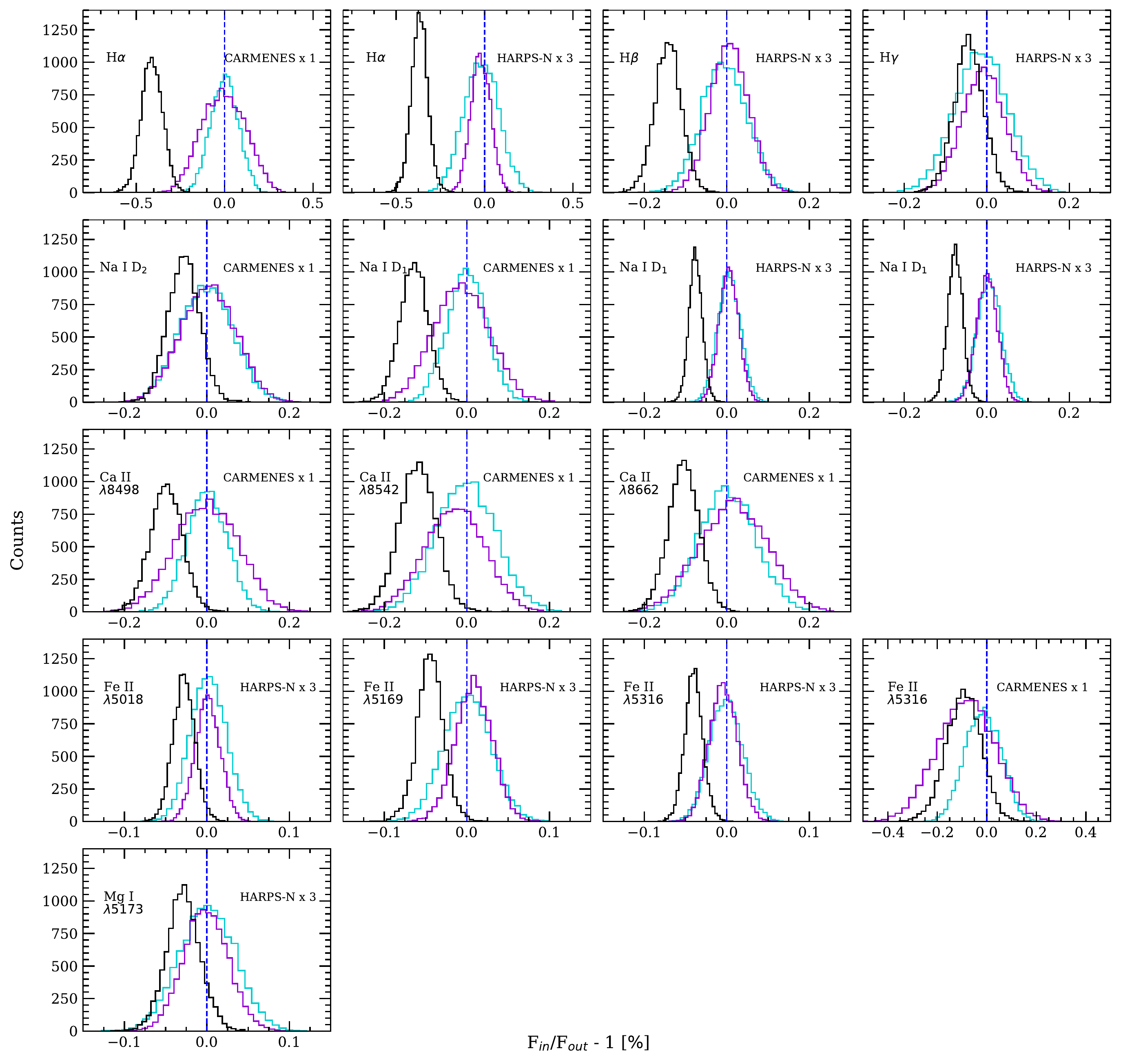}
\caption{Distributions of the Empirical Monte-Carlo analysis of the different lines, for a $1.5~\mathrm{\AA}$ passband. Each individual panel corresponds to the analysis of one line and one instrument, which are detailed in text. In violet we present the "out-out" scenario, in cyan the "in-in" scenario and in black the "in-out" scenario. The blue dashed vertical line shows the null absorption level. The CARMENES and HARPS-N analysis have different number of spectra, which means that the distributions will have different number of counts. This also happens for the three scenarios where the number of spectra considered is also different.}
\label{fig:hist}
\end{figure*}

\end{appendix}
\end{document}